\documentclass[12pt]{article}
\usepackage{a4wide}
\usepackage{amssymb}
\usepackage{graphicx}

\begin{document}
{\renewcommand{\thefootnote}{\fnsymbol{footnote}}
\medskip
\begin{center}
{\LARGE  Quantum Cosmology: Effective Theory}\\
\vspace{1.5em}
Martin Bojowald\footnote{e-mail address: {\tt bojowald@gravity.psu.edu}}
\\
\vspace{0.5em}
Institute for Gravitation and the Cosmos,
The Pennsylvania State
University,\\
104 Davey Lab, University Park, PA 16802, USA\\
\vspace{1.5em}
\end{center}
}

\setcounter{footnote}{0}

\newcommand{\lP}{\ell_{\mathrm P}}
\newcommand{\md}{{\mathrm{d}}}
\newcommand*{\R}{{\mathbb R}}
\newcommand*{\N}{{\mathbb N}}
\newcommand*{\Z}{{\mathbb Z}}
\newcommand*{\Q}{{\mathbb Q}}
\newcommand*{\C}{{\mathbb C}}

\begin{abstract}
  Quantum cosmology has traditionally been studied at the level of
  symmetry-reduced minisuperspace models, analyzing the behavior of wave
  functions. However, in the absence of a complete full setting of quantum
  gravity and detailed knowledge of specific properties of quantum states, it
  remained difficult to make testable predictions.  For quantum cosmology to
  be part of empirical science, it must allow for a systematic framework in
  which corrections to well-tested classical equations can be derived, with
  any ambiguities and ignorance sufficiently parameterized. As in particle and
  condensed-matter physics, a successful viewpoint is one of effective
  theories, adapted to specific issues one encounters in quantum
  cosmology. This review presents such an effective framework of quantum
  cosmology, taking into account, among other things, space-time structures,
  covariance, the problem of time and the anomaly issue.\footnote{Topical
    Review {\em Class. Quantum Grav.} 29 (2012) 213001}
\end{abstract}


\tableofcontents

\section{Introduction}

If quantum cosmology is ever to be part of empirical science, it must be
described by a good effective theory. There is no hope of exactly solving its
equations in realistic models or to tame conceptual quantum issues made even
more severe in the context of cosmology. The derivation of testable
predictions requires systematic approximations at semiclassical order and
beyond, and by experience with other areas of physics, effective actions or
equations are the best available tools.

While the speculative part of quantum cosmology, addressing for instance the
Planck regime or the status of multiverses, requires all subtleties of quantum
physics to be considered --- such as choices of Hilbert spaces,
self-adjointness properties of Hamiltonians or unitarity of evolution, an
understanding of deep conceptual issues and the measurement problem ---
physics can and must proceed without all these problems being
solved.\footnote{Especially in the loop-quantum-gravity community with its
  long and proud history of mathematical-physics primacy, it is sometimes
  said that one must become ``less rigorous'' in order to find useful and
  interesting physical results. This statement is, of course, incorrect; one
  does not become a physicist by being a mathematician first and then turning
  a little less rigorous. Physics requires as much rigor as mathematics, but a
  different kind of rigor.} Compare this situation for instance with quantum
field theory, for which no rigorous interacting and non-integrable version is
known. And yet, at the effective level it is the key tool behind the success
of elementary particle physics. Given the immensity of the Planck scale,
potentially observable effects in quantum cosmology are realized at low
energies where semiclassical quantum gravity, with the first few orders in
$\hbar$ taken into account, suffices. This feature makes effective theory in
quantum gravity and cosmology even more powerful than in other settings
\cite{EffectiveGR,BurgessLivRev}. As we will see in the course of this review,
somewhat surprisingly, even conceptual problems of quantum gravity can
advantageously be addressed with effective methods, especially with an
extension to effective constraints.

Given the amount of research on effective theories and their applications, one
may think that deriving an effective theory of quantum cosmology is a simple
and well-understood problem. However, this is not at all the case. Quite
general and powerful techniques of effective actions or potentials do exist,
employed with great success in particle physics and condensed-matter physics
alike. Quantum cosmology, on the other hand, is unique by virtue of several
features. It requires aspects of effective equations not encountered in other
fields, related for instance to the prevalence of canonical methods, the
generally covariant setting lacking evolution by a unique time parameter, or
the absence or inapplicability of non-perturbative ground states or other
distinguished classes of states. These technical problems will be discussed in
due course. For now, as a motivation of our detailed look into effective
theory, we state the following two general problems by which quantum cosmology
differs from other fields.

First, quantum cosmology is much like condensed-matter physics, with
microscopic quantum degrees of freedom manifesting themselves on length scales
far larger than their own. While the precise nature of microscopic degrees of
freedom (strings, loops, \ldots) remains unclear and disputed among the
different approaches, their presence in some form is widely agreed upon. By
considering large structures or space-time regions made from many building
blocks, quantum cosmology differs significantly from most situations
encountered in elementary-particle physics, where events with a comparatively
small number of particles in excited states close to the vacuum are
studied. Quantum cosmology is a many-body problem, a situation in which it is
difficult to derive and justify valid effective descriptions. The effective
view has been put to good use in condensed-matter physics, but only thanks to
rich and merciless experimental input to weed out wrong ideas and stimulate
new successful ones. Quantum cosmology is not (yet?) subject to experimental
pressure, and many (good and bad) ideas are sprouting. Effective cosmological
theory must be able to stand on its own, requiring a systematic and rigorous
formulation taking into account all features and consistency conditions to be
imposed in quantum gravity.

As the second problem, we observe that the theoretical foundation of quantum
cosmology is much weaker than that of condensed-matter physics to which it is
otherwise quite close. We know well which Hamiltonian we should use to find
all states and energies of excitations in a crystal, but mathematically the
problem is challenging and calls for the approximations of effective
theory. In quantum cosmology, we don't even know which precise Hamiltonian or
other underlying object to use for the dynamics of a universe. Even if we
choose one particular approach to quantum gravity, its mathematical objects or
its specializations to cosmology are incompletely known or understood, opening
wide the door for ambiguities and spurious constructions.

We need a well-understood theory to pinpoint places where best to look for
observational effects, and we need observations to guide our theoretical
constructions. Quantum cosmology, with an incompletely understood theory and
no current observations, is a slippery subject, depriving us of a good handle
to grasp its implications. In the absence of experiments, we can only rely on
conceptual arguments and internal mathematical consistency conditions which,
however, come along with their own problems. In this situation, effective
techniques have proven to be one of the few reliable approaches for physical
evaluations of the theory, allowing one to include all crucial quantum effects
in equations with clear physical meaning, and to take into account ambiguities
by sufficiently general parameterizations. This general framework and its
current status in the context of quantum cosmology are the topics of this
review.\footnote{Fundamental issues of quantum cosmology are reviewed in the
  companion article \cite{QCFund}.}

\section{Cosmological consistency}

One source of technical and conceptual problems in quantum cosmology and
quantum gravity in general is that the theory deals with relativistic
space-time, in the absence of a unique Hamiltonian to generate evolution in
time. Instead, many choices for time and corresponding Hamiltonians or
evolution equations are possible, and they must all lead to the same
physics. While this invariance is guaranteed classically, it implies a
complicated problem after quantization, presenting the strongest set of
consistency conditions to restrict possible choices of quantum
cosmologies. Unfortunately, these issues are often set aside in research on
quantum cosmology and even quantum gravity, owing to their complicated
nature. This intentional oversight implies a large number of ambiguities,
fixed in those contexts only by ad-hoc constructions. To keep this review
focused, we will not discuss the rather large body of works in such
directions, for instance those crucially using the distinction of a time
variable by gauge-fixing or deparameterization in canonical settings, and only
mention shortcomings in contexts in which they become apparent. Generally
speaking, results derived with a distinguished choice of time (or gauge)
cannot be considered physically reliable unless one can make sure that they do
not depend on one's choice of time.

In this section, we will discuss the main features that an effective theory of
quantum cosmology must deal with, which includes covariance and state
properties. The former ensures independence of choices of time, the latter
deals with additional freedom in quantum theories. By these considerations, we
will be guided toward suitable ingredients for the mathematical formulation of
effective theory.

\subsection{Covariance}
\label{s:Cova}

Effective equations of quantum cosmology are supposed to modify some
cosmological version of Einstein's equation by quantum corrections. In most
cases, such a cosmological version is a reduction to isotropic
Friedmann--Lema\^{\i}tre--Robertson--Walker space-times or homogeneous Bianchi
models (minisuperspaces), a restriction to some specific form of
inhomogeneous degrees of freedom such as Lema\^{\i}tre--Tolman--Bondi or Gowdy
geometries (midisuperspaces), or an inclusion of unrestricted but perturbative
inhomogeneity around a background in the former classes of models.

\subsubsection{Homogeneous models and automatic consistency}

In homogeneous models, the dynamics is completely determined by one equation
for gravitational degrees of freedom (such as the Friedmann equation) and one
for matter (such as the continuity equation). The Friedmann equation of
isotropic models,
\begin{equation} \label{Friedmann}
 \left(\frac{\dot{a}}{a}\right)^2 +\frac{k}{a^2}= \frac{8\pi G}{3}\rho\,,
\end{equation}
depends only on first-order derivatives and is therefore a constraint, to be
satisfied by initial values of second-order equations of motion. When
interpreted as a constraint (the Hamiltonian constraint), it is usually
written in the form of an energy-balance law:
\begin{equation} \label{constraint}
H:= -\frac{3}{8\pi G}(\dot{a}^2a+ka)+ E_{\rm matter}=0
\end{equation}
with the matter energy $E_{\rm matter}=\rho a^3$ contained in some region of
unit coordinate volume. (See \cite{Springer} for a detailed discussion of
coordinate factors when the volume is not fixed, especially in the context of
quantization.) In this form, the Hamiltonian constraint of gravity is obtained
by varying the action by the lapse function $N=\sqrt{-g_{00}}$. (For more on
constraints and canonical gravity, see \cite{CUP}.)

Given the Friedmann equation and the continuity equation 
\begin{equation}
 \dot{\rho}+3\frac{\dot{a}}{a}(\rho+P)=0\,,
\end{equation}
with pressure $P$, one can derive a second-order equation of motion by taking
a time derivative of (\ref{Friedmann}) and eliminating $\dot{\rho}$: the
Raychaudhuri equation
\begin{equation}
 \frac{\ddot{a}}{a} = -\frac{4\pi G}{3} (\rho+3P)\,.
\end{equation}
At this stage, we have all equations expected from the components of the
isotropic Einstein tensor: the time-time component providing the Friedmann
equation, the (identical) diagonal components of the spatial part amounting to
the Raychaudhuri equation, and all off-diagonal components vanishing
identically. The equations obtained are automatically consistent with each
other: By construction, the time derivative of the Friedmann equation vanishes
if the Raychaudhuri equation holds. Therefore, if the constraint imposed by
the Friedmann equation holds for initial values at some time, it holds at all
times. 

This latter property is realized for a large class of systems describing
versions of isotropic (or homogeneous) cosmology, not just for the classical
one resulting from Einstein's equation. As a phase-space function, $H(a,p_a)$
in (\ref{constraint}), with the momentum $p_a= -3(4\pi G)^{-1} a\dot{a}$ as it
follows from the variation $\partial L_{\rm grav}^{\rm iso}/\partial \dot a $
of the Einstein--Hilbert action reduced to isotropy, plays the role of the
Hamiltonian generating all equations of motion in proper time. The
Raychaudhuri equation indeed follows from the Hamiltonian equations of motion
$\dot{a}=\partial H/\partial p_a=\{a,H\}$ and $\dot{p}_a=-\partial H/\partial
a=\{p_a,H\}$ (the first of which is identical to the definition of the
momentum $p_a$). The matter terms $\rho$ and $P$ are realized by 
\begin{equation}
 \rho=\frac{E_{\rm  matter}}{a^3}\quad\mbox{ and }\quad 
 P=-\frac{1}{3a^2}\frac{\partial E_{\rm matter}}{\partial a}\,,
\end{equation}
the negative change of energy by volume change. Matter equations of motion
follow once $E_{\rm matter}$ in (\ref{constraint}) is expressed in terms of
canonical degrees of freedom.

The phase-space function $H(a,p_a)$ itself evolves according
to the same Hamiltonian law, $\dot{H}(a,p_a)=\{H(a,p_a),H\}=0$ and is
automatically constant in time, no matter what form $H(a,p_a)$ has. In
homogeneous situations, the single Hamiltonian constraint that determines
evolution is automatically preserved and consistent with evolution
equations. We can easily modify $H$ by any form of quantum corrections without
encountering consistency problems, issuing a powerful license to cosmological
model builders.

Consistency remains valid when we consider different choices of time. So far,
the equations were in proper time $\tau$. All other choices $t$ in homogeneous
models are related to $\tau$ by $\tau(t)=\int^{t} N(t'){\rm d}t'$, with a
lapse function $N$ that enters Hamiltonian equations as well: If ${\rm
  d}f(a,p_a)/{\rm d}\tau=\{f,H\}$, we have 
\[
 \frac{{\rm d}f(a,p_a)}{{\rm d} t} = \frac{{\rm
  d}\tau}{{\rm d}t}\frac{{\rm d}f(a,p_a)}{{\rm d}\tau} =N\{f,H\}\approx \{f,N
H\}
\]
for any other time with ${\rm d}\tau/{\rm d}t=N$.  In the last step, we are
allowed to pull the lapse function $N$ inside the Poisson bracket, keeping the
equality satisfied as a ``weak'' one, one that is valid provided the
constraint $H=0$ holds. Applying the general law to the constraint itself, we
again observe consistency: ${\rm d}H/{\rm d}t= \{H,NH\}\approx 0$.

The Hamiltonian constraint generates not only evolution with respect to a
given time choice, but also the transition between different choices as a
gauge transformation. Infinitesimally, with $N$ close to one, we have $\tau=
t+\epsilon$ with $\epsilon=\int (N-1){\rm d}t$, and for any function $f$,
\begin{equation}
 \delta_{\epsilon}f= f(\tau)-f(t)=\epsilon \frac{{\rm d}f}{{\rm d}t}
=\epsilon \{f,H\}\approx \{f,\epsilon
H\}\,.
\end{equation}
The Friedmann equation, amounting to the Hamiltonian constraint, is
automatically invariant under changes of time, and so are its solutions. Also
the evolution equations are invariant if we use the Jacobi identity for
Poisson brackets: 
\begin{eqnarray*}
 \left\{\frac{{\rm d}f}{{\rm d}t},\epsilon H\right\}&=& \{\{f,NH\},\epsilon H\}=
\{\{f,\epsilon H\},NH\}+ \{f,\{NH,\epsilon H\}\}\\
&=& \frac{{\rm d}\{f,\epsilon H\}}{{\rm
  d}t}- \{f,({\rm d}\epsilon/{\rm d}t) H\}+
\{f,(\delta_{\epsilon}N)H\}\,.
\end{eqnarray*}
Rearranging, we see that the gauge-transformed
$f$ evolves in agreement with the gauge-transformed ${\rm d}f/{\rm d}t$, with
a correction taking into account a possible gauge transformation of $N$ and
time dependence of $\epsilon$.  

The presence of a single Hamiltonian constraint therefore ensures dynamical
consistency and invariance, even if quantum modifications occur. For this
reason, homogeneous minisuperspace models are a simple and popular tool to
investigate possible consequences of quantum gravity and cosmology. But for
the very same reason, extreme care must be exercised when such models are used
for physical predictions: The trivialization of consistency conditions does
not hold in a more general context, and therefore spurious results can easily
be produced in their absence.

\subsubsection{Inhomogeneity and covariance}

Compared with minisuperspace models, inhomogeneous cosmology presents a very
different situation regarding consistency, even if inhomogeneity is small and
treated perturbatively. For gravity and matter to fit together in Einstein's
equation or a quantum modification thereof, a version of the contracted
Bianchi identity must hold. But this identity, relating different types of
dynamical equations, is easily destroyed if quantum corrections, for instance
those found in minisuperspace models, are inserted blindly. When amending
inhomogeneous equations by quantum corrections, one must face the problem of
anomaly freedom or covariance. Certain relations between different dynamical
equations and gauge generators, classically implemented by the contracted
Bianchi identity, must be preserved in the presence of quantum
corrections. 

This well-known classical fact is often overlooked in quantum treatments,
especially those making use of gauge fixing or deparameterization. Gauge
generators disappear when the gauge is fixed. A simple but illegitimate way
out of difficult consistency problems is therefore to fix the gauge before
quantum corrections are inserted. However, one then dispenses with ways to
check consistency and cannot be sure that results obtained are physically
viable. Most cases in which cosmological perturbations have been computed by
gauge fixing in loop quantum cosmology, for instance, have by now been shown
to be incorrect; important effects such as signature change have been
overlooked. See also the instructive discussion of \cite{GaugeInvTransPlanck}
in this context.  We will come back to this issue later and for now continue
with a classical discussion to provide more details.

The first implication of the Bianchi identity is the presence of
constraints. If we write $\nabla_{\nu}G^{\nu}_{\mu}=0$ in the form
\begin{equation} \label{ContractedSplit}
\partial_0G^0_{\mu}=
-\partial_aG^a_{\mu}-\Gamma^{\nu}_{\nu\kappa}G^{\kappa}_{\mu}+
\Gamma^{\kappa}_{\nu\mu}G^{\nu}_{\kappa}\,,
\end{equation}
with spatial indices ``$a$,'' it becomes evident that the components
$G^0_{\mu}$ of the Einstein tensor cannot contain second-order time
derivatives: On the right-hand side, all factors in the three terms are at
most second order in time, and there is one explicit time derivative on the
left-hand side, leaving only the option of first time derivatives in
$G^0_{\mu}$. These components of the Einstein tensor (minus $8\pi G$ times the
corresponding stress-energy components $T_{\mu}^0$ if there is matter) are
constraints on initial values, while the remaining components provide
evolution equations. In contrast to minisuperspace models, we are dealing with
a larger constrained system of four independent and functional constraints,
the Hamiltonian constraint $H=G^0_0-N^aG_a^0$ and the diffeomorphism
constraint $D=NG^0_a$, which are to be imposed pointwise or for all possible
multiplier functions $N$ and $N^a$ in
\begin{equation} \label{Constraints}
H[N]=-\int{\rm d}^3x N(G^0_0-N^aG^0_a)\quad\mbox{ and }\quad D[N^a]=
-\int{\rm   d}^3x N^a N G^0_a\,.
\end{equation}
We are dealing with an infinite number of constraints. (To define these
integrations, one introduces a foliation of space-time into spatial surfaces
$t={\rm const}$, also used to set up canonical variables. The time direction
$t^a$ at each point, used to define time derivatives in evolution equations,
may be different from the normal direction $n^a$ to spatial slices in
space-time, a freedom parameterized as $t^a=Nn^a+N^a$ with the lapse function
$N$ and the shift vector field $N^a$. The linear combinations of $G^0_0$ and
$G^0_a$ in (\ref{Constraints}), depending on the shift $N^a$ and lapse $N$,
take into account that constraints refer to directions normal and tangential
to spatial slices, not to coordinate directions such as the zero-index of the
Einstein tensor for a component along the time-evolution vector field $t^a$.)

As before, for consistency the constraints must always hold provided they are
imposed for initial values, a feature that is guaranteed by the Bianchi
identity as well. If we combine the Einstein tensor and the stress-energy
tensor $T_{\mu\nu}$ of matter in (\ref{ContractedSplit}), we see that
$\partial_0(G^0_{\mu}- 8\pi GT^0_{\mu})$ vanishes at any time provided the
constraints $G_{\mu}^0-8\pi G T_{\mu}^0$ themselves (and therefore their
spatial derivatives) vanish at that time and the evolution equations hold. By
virtue of the contracted Bianchi identity, the constraints are consistent with
evolution. Finally, as a third consequence, we see that all equations,
Einstein's equation and the contracted Bianchi identity as a consistency
condition, are covariant and independent of coordinates used. Therefore,
solutions will be covariant. All equations hold irrespective of the choice of
coordinates, a well-known feature in perturbative cosmology which allows one
to express all equations explicitly in terms of gauge-invariant variables
\cite{Bardeen,CosmoPert}.

As in our discussion of minisuperspace models, a canonical view is useful to
analyze which consistency conditions are satisfied automatically and which
ones are non-trivial. There is a Hamiltonian constraint, and therefore
Hamiltonian equations $\dot{f}=\{f,C\}$ are generated by a constraint
$C$. However, in the inhomogeneous context, the constraint is not unique (up
to a pre-factor $N$), nor is time evolution. We have a much larger choice of
possible time variables to generate evolution. The most general version of
equations of motion is obtained if we use all our constraints in a linear
combination, defining $H[N,N^a]:= H[N]+D[N^a]$. For fixed $N$ and $N^a$, the
Hamiltonian flow $\dot{f}=\{f,H[N,N^a]\}$ is then equivalent to Lie
derivatives $\dot{f}={\cal L}_tf$ along the time-evolution vector field
$t^a=Nn^a+N^a$ in space-time, foliated by spatial slices with unit normals
$n^a$.

For all constraints to be preserved by the evolution equations they generate,
we need $\{H[M,M^a],H[N,N^a]\}\approx 0$ for all $M$, $N$, $M^a$ and $N^a$. If
this condition is satisfied, the constraints are said to form a first-class
system. Unlike in homogeneous models, where $H[M]$ always commutes weakly with
$H[N]$, the general condition is highly non-trivial and provides strong
restrictions on consistent modifications of Einstein's equation. Quantum
corrections can no longer be inserted at will.

The same constraints that generate evolution provide gauge transformations,
classically equivalent to coordinate changes. We use the same general
combination as before, $H[\epsilon,\epsilon^a]$, but interpret the multipliers
$\epsilon$ and $\epsilon^a$ differently, not related to a time-evolution
vector field. Instead, the gauge transformation $\delta_{\epsilon^{\mu}}f=
\{f,H[\epsilon]+D[\epsilon^a]\}$ with the classical constraints is equivalent
to a coordinate transformation or the Lie derivative ${\cal L}_{\xi}f$ along
the space-time vector field $\xi^{\mu}$ with components such that
$\epsilon=N\xi^0$ and $\epsilon^a=\xi^a+N^a\xi^0$ \cite{LapseGauge}. The
factors of $N$ and $N^a$ again result because space-time coordinate changes
and the components of $\xi^{\mu}$ refer to coordinate directions, while
constraints refer to directions normal and tangential to spatial slices with
normal $n^a=N^{-1}(t^a-N^a)$. Also regarding gauge invariance, the condition
of a first-class constraint algebra is then sufficient for consistency: In
this case, all constraints are gauge invariant,
$\delta_{\epsilon}H[N,N^a]=\{H[N,N^a],H[\epsilon,\epsilon^a]\}\approx 0$, and
so are the evolution equations they generate. With this full set of gauge
transformations, one can freely change the constant-time spatial surfaces used
to define canonical variables and to integrate the constraints
(\ref{Constraints}). We are then dealing with a covariant theory of
space-time, not just with a theory on a fixed spatial foliation.

\subsubsection{Hypersurface-deformation algebra}

The crucial consistency condition for any classical constrained system, its
quantization, or an effective theory thereof, is therefore that it be first
class: all constraints $H[N,N^a]$ must have Poisson brackets that vanish when
the constraints are imposed,
\begin{equation}
 \{H[M,M^a],H[N,N^a]\}\approx 0\,.
\end{equation}
For gravity, or any generally covariant space-time theory, the specific
form is the hypersurface-deformation algebra \cite{DiracHamGR}
\begin{eqnarray}
 \{D[M^a],D[N^a]\}&=& D[{\cal L}_{N^b}M^a] \label{DD} \\
 \{H[M],D[N^a]\}&=& H[{\cal L}_{N^b}M] \label{HD}\\
 \{H[M],H[N]\}&=& D[q^{ab}(M\nabla_bN-N\nabla_bM)] \label{HH}
\end{eqnarray}
with the spatial metric $q^{ab}$. 

For a consistent quantization, an algebra of this form must be realized with
commutators for constraint operators instead of Poisson brackets, and an
effective constrained system must have quantum-corrected constraints such that
a first-class algebra holds with Poisson brackets. If this is realized, no
gauge transformations are broken by quantization and the quantum or effective
theory is called anomaly-free. If this condition is satisfied, all consistency
conditions that are classically implied by the contracted Bianchi identity
hold in the presence of quantum corrections, and the (quantum) theory
describes space-time rather than just a family of spatial slices. This
property may be achieved with exactly the same form of the algebra, or with
one that shows quantum corrections not just in the constraints but also in the
structure functions of the algebra, as long as two constraints still commute
up to another constraint. One universal example found in loop quantum gravity,
as the most prominent result regarding hypersurface deformations with quantum
corrections, has (\ref{DD}) and (\ref{HD}) unchanged, but (\ref{HH}) modified
to \cite{ConstraintAlgebra}
\begin{equation}
  \{H[M],H[N]\}= D[\beta q^{ab}(M\nabla_bN-N\nabla_bM)] \label{HHbeta}
\end{equation}
with some phase-space function $\beta$.  If the classical
hypersurface-deformation algebra is modified, gauge transformations no
longer correspond to Lie derivatives by space-time vector fields. Not just the
dynamics but even the structure of space-time may be modified by quantum
effects.  We will discuss specific examples and results in later parts of this
review.

By analyzing quantum or effective constraints and their algebra, one can draw
conclusions about quantum space-time structures. For instance, once the full
hypersurface-deformation algebra is known, one can specialize it to Poincar\'e
transformations by using linear $N$ and $N^a$. With $N(x)=\Delta t+v x$, for
instance, we have a combination of a time translation by $\Delta t$ and a
boost by $v$. With $\beta\not=1$ in (\ref{HHbeta}), the usual Poincar\'e
relations are modified. Although this may look like a version of deformed
special relativity \cite{DSR,DSR2,Rainbow}, there is no direct relation: In
deformed special relativity, one has non-linear realizations of the Poincar\'e
algebra, with structure constants depending on the algebra generators. In
(\ref{HHbeta}), we have corrections of structure functions depending on
phase-space degrees of freedom, not directly on the algebra generators $H[N]$
and $D[N^a]$. A deformed version of special relativity would require a
relation between phase-space variables, such as extrinsic curvature, and some
space-time generators, such as energy. Relations of this form do exist in some
regimes, for instance in asymptotically flat ones using the ADM energy, but
not in general.

\subsubsection{Consistent and inconsistent quantum cosmology}

In general terms, the problem of canonical quantum cosmology can be formulated
as follows: Find quantum corrections in $H[N,N^a]$, perhaps motivated by some
full theory of quantum gravity, such that these constraints remain first
class. This statement presents a well-defined mathematical problem of
classifying deformed algebras. For every consistent version that may exist, we
can compute and analyze equations of motion by standard means, as explicitly
written out above in the general classical case. At the present stage, several
examples of consistent deformations of constraints remaining first class are
known, mainly from loop quantum gravity, but there is no general
classification of these infinite-dimensional algebras.

As already mentioned, there are attempts to shortcut through the difficult
calculation of these algebraic structures by fixing the gauge before quantum
corrections or other modifications are put in. If the gauge is fixed, one
would no longer consider $H[\epsilon,\epsilon^a]$ as gauge generators, but
only use $H[N,N^a]$ as constraints and to generate evolution. Some consistency
conditions still need to be satisfied because the constraints are to be
preserved by evolution, but this can usually be achieved more easily than in
the non-gauge fixed case in which more fields are present. However, even if
formally consistent versions of preserved constraints can be found in this
way, such as those in \cite{ScalarHolEv}, they are not guaranteed to be
consistent because only a subclass of the constraint algebra can be tested
when some modes are eliminated beforehand. Even if these are classical gauge
modes, some consistency conditions and physical effects are
overlooked. Moreover, the procedure is intrinsically inconsistent because one
would first fix the gauge as it is determined by the classical constraints,
and then proceed to modify the constraints that generate the gauge.

A consistent scheme could be obtained only when the modified gauge structure
is taken into account from the very beginning, but for that one would have to
know a consistent version of non-gauge fixed constraints, not just of the
gauge-fixed ones. Finally, having fixed the gauge, there is no way of
calculating general gauge-invariant observables. Also here, one could only
refer to the classical invariant variables, whose form however must be
modified when quantum corrections are put into the constraints. Note that
quantum space-time structures and modified constraints in most cases imply
departures from classical manifold pictures, as we will see explicitly in the
examples provided later. One can no longer refer to the usual form of
coordinate transformations to compute gauge-invariant variables without using
the constraints explicitly. In modified space-times, the constraints are the
only means to compute gauge flows and invariants, but this can be done only if
the gauge has not been fixed.

Algebraic conditions can be simplified even more when gauge fixing is combined
with deparameterization. With the latter procedure, one chooses a phase-space
degree of freedom, for instance from matter, to rewrite constraint equations
as relational evolution equations with respect to this variable. Not just
gauge transformations but even constraints then disappear from the system, and
no strong consistency conditions remain. A popular example is the coupling of
a free, massless scalar field $\phi$, whose homogeneous mode $\bar{\phi}$ in
an expansion $\phi=\bar{\phi}+\delta\phi$ around some background can be
treated as a global time function. The canonical scalar Hamiltonian 
\begin{eqnarray}
 H_{\rm scalar}&=& \frac{1}{2}\int{\rm d}^3x \left(\frac{p_{\phi}^2}{\sqrt{\det q}}+
   \sqrt{\det q} q^{ab}(\partial_a\phi)(\partial_b\phi)\right)\\
 &=& \frac{1}{2} \left(\frac{\bar{p}_{\phi}^2}{\sqrt{\det
       \bar{q}}}+ \int{\rm d}^3x \delta(\cdots)\right)
\end{eqnarray}
with the momentum $p_{\phi}=\bar{p}_{\phi}+\delta p_{\phi}$ then has a purely
kinetic background term and is completely independent of $\bar{\phi}$. The
momentum $\bar{p}_{\phi}$ is therefore a constant of motion and never becomes
zero unless it vanishes identically. The background scalar $\bar{\phi}$ has no
turning points and is monotonic, serving as a global internal time along
classical trajectories. 

The Hamiltonian generating evolution with respect to this variable is obtained
by solving the Hamiltonian constraint $H_{\rm gravity}+H_{\rm scalar}=0$ (with
fixed $N$ as part of the gauge choice) for
$\bar{p}_{\phi}=H_{\bar{\phi}}(q_{ab},p^{ab},\delta \phi,\delta p_{\phi})$: we
have $\dot{\bar{\phi}}=\{\bar{\phi},H_{\bar{\phi}}\}=1$ and
$\dot{\bar{p}_{\phi}}=\{\bar{p}_{\phi},H_{\bar{\phi}}\}=0$, consistent with
evolution with respect to $\bar{\phi}$ where the dot stands for ${\rm d}/{\rm
  d}\bar{\phi}$, as well as Hamiltonian equations for the remaining variables,
such as $\dot{q}_{ab}=\{q_{ab},H_{\bar{\phi}}\}$ and
$\dot{\delta}\phi=\{\delta\phi,H_{\bar{\phi}}\}$.  There is just one
Hamiltonian generating all evolution equations, instead of a set of infinitely
many constraints.  Almost as in minisuperspace models one can then implement
in $H_{\bar{\phi}}$ any quantum corrections one may desire. (In the context of
cosmology, this approach has been suggested for instance in \cite{Hybrid} in
an analysis of Gowdy models and possible quantizations.)

Deparameterization is a powerful mathematical tool to derive properties of
physical Hilbert spaces, for which no general method exists in the absence of
deparameterizability. (See also Section~\ref{s:EffCons}.) In quantum
cosmology, this method has been applied in \cite{Blyth}, and used recently to
derive several aspects of self-adjointness and unitarity of evolution
\cite{SelfAdFlat,NonSelfAd,DensityOp}. But it cannot serve as a valid
procedure to evade the anomaly problem and do physical evaluations. To start
with, most realistic systems are not globally deparameterizable, with one
variable having no turning points at all. Moreover, also this procedure, on
its own, cannot weed out physical inconsistencies in spite of its formal
consistency. One has the same drawbacks as in the gauge-fixed approach, and on
top of that one has distinguished (or even introduced) one degree of freedom
as time. For physical consistency, one should then show that results for
observables do not depend on one's choice of time after quantization, but no
systematic procedures exist to this end, constituting part of the problem of
time in quantum gravity \cite{KucharTime,Isham:Time,AndersonTime}. (See
\cite{ReducedKasner} for a discussion in the context of deparameterization or
reduced phase-space quantization.)  And even if one thinks that the
distinguished time $\bar{\phi}$ should be sufficient for all physical purposes
and does not worry about relating results in different times, the fact that
there is no analog of the Bianchi identity to constrain modifications should
arouse suspicion.

Gauge fixing and deparameterization are not necessarily bad, but they provide
reliable results only when they are used {\em after} a consistent version of
quantum-corrected constraints has been found. When this is the case, it is
clear that all equations are consistent and gauge covariant, and instead of
computing complete gauge-invariant variables for the corrected constraints,
one may well pick a gauge or choose an internal time and work out physical
implications. There is also a possibility of providing consistent results with
gauge fixing or deparameterization before quantization, but then one would
have to show that all possible gauge fixings or deparameterizations would lead
to the same physical observables. This requirement can be achieved in cases of
simple constraints, such as the Gauss constraint of gravity in triad or
connection variables as described at the end of Section~\ref{s:SphSymm}, but
presents a much more involved problem for the complicated Hamiltonian
constraint. Problems and physical inconsistencies arise always when the gauge
or time is chosen according to the classical system, and modifications of the
constraints and thereby gauge transformations are inserted later.

\subsection{States}
\label{s:States}

Covariance or analogs of the contracted Bianchi identity severely constrain
consistent versions of quantum corrections, raising the manifold and metric
structures underlying space-time in general relativity to the effective or
quantum level. Although space-time structures and their covariance principles
may then differ significantly from well-known classical ones, there is still a
consistent dynamical theory independent of coordinates and gauge choices, and
the set of all equations has meaningful solutions available for well-defined
predictions. In addition to these restrictions on quantum corrections based on
gauge aspects, the quantum theory of gravitational degrees of freedom itself
should tell us what form of modifications we can have.

\subsubsection{Moments}

Effective equations, in general terms, describe quantum evolution by a
smaller, more manageable number of degrees of freedom compared with the full
quantum theory considered. When we go from classical physics to quantum
physics, every degree of freedom we have at first is replaced by infinitely
many parameters, for instance values the whole wave function takes at all
points in configuration space. It is difficult to deal with values of wave
functions in physical terms. Another parameterization, more convenient for
effective equations as it turns out, is the set of expectation values of basic
operators, $\langle\hat{a}\rangle$ and $\langle\hat{p}_a\rangle$ in
Wheeler--DeWitt quantum cosmology, together with fluctuations and higher
moments
\begin{equation}
 \Delta(a^bp_a^c):= \langle (\hat{a}-\langle\hat{a}\rangle)^b
 (\hat{p}_a-\langle\hat{p}_a\rangle)^c\rangle_{\rm Weyl}
\end{equation}
with operators in totally symmetric, or Weyl, ordering. For generic states,
all these moments with integer $b$ and $c$ such that $b+c\geq 2$, are
independent of one another and of the expectation values, and therefore
provide infinitely many degrees of freedom. (For $b+c=2$, for instance, we
have the two fluctuations $\Delta(a^2)=(\Delta a)^2$ and
$\Delta(p_a^2)=(\Delta p_a)^2$, and the covariance $\Delta(ap_a)=
\frac{1}{2}\langle\hat{a}\hat{p}_a+ \hat{p}_a\hat{a}\rangle-
\langle\hat{a}\rangle \langle\hat{p}_a\rangle$.)

Any set of effective equations provides a description between the classical
limit and the full quantum theory of the system considered, making use of
finitely many degrees of freedom for each classical one. There may be
additional degrees of freedom compared with the classical theory, but not
infinitely many more. By the new degrees of freedom included and their
coupling to expectation values --- the variables with a direct classical
analog --- quantum corrections result. This is our general definition of
effective theories, which we distinguish from coarse-grained theories. The
latter may remove some quantum degrees of freedom by integrating them out, but
typically leave whole towers of moments intact and therefore present
infinitely many variables for some classical degrees of freedom.

The behavior of the moments computed in simple unsqueezed Gaussian states with
fluctuation parameter $\sigma$ \cite{HigherMoments},
\begin{equation} \label{Gaussian}
\Delta(a^bp_a^c)=
2^{-(b+c)} \hbar^c\sigma^{b-c}\frac{b!\,c!}{(b/2)!
(c/2)!} 
\quad{\rm if}\, b \,{\rm and}\, c\, {\rm are}\,{\rm even}
\end{equation}
and $\Delta(a^bp_a^c)=0$ otherwise, suggest a natural organization of
effective theories of different degrees. A Gaussian state saturates the
uncertainty relation
\begin{equation} \label{Uncert}
 (\Delta a)^2(\Delta p_a)^2- \Delta(ap_a)^2\geq \frac{\hbar^2}{4}
\end{equation}
and, for position and momentum fluctuations of the same order, has
$\sigma=\Delta a=O(\hbar^{1/2})$. The product $\hbar^c\sigma^{b-c}\sim
O(\hbar^{(b+c)/2})$ then shows that moments of order $n=b+c$ are of the order
$n/2$ in $\hbar$. We use this observation to generalize the notion of
semiclassical states from simple Gaussians to a much larger class: If the
moments of a state behave as $\Delta(a^bp_a^c)\sim O(\hbar^{(b+c)/2})$ (said
to obey a $\hbar$ierarchy), the state is called semiclassical. Moments of
higher orders then affect only terms of high orders in $\hbar$ and can be
neglected in an approximation. Ignoring all moments of order higher than some
$n$, only finitely many quantum degrees of freedom are left, and we obtain an
effective theory. With $n+1$ moments of order $n$, an effective theory with
moments up to order $n$ has
$\sum_{k=2}^{n}(k+1)=\frac{1}{2}n^2+\frac{3}{2}n-2$ state parameters in
addition to the two basic expectation values.

\subsubsection{Dynamics and quantum back-reaction}

Effective solutions, describing the evolution of a quantum state, depend not
only on classical variables but also on the quantum state used, specified for
instance by initial values of its moments. In quantum cosmology, one first
promotes the constraint (\ref{constraint}) expressed in canonical variables
$a$ and $p_a$ as
\begin{equation}
 H(a,p_a)= -\frac{2\pi G}{3} \frac{p_a^2}{a}+E_{\rm matter}
\end{equation}
to an operator $\hat{H}$, choosing some ordering of $\hat{a}$ and
$\hat{p}_a$. Dirac quantization then implements the classical constraint
equation $H(a,p_a)=0$ by the condition $\hat{H}|\psi\rangle=0$ on physical
states. In particular, the expectation value $\langle\hat{H}\rangle$ must
vanish in all physical states. If we express this equation as a functional
equation on the space of expectation values and moments parameterizing states,
we obtain an expression such as
\begin{equation} \label{Qconstraint}
 \langle\hat{H}\rangle=  -\frac{2\pi G}{3}
 \left(\frac{\langle\hat{p}_a\rangle^2}{\langle\hat{a}\rangle}+ \frac{(\Delta
     p_a)^2}{\langle\hat{a}\rangle}+\cdots \right)+
\langle\hat{E}_{\rm matter}\rangle
\end{equation}
where we have used $\langle\hat{p}_a^2\rangle= \langle\hat{p}_a\rangle^2+
(\Delta p_a)^2$ and the dots indicate additional terms that contain the
covariance of $a$ and $p_a$ as well as higher moments, and depend on the
specific ordering chosen. If we take into account only expectation values in
$\langle\hat{H}\rangle=0$, the classical Hamiltonian and the classical
constraint surface are obtained. With fluctuations and higher moments, quantum
corrections result that couple moments to expectation values and change the
classical constraint surface. A systematic derivation and analysis of such
moment terms, starting with expectation values of Hamiltonians and
constraints, is the key ingredient of effective theories.

The moments couple to expectation values, thereby providing quantum
corrections to the classical motion, and their values therefore enter
effective equations. A complete effective theory cannot leave the moments
in equations as unknowns and instead provides evolution equations or other
conditions for them as well. But some freedom always remains, for instance in
the initial values chosen for evolution equations of moments. If effective
theories are formulated for expectation values without including additional
degrees of freedom such as moments to describe the evolution of classes of
states, the specific states must either be restricted to provide unique
effective equations, or be parameterized for a more general set. Often, such a
state dependence enters effective theories only implicitly, for instance in
the unique-looking low-energy effective action \cite{EffAcQM} free of any
state parameters. This effective action describes low-energy effects of states
near the vacuum of the interacting theory. In this way, the class of states is
specified, certainly a rather small set compared to all possible states.

In quantum cosmology, we are often interested in high-energy, Planckian
phenomena and cannot restrict attention to the low-energy effective
action. Even in low-energy regimes, which would be all we need to make contact
with potential observations, it is not clear what low-energy state should be
used. Quantum cosmology in its non-perturbative form, or quantum gravity in
general, does not have a vacuum or other distinguished low-energy
state. Effective theories of quantum cosmology must therefore be more general
than the low-energy effective action, leaving more freedom for states
parameterized in some suitable way. 

Even if we restrict attention to semiclassical regimes, the class of states to
be considered may be large: simple Gaussians provide at most a 2-parameter
family within a large set of semiclassical states when they are fully
squeezed. For uncorrelated Gaussians, wave functions
\begin{equation}
  \psi_{\sigma}(a)=
  N\exp\left(-\frac{(a-\langle\hat{a}\rangle)^2}{4\sigma^2}\right) 
  \exp(-ia\langle\hat{p}_a\rangle/\hbar)
\end{equation}
with a normalization constant $N$, we have, besides the two expectation values
$\langle\hat{a}\rangle$ and $\langle\hat{p}_a\rangle$ on which the state is
peaked, only one quantum parameter, the variance $\sigma$. The most general
Gaussian state is of the form $\psi_z(a)=\exp(-z_1a^2+z_2a+z_3)$ with three
complex numbers $z_i$ such that ${\rm Re}z_1>0$ for normalizability. Out of
these six real parameters, the two contained in $z_3$ do not matter for
moments because the real part is fixed by normalization and the imaginary part
contributes only a phase factor. For the remaining parameters, writing
$z_1=\alpha_1+i\beta_1$ and $z_2=\alpha_2+i\beta_2$ with real $\alpha_i$ and
$\beta_i$, we compute expectation values
\begin{equation}
 \langle\hat{a}\rangle= \frac{\alpha_2}{2\alpha_1}\quad,\quad
 \langle\hat{p}_a\rangle=
 \hbar\frac{\alpha_1\beta_2-\alpha_2\beta_1}{\alpha_1}
\end{equation}
and second-order moments
\begin{equation}
(\Delta a)^2 = \frac{1}{4\alpha_1}\quad,\quad
 (\Delta p_a)^2 = \hbar^2\alpha_1+\hbar^2\frac{\beta_1^2}{\alpha_1}\quad,\quad
 \Delta(ap_a) = -\hbar\frac{\beta_1}{2\alpha_1}
\end{equation}
One easily confirms that the uncertainty relation (\ref{Uncert}) is always
saturated. 

Deviations from saturation are much more difficult to parameterize, but easily
occur for evolved semiclassical states even if they start out as a
Gaussian. And even if one stays close to the saturation condition and does not
vary second-order moments much beyond the values they can take for Gaussians,
a Gaussian determines all higher moments in terms of the real and imaginary
parts of $z_1$. Our general semiclassicality condition $\Delta(a^bp_a^c)\sim
O(\hbar^{(b+c)/2})$ provides an infinite-parameter family instead of the
special 2-parameter one realized for Gaussians. Unless one can motivate
Gaussians by other means, for instance proximity to the Gaussian
harmonic-oscillator ground state or the vacuum of a free field theory, using
them exclusively may easily be too restrictive.  (Given the large parameter
space of states, one may be tempted to refer to probabilistic arguments to
pick ``likely'' states, following ideas that go back to an analysis of
inflation in quantum cosmology
\cite{MeasureUniverses,MeasureInflation}. However, such probability
considerations, though popular, are difficult, if not impossible, to make
sense of in quantum cosmology \cite{MeasureCosmo}.)

Not just the absence of a vacuum state but several other special
properties of quantum cosmology are important when we consider possible
states:
\begin{itemize}
\item Quantum cosmology considers long-term evolution. Even if we may be able
  to choose a specific form of semiclassical states at large volume and small
  curvature, it may change much when states are evolved back to high densities
  to infer possible implications at the big bang.
\item As already mentioned, even the form of semiclassical states is
  unclear. It is customary to explore semiclassical features using simple and
  nicely peaked Gaussian states. In quantum mechanics, such states provide
  interesting information, and they are realized in exactly this form as
  coherent states or the ground state of the harmonic oscillator. In quantum
  field theory, Gaussian states are then close to the perturbative vacuum even
  for interacting theories. The dynamics of quantum cosmology, however, is not
  near that of the harmonic oscillator (except for some special models), and
  the unquestioned use of Gaussians is more difficult to justify. But going
  beyond Gaussians is complicated in terms of wave functions, whose parameters
  then become much less controlled.
\item It is not clear how precisely quantum cosmology can be derived from some
  full theory of quantum gravity, but the number of degrees of freedom is
  certainly reduced either by exact symmetries or by perturbing around some
  background. In such situations, when degrees of freedom are eliminated, pure
  states easily become mixed. For general evolution equations able to model a
  full state, we should therefore allow for the possibility of mixed states,
  again going beyond pure Gaussians or other specific wave functions. Moments
  provide a parameterization of mixed states as well since they are based only
  on the notion of expectation values.
\item The question of covariance affects also the choice of classes of
  states. A state chosen in a minisuperspace model must have a chance of being
  the reduction of a full state that does not break covariance. This question
  may be difficult to analyze at the level of wave functions, but it also
  shows that a sufficiently large freedom in the choice of states must be
  included in considerations that aim to provide a consistent formulation of
  quantum cosmology.
\end{itemize}
Since state properties are important for effective actions and quantum
back-reaction, we should be as general as possible with the choice of states
we consider. With wave functions or density matrices for mixed states, such a
generality is difficult to achieve, but it is possible with parameterizations
such as the one by moments. Moments provide a general form of semiclassical
states, as already introduced. In effective theories, they are subject to
their own evolution equations which show how they may change as high-density
regimes are approached. They go well beyond the 2-parameter family of squeezed
Gaussians, and describe pure and mixed states alike.

\subsubsection{Quantum phase space and covariance}

Covariance at the level of effective equations with quantum back-reaction can
be addressed by combining the moment parameterization with the methods of the
previous section, deriving consistent constrained systems amended by quantum
corrections that include the moments. For the last question, we must be able
to fit moments into a phase-space structure, so as to be able to compute
Poisson brackets such as $\{H[M,M^a],H[N,N^a]\}$ with constraints that may
include moment terms, such as an inhomogeneous version of
(\ref{Qconstraint}). 

This construction is indeed possible: Together with expectation values, the
moments form a quantum phase space with Poisson brackets defined by the
commutator \cite{EffAc}, first for expectation values of arbitrary operators:
\begin{equation} \label{Poisson}
  \{\langle\hat{A}\rangle,\langle\hat{B}\rangle\}=
  \frac{\langle[\hat{A},\hat{B}]\rangle}{i\hbar}\,.
\end{equation}
This bracket satisfies the Jacobi identity and is linear. If we extend it to
polynomials of expectation values by imposing the Leibniz rule, all laws for a
Poisson bracket are satisfied, and we can apply the definition to moments. For
instance, we obtain $\{\langle\hat{a}\rangle,\langle\hat{p}_a\rangle\}=1$ and
$\{\langle\hat{a}\rangle,\Delta(a^bp_a^c)\}=0=
\{\langle\hat{p}_a\rangle,\Delta(a^bp_a^c)\}$ for all $b$ and $c$. The
moments, as defined here, are symplectically orthogonal to expectation values,
a convenient feature for calculations. Poisson brackets between different
moments have been computed explicitly but are lengthy; see \cite{EffAc} and
the correction of a typo in \cite{HigherMoments}. For low orders, it is
usually more convenient to compute Poisson brackets directly from the
definition (\ref{Poisson}). For instance, we calculate
\begin{eqnarray} \label{FluctBracket}
\{(\Delta q)^2,(\Delta p)^2\}&=&
 \{\langle\hat{q}^2\rangle-\langle\hat{q}\rangle^2,
 \langle\hat{p}^2\rangle-\langle\hat{p}\rangle^2\}\\
&=&
 \frac{\langle[\hat{q}^2,\hat{p}^2]\rangle}{i\hbar} - 2\langle\hat{p}\rangle
 \frac{\langle[\hat{q}^2,\hat{p}]\rangle}{i\hbar} - 2\langle\hat{q}\rangle
 \frac{\langle[\hat{q},\hat{p}^2]\rangle}{i\hbar} +
 4\langle\hat{q}\rangle\langle\hat{p}\rangle
 \frac{\langle[\hat{q},\hat{p}]\rangle}{i\hbar}\nonumber \\
& =&
 2\langle\hat{q}\hat{p}+\hat{p}\hat{q}\rangle-
 4\langle\hat{q}\rangle\langle\hat{p}\rangle= 4\Delta(qp)\nonumber
\end{eqnarray}
using the Leibniz identity and (\ref{Poisson}). 

If we know how a state enters effective constraints via its moments, a
question which we will address in due course, we can compute Poisson brackets
of effective constraints and see whether they provide a consistent deformation
of the classical constraint algebra. If a consistent deformation is realized,
the system can be analyzed further by standard canonical means to arrive at
observables and dynamical equations. Note, however, that the Poisson tensor
for moments truncated to some order is in general not invertible, for instance
on the three second-order moments which form an odd-dimensional Poisson
manifold: With (\ref{FluctBracket}) and 
similar calculations for the other second-order moments, we
obtain 
\begin{eqnarray}\label{FluctAlg}
\{(\Delta q)^2,(\Delta p)^2\}&=& 4\Delta(qp)\,,\\ 
\{(\Delta q)^2,\Delta(qp)\}&=& 2(\Delta q)^2\,,\\ \{(\Delta
p)^2,\Delta(qp)\}&=&-2(\Delta p)^2\,,
\end{eqnarray}
the 3-dimensional Poisson manifold of second-degree polynomials.
Symplectic geometry therefore cannot be used for effective theories,
while Poisson geometry is available from the definition (\ref{Poisson}). All
relevant properties of constrained systems, such as the distinction between
first and second class or properties of gauge transformations, can be
formulated at the level of Poisson geometry \cite{brackets}. 

\subsubsection{Quantum-gravity states}

In summary of this section, we note that there are several specific issues in
the derivation of effective theories for quantum cosmology, compared to other
fields in which such methods are in use. The central theme is covariance, an
issue which can be addressed only if one goes beyond the traditional
quantum-cosmological realm of minisuperspace models. The notion of covariance
and the question whether it is realized consistently may be affected by new
quantum space-time structures introduced by the specific form of quantum
geometry in one's approach to quantum gravity. 

The second, less obvious source of potential modifications of covariance is
the form of dynamical quantum states used. Effective actions or equations
depend on the classes of quantum states whose evolution they approximate,
which should be specific solutions of some underlying quantum theory of
gravity. Even if a theory such as quantum gravity is covariant, the selection
of specific solutions may always break this symmetry. Covariance may then be
realized in a deformed way, or only partially within one effective theory. For
instance, even if the states used are peaked on covariant classical fields,
giving rise to covariant effective terms depending on expectation values,
their fluctuations or higher moments may not be fully covariant. Changing the
gauge in quantum gravity may then require the transition to a different
effective theory, in which different state parameters have been used
(reminiscent of the application of background-field methods in standard
quantum field theory). On the other hand, if the class of states is
sufficiently large, all gauge-related parameters could be encompassed within
one effective theory. These considerations highlight the importance of the
selection of states in the derivation of effective theories.

Both quantum geometry and quantum dynamics must be part of one consistent
quantum theory of gravity; in a complete treatment, their effects therefore
cannot be separated from each other. However, they are derived by different
means, making use of different expansions of the expectation-value function
$\langle\hat{H}\rangle$ of the Hamiltonian or Hamiltonian constraint in the
class of states used. In what follows, as in most derivations in the
literature, we will split the treatment into one of quantum-geometry
corrections first, as they are easier to see, followed by a discussion of
quantum-dynamics corrections. When their expressions are known, we can compare
implications of different effects and see if some are more relevant than
others in specific regimes.

In the canonical setting, quantum-geometry corrections are specific to loop
quantum gravity which has given rise to many results regarding
background-independent quantization \cite{Rov,ThomasRev,ALRev}. The effects in
$\langle\hat{H}\rangle$ are not unique, but characteristic enough to show
implications for quantum space-time structure. After an overview of these
terms in the next section, we will discuss dynamical quantum back-reaction,
obtained from a further expansion of $\langle\hat{H}\rangle$ by the moments
parameterizing states. We will see the relation of canonical formulations to
the low-energy effective action used in particle physics and the role of
higher time derivatives in effective equations. Finally, we will put together
our results to find properties of general effective actions taking into
account all possible effects of canonical quantum gravity, and thereby shed
light on quantum space-time structure.

\section{Quantum geometry of space}
\label{QGS}

Canonical gravity implements space-time structure by imposing the constraints
$H[N]+D[N^a]$ and requiring invariance under the gauge flow they
generate. (There may be additional constraints, such as the Gauss constraint
if triad variables are used. Such constraints, however, restrict auxiliary
degrees of freedom not related to transformations of space-time.) The
diffeomorphism (and Gauss) constraint generates a simple flow by Lie
derivatives, and has a direct action on quantum states. If we use states
$\psi[q_{ab}]$ as in Wheeler--DeWitt quantum gravity (leaving aside the
complicated question of how to define an inner product) we have a formal
action $\hat{D}[N^a]\psi[q_{bc}]= \psi[{\cal L}_{N^a}q_{bc}]$. States are
annihilated by the diffeomorphism constraint if they depend only on spatial
invariants, of the same form as classically, and expectation values in
non-invariant states transform according to the classical gauge
transformations. In loop quantum gravity, the action on states is different
because discrete spatial structures do not allow infinitesimal diffeomorphisms
to be represented. But for finite diffeomorphisms, the situation is exactly as
just described. Therefore, we do not expect significant quantum modifications
of the spatial diffeomorphism constraint (but see
Section~\ref{s:Diffeo}). Similarly, the Gauss constraint, if there is one, has
a simple and direct action that does not suggest modifications; see also the
end of Section~\ref{s:SphSymm}.

\subsection{Hamiltonian constraint}

The Hamiltonian constraint $\hat{H}[N]$ presents a different story. It cannot
be quantized directly by promoting its (complicated) classical gauge flow to
an action on states. The only procedure to arrive at a constraint operator is
tedious: inserting basic operators quantizing the classical kinematical phase
space into the expression for the Hamiltonian constraints. As a phase-space
function, the constraint contains rather involved combinations of the basic
fields, which require regularization and sometimes even other modifications
for a well-defined operator to result \cite{RS:Ham,QSDI}. Also on general
grounds, we do expect quantum corrections in the Hamiltonian constraint
because it includes all about the dynamics of the theory. The theory being
interacting, quantum corrections must result. The Hamiltonian constraint is
therefore the place where we should look for characteristic quantum
corrections of different theories, as well as possible restrictions by
consistency requirements.

\subsubsection{Gauge flow}
\label{s:Gauge}

A constraint operator $\hat{H}[N]$ restricts states by
$\hat{H}[N]|\psi\rangle_{\rm phys}=0$ and generates a gauge flow
$|\psi\rangle_{\epsilon}=
\exp(-i\hat{H}[\epsilon]/\hbar)|\psi\rangle$. Physical states, on which
$\exp(-i\hat{H}[\epsilon]/\hbar)$ acts as the identity, are gauge-invariant,
and the gauge flow need not be considered separately if first-class constraint
operators are solved. But exact solutions are complicated to find, and when
quantum corrections are computed by systematic approximation schemes, the
situation is quite different. Expectation values are constrained by
$\langle\hat{H}[N]\rangle$, a weaker condition than
$\hat{H}[N]|\psi\rangle=0$. We have additional constraints
$\langle(\hat{f}-\langle\hat{f}\rangle)\hat{H}[N]\rangle=0$, an enlarged set
of constraints which all vanish in physical states, for arbitrary
$\hat{f}$. These constraints in general are all independent, constraining not
just expectation values but also moments. In an effective theory, as spelled
out in detail later, one solves this infinite set of quantum phase-space
constraints order by order in the moments. To any given order, the sharp
condition $\hat{H}[N]|\psi\rangle_{\rm phys}=0$ for physical states is not
fully implemented, and on the corresponding solution space there is a
non-trivial gauge flow by quantum constraints. Moreover, the effective
treatment at this stage is more general because one does not assume a
(kinematical) Hilbert-space structure when solving the constraints for
moments; therefore the standard argument of a trivial gauge flow does not
apply. At the effective level, there are non-trivial gauge transformations by
quantum constraints, bringing solution procedures closer to classical ones:
one solves phase-space constraints and factors out their gauge.

For an effective theory, the key ingredient is therefore the expectation value
$\langle\hat{H}[N]\rangle$ in a sufficiently large class of states, or the
general expression parameterized by the moments of states. As in
(\ref{Qconstraint}), moments then appear in quantum corrections that change
the constraint surface. With a modified constraint surface, the gauge flows
must receive quantum corrections as well for them to be tangential to the
constraint surface mapping solutions to constraints into other solutions, as
required for a first-class system. Indeed, the expectation value of the
constraint, an expression that includes quantum corrections, determines the
gauge flow on the quantum phase space with Poisson brackets (\ref{Poisson}) by
\begin{equation} \label{dOdt}
\delta_{\epsilon}\langle\hat{O}\rangle(\epsilon)= \epsilon
 \frac{\delta}{\delta\epsilon} \langle\psi |\exp(i\hat{H}[\epsilon]/\hbar)
 \hat{O} \exp(-i\hat{H}[\epsilon]/\hbar)|\psi\rangle=
 \frac{\langle[\hat{O},\hat{H}[\epsilon]]\rangle}{i\hbar}=
 \{\langle\hat{O}\rangle,\langle\hat{H}[\epsilon]\rangle\}
\end{equation}
for expectation values $\langle\hat{O}\rangle(\epsilon)=
{}_{\epsilon}\langle\psi|\hat{O}|\psi\rangle_{\epsilon}$, and therefore for
any quantum phase-space function such as the moments by using the Leibniz
rule. If the constraint surface changes by corrections in $\hat{H}$, so do the
gauge transformations. For the quantum corrected constraint surface and the
gauge flow we are therefore required to compute
$\langle\hat{H}[\epsilon]\rangle$ in general kinematical states.

If quantum constraint operators are represented in an anomaly-free way, such
that any commutator $[\hat{C}_1,\hat{C}_2]$ is an operator of the form
$f(\hat{q},\hat{p})\hat{C}_3$ with another constraint operator $\hat{C}_3$ and
structure functions $f(q,p)$, the quantum constraints
$\langle\hat{C}_i\rangle$ form a consistent first-class system:
$\{\langle\hat{C}_1\rangle,\langle\hat{C}_2\rangle\}=
\langle[\hat{C}_1,\hat{C}_2]\rangle/i\hbar= \langle
f(\hat{q},\hat{p})\hat{C}_3\rangle$. Since all expressions such as $\langle
f(\hat{q},\hat{p})\hat{C}_3\rangle$ vanish when computed in physical states,
they provide effective constraints (in general independent of
$\langle\hat{C}_3\rangle$ as phase-space functions). If all these constraints
are imposed, the effective constrained system is first class and has a
consistent gauge flow generated by all these constraints
\cite{EffCons,EffConsRel,EffConsComp}.

Such systems of infinitely many constraints for each local classical degree of
freedom can be difficult to analyze. However, just as the moments obey a
$\hbar$ierarchy in semiclassical regimes, the effective constraints can be
truncated to finite sets to any given order in $\hbar$ or in the moments. The
leading corrections can be found in direct expectation values
$\langle\hat{C}_i\rangle$, restricting expectation values with quantum
corrections that depend on the moments. For instance, (\ref{Qconstraint}),
when imposed as a constraint, shows how the classical constraint surface and
the gauge flow of $\langle\hat{a}\rangle$ and $\langle\hat{p}_a\rangle$ are
changed by fluctuations $(\Delta p_a)^2$. These corrections depend on the value
of $(\Delta p_a)^2$, which is constrained and subject to gauge flows by
higher-order constraints such as $\langle\hat{p}_a\hat{H}\rangle$. When all
constraints have been solved and gauge flows factored out to a certain order,
the condition $\hat{H}|\psi\rangle=0$ has been implemented for states used in
expectation values and moments. One has then computed observables in physical
states, sidestepping the complicated problem of computing an integral form of
the physical inner product. This is one example for the use of effective
methods to tame complicated technical and conceptual issues in quantum
gravity. Even the problem of time can be solved, at least at the semiclassical
level: Different choices of time are related by mere gauge transformations in
the quantum phase space \cite{EffTime,EffTimeLong,EffTimeCosmo}. We will come
back to these conceptual question in Section~\ref{s:EffCons}, and for now note
the lesson that expectation values of constraints supply the key ingredient
for effective gauge theories.

\subsubsection{Quantum-geometry  effects}

We should then apply effective techniques to the Hamiltonian constraint of
gravity, computing $\langle\hat{H}[N]\rangle$.
Expressions for the Hamiltonian constraint are rather complicated even
classically, given by 
\begin{equation} \label{HADM} 
H_{\rm grav}(q_{ab},\pi^{cd})= -\frac{16\pi G}{\sqrt{\det q}}
  \left(\pi_{ab}\pi^{ab}- \frac{1}{2}(\pi^a_a)^2\right)+ \frac{\sqrt{\det
      q}}{16\pi G} {}^{(3)}R
\end{equation}
in ADM variables, the spatial metric $q_{ab}$ and its momentum
\[
 \pi^{cd}= \frac{\sqrt{\det q}}{16\pi G} (K^{cd}-K^a_aq^{cd})
\]
in terms of extrinsic curvature (with the spatial Ricci scalar
${}^{(3)}R$). 

Loop quantum gravity uses a different set of variables,
the densitized triad $E^a_i$ instead of the spatial metric
$q^{ab}=E^a_iE^b_i/\det (q_{cd})$ and the Ashtekar--Barbero connection
$A_a^i=\Gamma_a^i+\gamma K_a^i$ \cite{AshVar,AshVarReell}, defined by
combining the spin connection $\Gamma_a^i$ compatible with the densitized
triad and extrinsic curvature $K_a^i=K_{ab}E^{bi}/\sqrt{|\det (E^c_i)|}$. The
basic Poisson brackets are
\begin{equation} \label{AE}
 \{A_a^i(x),E^b_j(y)\}= 8\pi\gamma G \delta_a^b \delta_j^i \delta(x,y) \,.
\end{equation}
The Barbero--Immirzi parameter $\gamma>0$ \cite{AshVarReell,Immirzi} does not
play a role classically but appears in quantum spectra and corrections. In
these variables, the Hamiltonian constraint is
\begin{equation} \label{HAsh}
H_{\rm grav}(A_a^i,E^b_j) = 
-\frac{E^a_iE^b_j\epsilon^{ij}{}_k}{16\pi G \sqrt{|\det (E^c_l)|}}
  \left(F_{ab}^k+(1+\gamma^{-2}) \epsilon^k{}_{mn}
   (A_a^m-\Gamma_a^m)(A_b^n-\Gamma_b^n)\right)
\end{equation}
with the curvature $F_{ab}^k$ of the Ashtekar--Barbero
connection. 

Both expressions are rather complicated to quantize, owing for instance to the
expressions of $^{(3)}R$ or $\Gamma_a^i$, to be written in terms of the metric
or densitized-triad operators.  Fortunately, there are several characteristic
features, common to both versions: The Hamiltonian constraint is quadratic in
the connection or extrinsic curvature, and it requires an inverse of the
spatial metric or the densitized triad. These two features, when combined with
quantum-representation properties, imply characteristic structures of quantum
geometry and associated corrections to classical equations. These corrections,
in turn, can be analyzed for potential implications even if their precise form
cannot be determined from a complete calculation of $\langle\hat{H}\rangle$ in
semiclassical states. And even if expectation values could be computed for a
specific $\hat{H}$ and some states, ambiguities in the construction of
$\hat{H}$ or the choice of states would be so severe that only general
features and characteristic effects could be trusted.

In a Wheeler--DeWitt quantization, we have formal quantum constraint equations
with $\pi_{ab}$ in (\ref{HADM}) replaced by functional derivatives with
respect to $q_{ab}$, acting on wave functions $\psi[q_{ab}]$. The formal
nature leaves precise representation properties unclear, and therefore does
not show specific quantum effects; one simply takes the classical expression
and performs the usual substitution of momenta by derivative operators. One
does not expect strong quantum-geometry effects, simply because quantum
geometry has not been completed in this setting. Loop quantum gravity, on the
other hand, has celebrated its greatest success so far at this level of
quantum representations, and indeed sheds considerable light on questions of
quantum corrections resulting from quantum geometry.

\subsubsection{Loop representation and background independence}

Loop quantum gravity looks closely at representation properties of basic
operators. To eliminate the need for formal functional derivatives and the
associated delta functions in quantizations of (\ref{AE}), the classical
fields $A_a^i$ and $E^b_j$ are smeared or integrated over suitable sets in
space: holonomies $h_e(A_a^i)= {\cal P}\exp(\int_e {\rm d}\lambda t_e^a A^i_a
\tau_i)$ and fluxes $F^{(f)}_S(E^b_j)= \int_S{\rm d}^2y n^S_a E^a_if^i$ with
the tangent vector $t_e^a$ to a curve $e$ and the co-normal $n^S_a$ to a
surface $S$ (on which an su(2)-valued smearing function $f^i$ is
chosen). Allowing for all curves and surfaces, all information about the
fields can be recovered. By the integrations, the delta function in
$\{A_a^i(x),E^b_j(y)\}=8\pi\gamma G\delta^b_a\delta^i_j\delta(x,y)$ is
eliminated, and a well-defined holonomy-flux algebra results:
\begin{equation} \label{HolFlux}
 \{h_e(A_a^i),F_S(E^b_j)\}= 8\pi\gamma G h_{e\rightarrow x}\tau_if^i(x)
 h_{x\rightarrow e}
\end{equation}
if there is only one intersection point $\{x\}=e\cap S$, denoting by
$h_{e\rightarrow x}$ and $h_{x\rightarrow e}$ the holonomy along $e$ up to $x$
and starting at $x$, respectively.  With strong uniqueness properties of
possible quantum representations \cite{FluxAlg,Meas,LOST,WeylRep}, spatial
quantum geometry in this setting is under excellent control.

Representations of the holonomy-flux algebra then provide operators to be
inserted in (\ref{HAsh}) to obtain a quantized Hamiltonian constraint. The
kinematical Hilbert space is spanned by cylindrical states $\Psi(A_a^i)=
\psi(h_{e_1}(A_a^i),\ldots, h_{e_n}(A_a^i))$, each of which depends on the
connection via a finite number of holonomies. The full state space, has no
restriction on the number of holonomies that may appear, thereby representing
the continuum theory rather than some lattice model. The inner product is
obtained by integrating the product of two cylindrical functions over as many
copies of SU(2) as there are non-trivial dependencies on holonomies in both
states, using the normalized Haar measure. By completion of the space of
cylindrical functions, the kinematical Hilbert space is obtained
\cite{ALMMT}. Holonomies $\hat{h}_e$ then act as multiplication operators,
changing the dependence of a cylindrical state on $h_e$, or creating a new
dependence if the state was independent of $h_e$ before acting. Flux
operators, representing the densitized triad, become derivative operators in
the connection representation used, and can be expressed in terms of invariant
derivative operators on SU(2) (or angular-momentum operators). They have
discrete spectra, indicating modifications to the classical spatial structure
\cite{AreaVol,Area,Vol2}: distance, areas or volumes computed from $E^a_i$
can, after quantization, no longer increase continuously even if the
underlying curves, surfaces, or regions are deformed by homotopies.

As a further consequence of discreteness, it turns out that holonomy operators
do not continuously depend on the curves used. A cylindrical state depending
only on $h_e$ and one depending only on $h_{e\circ e'}$ with a piece $e'$
appended to the curve refer to two different holonomies, and are orthogonal
according to the inner product just described, even if $e$ is a 1-point
set. While we classically have $t_{e'}^a A_a^i(x)\tau_i= \lim_{e'\to\{x\}}
(h_{e'}(A_a^i)-1)/|e'|$, where $t_{e'}^a$ is the tangent vector of $e'$ at $x$
and $|e'|$ its coordinate length, the sequence $\hat{h}_{e'}|\psi\rangle$ of
states after quantization does not converge for any $|\psi\rangle$ if $e'$ is
changed. The edge dependence disappears when one implements the diffeomorphism
constraint, which has not been assumed in the previous constructions. States
$\hat{h}_{e'}|\psi\rangle$ for different $e'$ are no longer orthogonal, but
they all give rise to the same state when diffeomorphism are factored out. The
limit can then be taken, but always equals zero. Again, there is no way of
deriving a connection operator.

Unlike with classical expressions, it is not possible to take a
derivative of $h_e(A_a^i)$, for instance by the endpoint of $e$, to obtain an
expression for a quantized $A_a^i$. Loop quantum gravity does not offer
connection operators; all connection dependence in the Hamiltonian constraint
must be expressed in terms of holonomies. No quadratic function as it appears
in the constraint can exactly agree with a linear combination of exponentials,
and modifications arise, motivated by background-independent quantum geometry.

\subsubsection{Holonomy corrections}

When the Hamiltonian constraint is quantized in loop quantum gravity,
holonomies are used instead of connection components \cite{RS:Ham}, providing
a ``regularization'' necessary to render the constraint expressible by basic
operators. However, the limit in which the ``regulator'' --- the specific
curves used for holonomies --- is removed does not exist; we are not dealing
with a proper regularization. (In \cite{QSDI}, the limit is argued to exist
and be trivial if spatially diffeomorphism-invariant states are used. But with
this assumption there is no handle on the full off-shell algebra and its
anomaly problem.) Instead, one usually interprets the difference of a
holonomy-modified constraint or its effective versions with the classical
expression as a series of higher-order corrections, amending the classical
Hamiltonian by higher powers of the connection, or intrinsic and extrinsic
curvature.  In this viewpoint, the modification is similar to expected
higher-curvature corrections --- except for the issue of covariance that we
will have to address. (In some models of loop quantum cosmology, it is
possible to represent an exponentiated version of the constraint in terms of
holonomy operators without introducing modifications to the classical
expression \cite{NonExpLQC}. However, the procedure seems to depend
sensitively on specific properties of the model used and is not available in
general.)

Holonomy corrections therefore appear as higher-order terms  such as
\begin{equation} \label{DerExp}
-2 {\rm tr}(\tau_ih_e(A_b^j))-t_e^aA_a^i(x)=
-\frac{1}{24}(t_e^aA_a^i(x)(t_e^bA_b^j(x))(t_e^cA_{cj}(x))+ 
t_e^at_e^b \partial_bA_a^i(x)+\cdots
\end{equation}
obtained from a Taylor expansion of the exponential and a derivative expansion
of the integration. These terms depend on the routing of the curve $e$, to be
chosen suitably for a quantization of the Hamiltonian constraint. The
condition of anomaly-freedom puts strong restrictions on the possible routings
\cite{TwoPlusOneDef} which, however, are difficult to evaluate. Moreover, the
expansion is done in an effective derivation, requiring the calculation of
expectation values that lead to additional moment terms. All these
calculations are difficult to perform explicitly, but the form of corrections
is clear and quite characteristic: higher orders as well as higher spatial
derivatives in the connection. With these types of corrections, suitably
parameterized, one can look for consistent deformations of the constraint
algebra to find versions for a physical evaluation of these corrections, or to
derive further restrictions on the quantization choices made.

In effective equations, the use of holonomies as basic operators of the
quantum theory has another consequence: The holonomy-flux algebra
(\ref{HolFlux}) is not canonical, with non-constant Poisson brackets. Moments
of states used for effective descriptions are based on expectation values of
basic operators, holonomies and fluxes in loop-quantized models. The general
constructions of effective theories still go through: Poisson brackets on the
quantum phase space, the quantum Hamiltonian or constraints, and so
on. However, the explicit Poisson relations (\ref{Poisson}) evaluated between
individual moments of the form $\Delta(h^bF^c)$ are different from the
canonical case. In particular, moments no longer Poisson-commute with
expectation values. This feature requires care and complicates some
calculations. These problems, however, are only technical; see the examples
provided later in Section~\ref{s:HarmLQC} and in
\cite{BouncePert,BounceCohStates}.

\subsubsection{Inverse-triad corrections}

Fluxes are linear in the densitized triad and do not suggest the same kind of
modification as holonomies. Nevertheless, there is a characteristic effect
associated also with them. Flux operators have discrete spectra, containing
zero as an eigenvalue. Such operators do not have densely-defined inverses,
and yet we need an inverse of the densitized triad for the classical
Hamiltonian constraint. To obtain such quantizations, a more indirect route is
taken in loop quantum gravity, which does result in well-defined operators but
introduces another kind of correction, called inverse-triad correction. 

To see the form of these corrections, we show a lattice calculation of
inverse-triad operators. As proposed in \cite{QSDI,QSDV}, the combination of
triad components required for the Hamiltonian constraint (\ref{HAsh}) is first
rewritten as
\begin{equation} \label{Inverse}
 2\pi\gamma G
 \epsilon^{ijk}\epsilon_{abc} \frac{E^b_jE^c_k}{{\sqrt{|\det
       E|}}}=\left\{A_a^i,\int{\sqrt{|\det E|}}\mathrm{d}^3x\right\}\,.
\end{equation}
In the new form, no inverse is needed, $A_a^i$ can be expressed by holonomies,
the volume operator can be used for $\int{\sqrt{|\det E|}}\mathrm{d}^3x$, and
the Poisson bracket be turned into a commutator divided by $i\hbar$. The
resulting operators are rather contrived, especially with SU(2) holonomies and
derivatives involved. But the presence of corrections and their qualitative
form can be illustrated by a U(1)-calculation, for which we also assume
regular cubic lattices. 

On a cubic lattice, we can assign a unique plaquette to each link $e$, by
which we then label fluxes. Our basic operators are $\hat{h}_e$, which as a
multiplication operator by $\exp(i\int_eA_at_e^a{\rm d}\lambda)$ takes values
in U(1), and $\hat{F}_e$ for a fixed set of edges $e$ in a regular cubic
lattice. We are therefore computing inverse-triad operators and their
expectation values for a fixed subset of cylindrical states, but the lattice
can be as fine as we want, allowing us to capture all continuum degrees of
freedom. For this U(1)-simplification, the holonomy-flux algebra, quantizing
an Abelian version of (\ref{HolFlux}), reads $[\hat{h}_e,\hat{F}_e]=
-8\pi\gamma\ell_{\rm P}^2\hat{h}_e$ while all operators commute if they belong
to different edges. Moreover, the U(1)-valuedness of holonomies implies the
reality condition $\hat{h}_e\hat{h}_e^{\dagger}=1$, which we will make use of
below.

We first rewrite (\ref{Inverse}) in terms of holonomies instead of connection
components, and express the volume $V=\int\sqrt{|\det E|}{\rm d}^3x$ by
lattice fluxes $\sqrt{|F_1F_2F_3|}$ per vertex, with the three fluxes through
plaquettes in all three directions around the vertex:
$t_e^a\{A_a,V\}=ih_e\{h_e^{-1},\sqrt{|F_1F_2F_3|}\}$ or, more symmetrically,
$\frac{1}{2}i(h_e\{h_e^{-1},\sqrt{|F_1F_2F_3|}\}-
h_e^{-1}\{h_e,\sqrt{|F_1F_2F_3|}\})$, computes the Poisson bracket at the
vertex. Since $h_e$ commutes with all but one of the $F_I$, we can focus on
one of them, $\sqrt{|\hat{F}_e|}$.  Keeping the power more general, the
quantization of some inverse power of flux takes the form
\begin{equation}
 \widehat{(|F|^{r-1} {\rm sgn}F)_e}=
 \frac{\hat{h}_e^{\dagger}|\hat{F}_e|^r\hat{h}_e-
   \hat{h}_e|\hat{F}_e|^r\hat{h}_e^{\dagger}}{16\pi Gr\gamma\ell_{\rm P}^2}
 =: \hat{I}_e\,.
\end{equation}
For any $0<r<1$ we quantize an inverse power of $F$ but need not use any
inverse in the commutator.  

Following \cite{InflTest}, we can now easily
simplify these operators, if we observe the relations of the
U(1)-holonomy-flux algebra, together with the reality condition. These
relations imply
\[
 \hat{h}_e^{\dagger}|\hat{F}_e|^r\hat{h}_e= |\hat{F}_e+8\pi\gamma\ell_{\rm
   P}^2|^r \quad,\quad \hat{h}_e|\hat{F}_e|^r\hat{h}_e^{\dagger}=
 |\hat{F}_e-8\pi\gamma\ell_{\rm P}^2|^r\,,
\]
such that
\begin{equation}
 \hat{I}_e= {\frac{|\hat{F}_e+8\pi\gamma\ell_{\rm
     P}^2|^r- |\hat{F}_e-8\pi\gamma\ell_{\rm
     P}^2|^r}{16\pi Gr\gamma\ell_{\rm P}^2}}\,.
\end{equation}
Eigenvalues of this operator can easily be computed, with eigenstates equal to
flux eigenstates \cite{InvScale,Ambig}. All eigenvalues are finite, as
required for a densely-defined operator, and show how the classical divergence
of $|F|^{r-1}$ at $F=0$ is cut off. For inverse-triad corrections in effective
Hamiltonians, however, we need expectation values of $\hat{I}_e$ in
semiclassical states. Explicit calculations would require good knowledge of
semiclassical wave functions or coherent states.

For general effective equations it is sufficient, and even more useful, to
perform a moment expansion, keeping the specific state free and parameterized
by moments. Staying at the expectation-value order of effective expressions,
we have
\begin{equation} \label{IExp}
 \langle\hat{I}_e\rangle= {\frac{|\langle\hat{F}_e\rangle+8\pi\gamma\ell_{\rm
     P}^2|^r- |\langle\hat{F}_e\rangle-8\pi\gamma\ell_{\rm
     P}^2|^r}{16\pi Gr\gamma\ell_{\rm P}^2}}+ \mbox{moment terms}\,.
\end{equation}
Already to this order we see characteristic corrections (depending on $\hbar$
via the Planck length). Inverse-triad corrections therefore have a
contribution independent of quantum back-reaction.

Interpreting $\langle\hat{F}\rangle=:L^2$ as the discrete quantum-gravity
scale (the lattice spacing as measured by flux operators), we find the
correction function
\begin{equation} \label{alpha}
\alpha_r(L):= \frac{\langle\hat{I}\rangle}{I_{\rm class}}=
 \frac{|L^2+8\pi\gamma\ell_{\rm P}^2|^r-
  |L^2-8\pi\gamma\ell_{\rm P}^2|^r}{16\pi\gamma r\ell_{\rm P}^2}L^{2(1-r)}
\end{equation}
that will appear in an effective Hamiltonian constraint. To leading order in
an expansion by $\hbar$ (or $\ell_{\rm P}^2/L^2$), the correction function
equals one. But even if no moment terms are included, there are quantum
corrections in the full form of $\alpha(L)$. Corrections are strong for
$L^2\sim 8\pi\gamma\ell_{\rm P}^2$ or smaller, typically in the deep quantum
regime, where $\alpha(L)$ drops to zero at $L=0$. However, even for larger
$L$, $\alpha_r(L)$ is not identical to one and implies interesting
corrections.

In addition to corrections contained in (\ref{alpha}) and quantum
back-reaction from moment terms, the flux dependence implies corrections from
a derivative expansion of the integrations involved, as already seen for
holonomies. Moreover, non-Abelian holonomies do not lead to exact
cancellations in the substitution of $h_e\{h_e^{-1},V\}$ for
$t_e^a\{A_a^i,V\}$ and rather imply additional higher-order corrections by
powers of $A_a^i$ \cite{DegFull}. As noted in the context of holonomy
corrections, such extra terms mix with higher-curvature corrections. The
leading term in (\ref{alpha}), on the other hand, shows a different dependence
on parameters that distinguish a given cosmological regime and are more
characteristic. Their effects can thus be studied in isolation.

\subsection{Diffeomorphism constraint}
\label{s:Diffeo}

We have already stated that the diffeomorphism constraint can be quantized by
its direct action on spatial functions or other objects such as curves and
surfaces. In loop quantum gravity, for instance, a diffeomorphism $\Phi$ acts
by shifting all arguments of a cylindrical function by $h_e\mapsto
h_{\Phi(e)}$, the usual pull-back of functions. The representation of the
holonomy-flux algebra is diffeomorphism covariant under this action, showing
that no quantum corrections to classical diffeomorphisms result. It is not
possible to compute or represent an infinitesimal action or the diffeomorphism
constraint because two states that differ by a non-trivial diffeomorphism are
either identical (if the diffeomorphism does not change the underlying graph)
or orthogonal. But finite diffeomorphisms suffice to remove the related gauge,
which is done without quantum corrections.

Nevertheless, the situation is not completely satisfactory because the
diffeomorphism constraint is a crucial ingredient of the
hypersurface-deformation algebra. If diffeomorphisms are represented without
quantum corrections, there should be no deformations of the relations
(\ref{DD}) and (\ref{HD}) of the hypersurface-deformation algebra for
commutators involving at least one spatial deformation. However, the
diffeomorphism constraint also appears on the right-hand side of (\ref{HH}),
the crucial part for space-time structure. On the left-hand side, we have two
Hamiltonian constraints, which we do quantize in loop quantum gravity and
whose commutators we can, in principle, compute. The result should be a
well-defined operator, which must vanish on physical states for the
quantization to be anomaly-free. However, classically it corresponds to a
diffeomorphism constraint, which cannot be represented directly. 

To check for anomaly freedom, one must then find an operator version of the
right-hand side of (\ref{HH}), taking into account the structure function
$q^{ab}$, to be turned into an operator as well. This is one of the most
important but still outstanding issues in loop quantum gravity, which was
evaded by the arguments of \cite{QSDI} and only partially addressed by the
advanced constructions of \cite{LM:Vertsm,Consist}. More recently, the issue
has been revisited in several models \cite{TwoPlusOneDef}, with encouraging
results. At least in U(1)-versions of $2+1$-dimensional gravity, one can
indeed make sense of the right-hand side of (\ref{HH}) as an operator, in such
a way that the quantum constraint algebra is anomaly-free. As a side product,
the same deformation (\ref{HHbeta}) with inverse-triad corrections as seen by
effective methods \cite{ConstraintAlgebra} appears. (Holonomy corrections and
their deformation of the constraint algebra could not be seen by the methods of
\cite{TwoPlusOneDef}, going back to \cite{LM:Vertsm,Consist}, because the
consistency conditions of anomaly freedom are tested only at vertices.)

In \cite{TwoPlusOneDef}, the diffeomorphism constraint itself did not have to
be amended by quantum corrections. However, other considerations in the same
context have been put forward that may suggest such terms \cite{DiffeoOp}. At
present, the status regarding quantum corrections in the diffeomorphism
constraint is incomplete, but a consistent implementation does not appear to
be easy. From the point of view of effective theory, corrections to
diffeomorphisms do not seem required because, in any canonical space-time
theory, one is dealing with fields as functions on space. These functions are
represented using some coordinates, but physics as always must be independent
of the choice. There must therefore be a part of the gauge content of the
theory that requires independence under arbitrary changes of spatial
coordinates or, infinitesimally, invariance under spatial Lie derivatives. But
then, a gauge transformation that amounts to a Lie derivative of all fields
must have a generator identical to the diffeomorphism constraint uniquely
associated with the fields \cite{KucharHypI}. The spatial structure assumed in
canonical formulations leaves no room for corrections in the diffeomorphism
constraint.

The space-time structure is not presupposed in canonical quantum gravity and
may well change, as indicated by some quantum corrections in the Hamiltonian
constraint. Space-time, unlike space, has dynamical content and can easily
receive quantum corrections, as borne out in loop quantum gravity. Having the
classical structure of space but modified space-time is therefore
consistent. Nevertheless, in an effort to relax some of the general
assumptions of canonical formulations, one could expect changes to the spatial
manifold structure as well, as perhaps indicated by potential corrections in
the diffeomorphism constraint such as those in \cite{DiffeoOp}.

\subsection{Quantum-geometry effects}

Comparing holonomy and inverse-triad corrections, we have several important
properties:
\begin{itemize}
\item Holonomy corrections crucially add higher powers of the connection to
  the classical quadratic form of the Hamiltonian constraint. In flat
  isotropic cosmological models, the connection is proportional to the Hubble
  parameter ${\cal H}$, which in turn is proportional to the square root of
  the energy density. Holonomy corrections in cosmological models therefore
  depend on the dimensionless parameters $\ell_{\rm P}{\cal H}$ or
  $\sqrt{\rho/\rho_{\rm P}}$, both of which are tiny in observationally
  accessible regimes. 

  Inverse-triad corrections, on the other hand, depend on
  the ratio $\ell_{\rm P}^2/L^2$ with the discrete quantum-gravity scale $L$
  in (\ref{alpha}). It is not easy to estimate $L$, but the dimensionless
  ratio associated with it certainly need not be small. Inverse-triad
  corrections can be more significant than holonomy corrections in
  observationally accessible regimes. (The scale $L$ may change in time,
  depending on the form of lattice refinement realized
  \cite{InhomLattice,CosConst}.)
\item Holonomy corrections and inverse-triad corrections are both obtained
  from properties of integrated objects, holonomies and fluxes. One should
  therefore expect not just higher-order terms as in the expansions already
  discussed, but also higher spatial derivatives in a derivative
  expansion of inhomogeneous models. For holonomy corrections,
  higher-derivative terms are crucial because they should be part of
  higher-curvature corrections together with higher powers of the connection
  that immediately arise from expanded holonomies. Only a suitable combination
  of higher powers and derivatives can result in consistent covariant
  versions.
\item Following up on the last item, we also need higher time derivatives to
  complete higher-order corrections to covariant objects related to
  curvature. Such corrections should be present even in homogeneous models,
  but are not easy to see directly from the form of holonomies. However, such
  terms cannot be ignored, because high-curvature regimes have significant
  contributions from higher-order and higher-derivative terms. In isotropic
  models, ${\cal H}^2$ and $\dot{\cal H}$ are of similar orders, both related
  to linear combinations of stress-energy components by the Friedmann and
  Raychaudhuri equations. An expansion of holonomies only by ${\cal H}$
  (related to the isotropic connection; see Section~\ref{s:ModFried}) but
  ignoring higher time derivatives would be inconsistent. To see how higher
  time derivatives arise in canonical quantum theories, we have to pause our
  description of loop quantum gravity and return to more details of quantum
  back-reaction.
\end{itemize}

\section{Quantum back-reaction}
\label{s:QBR}

For a canonical effective theory, quantum Hamiltonians and quantum constraints
$\langle\hat{H}\rangle$, generating evolution or gauge flows by (\ref{dOdt}),
must be expanded systematically by moments of states to see all quantum
effects. This is also the case for individual non-linear correction functions
such as $\langle\hat{I}\rangle$ of inverse triads (\ref{IExp}) or
$\langle\hat{h}\rangle$ of holonomies as they may be implied by
quantum-geometry effects of loop quantum gravity. Additional terms, products
of expectation values and moments, are then added to the constraints.

\subsection{Effective quantum mechanics}

The correctness of the quantum dynamics resulting from a moment-expanded
$\langle\hat{H}\rangle$ can be illustrated with a quantum-mechanical
example. We start with the well-known Ehrenfest equations
\begin{equation}\label{Ehrenfest}
 \frac{{\rm d}\langle\hat{q}\rangle}{{\rm d}t}=\langle\hat{p}\rangle/m
 \quad,\quad \frac{{\rm
     d}\langle\hat{p}\rangle}{{\rm d}t}= -\langle V'(\hat{q})\rangle
\end{equation}
for basic expectation values, computed using (\ref{dOdt}). These equations
have been analyzed by \cite{Hepp} in the limit $\hbar\to 0$ to prove that
quantum mechanics has the correct classical limit. Effective equations go
beyond this limit by performing a systematic expansion in $\hbar$. 

The first Ehrenfest equation takes exactly the classical form, while the
momentum expectation value is subject to quantum corrections: ${\rm
  d}\langle\hat{p}\rangle/{\rm d}t=-\langle V'(\hat{q})\rangle$ does not equal
the classical force $F(\langle\hat{q}\rangle)= -V'(\langle\hat{q}\rangle)$ at
position $\langle\hat{q}\rangle$ (unless the potential is at most
quadratic). Moments as quantifiers of corrections arise when we expand the
quantum force $F_Q={\rm d}\langle\hat{p}\rangle/{\rm d}t$ as
\begin{eqnarray} 
-\langle V'(\hat{q})\rangle= -\langle V'(\langle\hat{q}\rangle+
 (\hat{q}- \langle\hat{q}\rangle))\rangle= -V'(\langle\hat{q}\rangle)-
\sum_{n=2}^{\infty} \frac{1}{n!}\frac{\partial^{n+1}
 V(\langle\hat{q}\rangle)}{\partial\langle\hat{q}\rangle^{n+1}}
 \Delta(q^n)\nonumber\\
 = F_Q(\langle\hat{q}\rangle, \Delta(q^n))\,, \label{FQ}
\end{eqnarray}
or the quantum potential as
\begin{equation} \label{QPot}
V_Q(\langle\hat{q}\rangle, \Delta(q^n))= \langle V(\langle\hat{q}\rangle+
 (\hat{q}- \langle\hat{q}\rangle))\rangle= V(\langle\hat{q}\rangle)+
\sum_{n=2}^{\infty} \frac{1}{n!}\frac{\partial^{n}
 V(\langle\hat{q}\rangle)}{\partial\langle\hat{q}\rangle^n}
 \Delta(q^n)
\end{equation}
such that $-\langle V'(\hat{q})\rangle= -\partial
V_Q/\partial\langle\hat{q}\rangle$. 

\subsubsection{Quantum Hamiltonian}

The quantum potential is defined as a function on the infinite-dimensional
quantum phase space of expectation values and moments, whose Poisson structure
is given by (\ref{Poisson}). In a quantum Hamiltonian
\begin{equation}\label{HQ}
 H_Q=\langle\hat{H}\rangle = \frac{1}{2m}(\langle\hat{p}\rangle^2+
 \Delta(p^2))+ V_Q(\langle\hat{q}\rangle, \Delta(q^n))\,,
\end{equation} 
we therefore have terms generating a dynamical flow of the moments by
\begin{equation}\label{Deltadot}
 \dot{\Delta}(q^bp^c)= \{\Delta(q^bp^c),H_Q\}\,.
\end{equation}
The coupled set of equations for expectation values and moments,
(\ref{Ehrenfest}) and (\ref{Deltadot}), is equivalent to the Schr\"odinger
flow of quantum mechanics, but its solutions do not provide wave functions but
rather variables directly related to observations. It can be solved with
different approximations, most importantly a semiclassical one by the order of
moments, sometimes combined with an adiabatic one. In the latter case, applied
to anharmonic oscillators, effective equations are equivalent to those of the
low-energy effective action \cite{EffAc}. The validity and usefulness of the
canonical effective scheme is thereby established.

In the context of quantum gravity, the feature of higher time derivatives in
effective equations, a crucial ingredient of higher-curvature corrections, is
of particular interest. The moments are related to such terms, although not in
a direct way. Equation (\ref{Ehrenfest}) combined with (\ref{FQ}) already
shows that a specific linear combination of the moments, with coefficients
depending on expectation values, amounts to the time derivative of the
momentum, or the second derivative of the position expectation value. Higher
than second time derivatives of $\langle\hat{q}\rangle$ can be computed by
taking further derivatives of (\ref{Ehrenfest}) and inserting
$\dot{\Delta}(q^n)$ and, for higher than third order, $\dot{\Delta}(q^bp^c)$
according to equations of motion (\ref{Deltadot}) generated by the quantum
Hamiltonian. Different combinations of the moments therefore provide all
higher time derivatives of $\langle\hat{q}\rangle$. With the scheme just
sketched, it is difficult to invert the equations to find expressions for
moments in terms of higher time derivatives, or to eliminate all moments and
end up with a higher-derivative equation just for $\langle\hat{q}\rangle$
instead of the moment-coupled (\ref{Ehrenfest}), (\ref{FQ}) and
(\ref{Deltadot}). But with more-refined methods, as well as an adiabatic
expansion, this task can be performed. For quantum cosmology, we learn that we
must study quantum back-reaction to see all terms relevant for
higher-curvature corrections.

In semiclassical regimes, the moments by definition obey the $\hbar$ierarchy
$\Delta(q^bp^c)\sim O(\hbar^{(b+c)/2})$, as can easily be verified in
Gaussians; see (\ref{Gaussian}) and \cite{HigherMoments}. We can therefore
consider the first term $\frac{1}{2} V''(\langle\hat{q}\rangle) (\Delta q)^2$
for $n=1$ in (\ref{QPot}) as the leading semiclassical correction, providing a
quantum Hamiltonian 
\begin{equation} \label{HQsecond}
 H_Q=\langle\hat{H}\rangle = \frac{1}{2m}\langle\hat{p}\rangle^2+
 V(\langle\hat{q}\rangle)+\frac{1}{2m} 
 (\Delta p)^2+ \frac{1}{2} V''(\langle\hat{q}\rangle) (\Delta q)^2\,.
\end{equation}
(The kinetic term contributes $(\Delta p)^2/2m$, potentially of the same order
as $\frac{1}{2}V''(\langle\hat{q}\rangle)(\Delta q)^2$. But it does not appear
in a product with expectation values and therefore does not cause quantum
back-reaction.)  For equations of motion of expectation values and second-order
moments, relevant to this order, we use the Poisson brackets (\ref{FluctAlg}).
Applied to our second-order quantum Hamiltonian, we find
\begin{eqnarray}
 \frac{{\rm d} \langle\hat{q}\rangle}{{\rm d}t} &=&
 \frac{\langle\hat{p}\rangle}{m}\label{dqdt}\\
\frac{{\rm d} \langle\hat{p}\rangle}{{\rm d}t} &=& -V'(\langle\hat{q}\rangle)-
\frac{1}{2}V'''(\langle\hat{q}\rangle) (\Delta q)^2\label{dpdt}\\
\frac{{\rm d}(\Delta q)^2}{{\rm d}t} &=&
\frac{2}{m}\Delta(qp) \label{dqqdt} \\
\frac{{\rm d}\Delta(qp)}{{\rm d}t} &=& \frac{1}{m} (\Delta p)^2-
V''(\langle\hat{q}\rangle) (\Delta q)^2\label{dqpdt}\\
\frac{{\rm d}(\Delta p)^2}{{\rm d}t} &=&
-2V''(\langle\hat{q}\rangle)\Delta(qp)\,. \label{dppdt} 
\end{eqnarray}
For a given potential, one may solve these equations numerically. However, it
would be more instructive to compute $(\Delta q)^2$ and insert it in
(\ref{dpdt}) to see what quantum corrections result. So far, all equations are
coupled to one another and one cannot solve independently for $(\Delta q)^2$
(unless $V''$ is constant, the case of the harmonic oscillator, a constant
force or a free particle). But with an additional adiabatic approximation for
the moments, decoupling can be achieved.

\subsubsection{Adiabatic approximation}
\label{s:Adia}

To zeroth order in an adiabatic approximation, we assume the moments (but not
expectation values) to be time independent. We will denote the adiabatic order
by an integer subscript. Eqs. (\ref{dqqdt}) and (\ref{dppdt}) then imply
$\Delta_0(qp)=0$ at zeroth adiabatic order, and (\ref{dqpdt}) shows that
$(\Delta_0 p)^2=mV''(\langle\hat{q}\rangle) (\Delta_0 q)^2$. With the last
equation, we see that the zeroth-order adiabatic approximation cannot be valid
unless $\langle\hat{q}\rangle$ is constant in time as well. To avoid such a
restrictive condition, we proceed to higher adiabatic orders, from order $i$
to order $i+1$ by inserting time derivatives of $\Delta_i(q^bp^c)$ on the
left-hand sides of (\ref{dqqdt})--(\ref{dppdt}) to compute
$\Delta_{i+1}(q^bp^c)$ on the right-hand sides. (For a systematic
implementation of the adiabatic approximation, see \cite{EffAc,HigherTime}.)
With time derivatives known from preceding orders, the equations to solve for
the moments are initially algebraic, but additional consistency conditions
relating different orders sometimes imply differential equations for
coefficients, as we will see in this example.

To first adiabatic order,
\begin{eqnarray}
 \Delta_1(qp)&=&\frac{1}{2}m \frac{{\rm
  d}(\Delta_0 q)^2}{{\rm d}t}
= -\frac{1}{2V''(\langle\hat{q}\rangle)} \frac{{\rm
  d}(\Delta_0p)^2}{{\rm d}t} \label{Delta1qp}\\
&=& -\frac{1}{2}m\left(
\frac{V'''(\langle\hat{q}\rangle)}{V''(\langle\hat{q}\rangle)} \frac{{\rm
    d}\langle\hat{q}\rangle}{{\rm d}t} (\Delta_0q)^2+  \frac{{\rm d}
  (\Delta_0q)^2}{{\rm d}t}\right) 
\end{eqnarray}
using (\ref{dqqdt}), (\ref{dppdt}) and our zeroth-order condition relating
$(\Delta_0p)^2$ to $(\Delta_0q)^2$. The two lines can both hold
only if 
\begin{equation}
\frac{{\rm d}(\Delta_0q)^2}{{\rm d}t} =- \frac{1}{2}
\frac{V'''(\langle\hat{q}\rangle)}{V''(\langle\hat{q}\rangle)} \frac{{\rm 
    d}\langle\hat{q}\rangle}{{\rm d}t} (\Delta_0q)^2\,,
\end{equation}
solved by 
\begin{equation} \label{Delta0q}
 (\Delta_0q)^2= \frac{C}{\sqrt{V''(\langle\hat{q}\rangle)}}
\end{equation}
with a constant $C$.  Our zeroth-order adiabatic relation between the moments
then shows that $(\Delta_0p)^2= mV''(\langle\hat{q}\rangle) (\Delta_0q)^2= mC
\sqrt{V''(\langle\hat{q}\rangle)}$. Inserting these solutions in the quantum
Hamiltonian (\ref{HQsecond}), we obtain a correction
\begin{equation} \label{EffPotsecond}
 \frac{1}{2m}(\Delta_0p)^2+ \frac{1}{2}V''(\langle\hat{q}\rangle)
 (\Delta_0q)^2= C\sqrt{V''(\langle\hat{q}\rangle)}
\end{equation}
to the classical Hamiltonian. 

As one goes to higher orders in the adiabatic approximation, one takes more
and more time derivatives of $\Delta_0(q^bp^c)$. We can see this feature
already with the low-order equations found here. So far, we have used the
first adiabatic order only to restrict the zeroth-order solutions. But with
the solution (\ref{Delta0q}) found for $(\Delta_0q)^2$, we obtain from
(\ref{Delta1qp}) the moment
\begin{equation}
 \Delta_1(qp)= \frac{1}{2}m\frac{{\rm d}(\Delta_0q)^2}{{\rm d}t}= -\frac{1}{4}
 Cm \frac{V'''(\langle\hat{q}\rangle)}{V''(\langle\hat{q}\rangle)^{3/2}}
 \frac{{\rm d}\langle\hat{q}\rangle}{{\rm d}t}\,,
\end{equation}
depending on a first-order derivative of $\langle\hat{q}\rangle$. We do not
need $\Delta(qp)$ in the quantum Hamiltonian, but this pattern continues for
all moments at higher adiabatic orders, including $\Delta_i(q^n)$. When we go
beyond second adiabatic order and insert solutions into expectation-value
equations, higher-derivative effective equations will be obtained; see
\cite{HigherTime} for explicit derivations. Quantum back-reaction by moments
is responsible for these higher-derivative corrections, but there is no direct
correspondence between the moments as independent quantum degrees of freedom
and new degrees of freedom that appear in higher-derivative equations because
more initial values need to be specified. It is not the moment expansion
itself which gives rise to higher derivatives, but rather the adiabatic
expansion of individual moments. The order of moments corresponds to a
semiclassical expansion, according to $\Delta(q^bp^c)\sim O(\hbar^{(b+c)/2})$
in semiclassical states, not to a derivative expansion. Any fixed order in
$\hbar$ can produce arbitrarily high orders of time derivatives if the
adiabatic expansion is pushed further.

\subsubsection{State dependence} 

The parameter $C$ in (\ref{Delta0q}), related to second-order moments, is of
the order $\hbar$ in semiclassical states; the correction (\ref{EffPotsecond})
is therefore the first-order semiclassical correction under the assumption of
zeroth adiabatic order for the moments.  We cannot choose arbitrary values for
$C$ because the uncertainty relation (\ref{Uncert}) must be obeyed, such that
$C=m^{-1/2} \Delta_0q\Delta_0p\geq \frac{1}{2}\hbar/\sqrt{m}$. Requiring the
uncertainty relation to be saturated determines $C$. In general, this
condition may be too strong because we would assume saturation at all times,
amounting to the existence of a dynamical coherent state which is not
guaranteed for general potentials. But to zeroth adiabatic order, with the
solutions found here, such an assumption is consistent: all dependence on
$\langle\hat{q}\rangle$ drops out in the product of
$(\Delta_0q)^2(\Delta_0p)^2$ (and we have $\Delta_0(qp)=0$).

Without additional assumptions on the states solved for, or initial conditions
for the moment equations (\ref{dqqdt})--(\ref{dppdt}), the constant $C$
remains undetermined. One possibility to fix $C$, in the class of models of
this example, is to assume that solutions are close to the harmonic-oscillator
vacuum or some other specific state. If the potential
$V(q)=\frac{1}{2}m\omega^2q^2$ is harmonic, $(\Delta_0q)^2=
C/\sqrt{m\omega^2}$ is constant --- in this case there are states for which
the adiabatic approximation is exact --- and equals the Gaussian spread
$\sigma^2$ in a coherent state: we may write $C=\sigma^2\sqrt{m\omega^2}$. For
the harmonic oscillator, dynamical coherent states do exist and the
uncertainty relation may be satisfied at all times. In this case,
$C=\frac{1}{2}\hbar/\sqrt{m}$, or $(\Delta_0q)^2= \frac{1}{2}\hbar/m\omega$,
the correct relation for position fluctuations in the ground state. With
$(\Delta_0p)^2=\frac{1}{2}m\hbar\omega$, the non-classical terms in the
quantum Hamiltonian amount to the zero-point energy $\frac{1}{2}\hbar\omega$.

For a general potential $V$, we do not have the frequency parameter $\omega$
to refer to, but we can define it as the square root of $2/m$ times the
coefficient of the quadratic term in a Taylor expansion $V(q)=V_0+
V_1q+\frac{1}{2}m\omega^2q^2+\cdots$, assuming that the coefficient is not
zero. In this way, we treat higher than second-order terms in the potential as
an anharmonicity.  Specifying the class of states solved for as those that are
close to a harmonic-oscillator ground state, we can therefore write
$(\Delta_0q)^2= \frac{1}{2}\hbar/\sqrt{mV''(\langle\hat{q}\rangle)}$. The
correction $\frac{1}{2}\hbar\sqrt{V''(\langle\hat{q}\rangle)/m}$ in the
effective Hamiltonian (\ref{EffPotsecond}) then agrees with that found for the
low-energy effective action \cite{EffAcQM}, a relation that holds to higher
adiabatic orders as well \cite{EffAc}.

The canonical picture of quantum back-reaction provides an interpretation
of moment-coupling terms as an analog of loop diagrams in quantum field
theory, with moments taking the place of $n$-point functions. A formulation of
the canonical effective scheme for quantum field theory is not fully worked
out yet, but its implications for quantum gravity and cosmology can
nevertheless be seen. Already in minisuperspace models there are
characteristic effects which show cosmological implications of quantum
corrections.

\subsubsection{Notes on the WKB approximation}

The WKB approximation is often seen as implementing a semiclassical regime, in
the sense that leading terms in powers of $\hbar$ are considered in the
quantum evolution equation for states, expanded as $\psi(q)=
\exp\left(i\hbar^{-1}\sum_{n=0}^{\infty} \hbar^n S_n(q)\right)$ with an
asymptotic series. With this ansatz in the Schr\"odinger equation, one can
solve order by order in $\hbar$ to find expressions for the $S_n$: in quantum
mechanics, 
\begin{equation}
 \frac{1}{2m} \left(\frac{{\rm d}S_0}{{\rm d}q}\right)^2+ V(q)=E \quad,\quad
 i\frac{{\rm d}^2S_0}{{\rm d}q^2}+ 2\frac{{\rm d}S_0}{{\rm d} q}
 \frac{{\rm d}S_1}{{\rm d}q}=0
\end{equation}
for zeroth and first order in $\hbar$ implies $S_1=-\frac{1}{4}i
\log(2m(E-V(q))+{\rm const}$, while $S_0$ satisfies the classical
Hamilton--Jacobi equation.

Solutions obtained by the WKB approximation do not directly provide
observables such as expectation values, for which additional integrations
would be necessary. Such integrations are usually complicated to perform not
just analytically but also numerically, given the strongly oscillating nature
of WKB solutions in semiclassical regimes.  Moreover, WKB solutions do not
show how quantum corrections can be included in classical equations as the
dominant quantum effects. In particular, although quantum back-reaction is
implicitly contained in solutions to the WKB equations, it does not appear in
the form of effective potentials or quantum forces useful for intuitive
explanations of quantum effects. In the WKB approximation, $S_0$ satisfies
exactly the classical Hamilton--Jacobi equation, without any quantum
corrections. Corrections to the dynamics arise by higher orders of $S_n$ in
the wave function, but they do not appear in a form added to the
Hamilton--Jacobi (or another classical) equation.

While the WKB approximation, as an expansion in $\hbar$, does have a
semiclassical flavor, it can more generally be viewed as a formal expansion to
produce solutions for wave functions. The WKB equations are obtained by
solving the Schr\"odinger equation exactly at every order of $\hbar$: An
equation $\sum_{n=0}^{\infty}E_n\hbar^n=0$ is interpreted as implying $E_n=0$
for all $n$. From a semiclassical perspective, on the other hand, one would
interpret an equation $\sum_{n=0}^{\infty}E_n\hbar^n=0$ as providing a tower
of quantum corrections $\sum_{n=1}^{\infty}E_n\hbar^n$ to the classical
expression $E_0$, and then be interested in solutions to the equations
$\sum_{n=0}^{N}E_n\hbar^n=0$ cut off at finite orders of $\hbar$. Additional
consistency conditions are needed to determine the $E_n$ showing up in quantum
corrections. Usually, the $E_n$ for $n>0$ depend on state parameters such as
fluctuations, while $E_0$ depends only on expectation values and equals the
classical expression. A dynamical equation $\sum_{n=0}^{N}E_n\hbar^n=0$ then
encodes the quantum back-reaction of state parameters on the expectation
values, implying deviations from classical behavior, as derived systematically
by effective equations. 

In principle, one could derive such quantum corrections from WKB solutions by
computing expectation values of the $\hbar$-expanded wave functions. But the
WKB approximation does not automatically arrange the terms in its equation by
semiclassical relevance. While canonical effective equations have a direct
correspondence to the low-energy effective action, as already seen, the WKB
approximation does not produce all terms \cite{EffAcWKB}. Another question,
important in the context of quantum gravity and quantum cosmology, is the
treatment of quantum constraints (or the physical Hilbert space), which
remains open in the context of WKB solutions. (For instance, one may solve
$\hat{H}|\psi\rangle=0$ with WKB techniques, but for approximate solutions,
the gauge flow $\exp(-i\hat{H}[\epsilon]/\hbar)|\psi\rangle_{\rm WKB}$ does
not automatically vanish.)  Canonical effective techniques, on the other hand,
apply to constrained systems as well and even help to solve some long-standing
conceptual problems of quantum gravity related to constraints and gauge.

\subsubsection{Effective constraints and the problem of time}
\label{s:EffCons}

As already indicated in Section~\ref{s:Gauge}, a quantum constrained system
with constraint operators $\hat{C}$ produces quantum constraints
$C_Q:=\langle\hat{C}\rangle$, defined just like a quantum Hamiltonian
(\ref{HQ}), but also independent quantum phase-space functions $C_f=\langle
(f(\hat{q},\hat{p})-\langle f(\hat{q},\hat{p})\rangle)\hat{C}\rangle$ (in this
ordering) constrained to vanish in physical states
\cite{EffCons,EffConsRel}. In semiclassical expansions, calculating order by
order in the moments, polynomial $f(q,p)$ are sufficient. To fixed order in
the moments, only finitely many constraints are then present. Their number is
larger than the number of classical constraints because they remove not only
expectation values of constrained degrees of freedom but also the
corresponding moments.

With the ordering of constraint operators to the right of $f(q,p)$ chosen in
effective constraints, they are automatically first class if the constraint
operators are first class. There are then constraint equations to be solved,
and gauge flows to be factored out. The gauge flow is computed using the
Poisson brackets (\ref{Poisson}), affecting also the moments. Standard
techniques of constrained systems can then be used, except that moments
truncated to a fixed order usually define a non-symplectic Poisson
manifold. This feature requires some care and may have several consequences,
for instance that the number of independent gauge flows does not equal the
number of first-class constraints. Nevertheless, the usual classification of
constraints and gauge flows applies \cite{brackets}.

To see the treatment of effective constraints we consider a Hamiltonian
constraint operator $\hat{C}=\hat{p}_{\phi}^2- \hat{p}^2+W(\hat{\phi})$ for a
free, massless relativistic particle $(q,p)$ coupled to a second degree of
freedom $(\phi,p_{\phi})$ with an arbitrary $\phi$-dependent potential
$W(\phi)$. Depending on the form of $W(\phi)$, $p_{\phi}$ may become zero
along trajectories generated by the Hamiltonian constraint, in which case
$\phi$ does not serve as global internal time. On the other hand, with a
$q$-independent Hamiltonian constraint, we could deparameterize by $q$,
obtaining evolution by the
classical Hamiltonian $p= \pm \sqrt{ p_{\phi}^2+W(\phi)}$. We have equations
of motion ${\rm d}\phi/{\rm d}q= \pm p_{\phi}/\sqrt{ p_{\phi}^2+W(\phi)}$ and
${\rm d}p_{\phi}/{\rm d}q= \mp \frac{1}{2}W'(\phi)/\sqrt{
  p_{\phi}^2+W(\phi)}$. The momentum $p_{\phi}$ evolves, and could indeed
become zero.

The constraint operator gives rise to the effective constraints
\cite{EffTime,EffTimeLong}
\begin{eqnarray}
C_Q&=&\langle\hat{p}_{\phi}\rangle^2-\langle\hat{p}\rangle^2+
 (\Delta p_{\phi})^2-(\Delta p)^2+W(\langle\hat{\phi}\rangle)+
 {\textstyle\frac{1}{2}}W''(\langle\hat{\phi}\rangle)(\Delta \phi)^2
 \label{Cnon}\\
C_{\phi}&=& 2\langle\hat{p}_{\phi}\rangle\Delta(\phi p_{\phi})+ i 
 \hbar \langle\hat{p}_{\phi}\rangle
 -2p\Delta(\phi p)+W'(\langle\hat{\phi}\rangle) (\Delta \phi)^2\\
C_{p_{\phi}}&=& 2\langle\hat{p}_{\phi}\rangle(\Delta p_{\phi})^2-
 2\langle\hat{p}\rangle\Delta(p_{\phi}p)+
 W'(\langle\hat{\phi}\rangle)(\Delta(\phi p_{\phi})-
{\textstyle\frac{1}{2}}i\hbar)
\end{eqnarray}
expanded to second order in the moments, together with additional constraints
$C_q$, $C_p$, $C_{qp}$ and so on, which we will not make use of. These
constraints can be solved to find the quantum-corrected constraint surface,
and their gauge flows can be computed to find moments of observables in
physical states. Once the non-symplectic nature of the Poisson manifold of
second-order moments is taken into account, these calculations are not very
different from standard procedures.

The effective constraints shown here illustrate another important feature: the
complexity of constraints and their solutions. It comes about because
effective constraints, to be first class, are defined in a non-symmetric
ordering, while moments are by definition Weyl ordered. Reorderings required
to express effective constraints as functions of the moments then introduce
imaginary contributions by the commutator $[\hat{q},\hat{p}]=i\hbar$. For
$C_{\phi}$ and $C_{p_{\phi}}$ to vanish, some moments must be complex. While
moments before the imposition of constraints, belonging to a kinematical
Hilbert space, should be real as the expectation values of Weyl-ordered
operators, after solving the constraints one moves to the physical Hilbert
space, in general not related to a subspace of the kinematical one. After
solving the constraints, the original kinematical moments may therefore take
complex values, as long as physical observables of the quantum constrained
system are subject to reality conditions.

For further consequences, we study the problem of time in this system, using
the variable $\phi$ as internal time even though it does not deparameterize
the system globally. At the full quantum level, local internal times, free of
turning points only for finite ranges of evolution, cannot easily be made
sense of: if internal time exists only for a finite range, evolution cannot be
unitary even in this range. (See for instance the discussion in
\cite{RovelliTimeModel,RovelliTime,RovelliTimeReply}.  If states are evolved
by local internal times past their turning points, evolution freezes:
expectation values are stuck at constant values
\cite{WaldTime,WaldTimeModels}.)  This consequence is the reason why the
problem of time is much more severe at the quantum level, compared to the
classical one. At the effective level, as we will see, the problem of time can
be overcome, allowing consistent derivations of observables without using
artificial deparameterizations \cite{EffTime,EffTimeLong,EffTimeCosmo}.

If $\phi$ is used as (local) internal time, it is not represented as an
operator on the resulting physical Hilbert space, whatever it may be. No
generally manageable techniques are known to derive physical Hilbert spaces
and evolution in non-deparameterizable systems (for some possibilities of
Hilbert-space derivations, see e.g.\
\cite{WaldTime,WaldTimeModels,Master,MasterTesting}). At the effective level,
it is sufficient to distinguish $\phi$ as non-operator time by requiring that
its moments in effective constraints vanish,
\begin{equation} \label{Zeitgeist}
 (\Delta \phi)^2=\Delta(\phi q)=\Delta(\phi p)=0
\end{equation}  
while its expectation value $\langle\phi\rangle$ (denoted without the hat to
indicate that $\hat{\phi}$ no longer acts as an operator) will become the time
parameter. In fact, the conditions (\ref{Zeitgeist}) implement a good gauge
fixing of the second-order constraints $C_{\phi}$, $C_{p_{\phi}}$, $C_q$ and
$C_p$ after quantization. (With constraints on a non-symplectic Poisson
manifold, only three gauge-fixing conditions are required for four
constraints. The remaining second-order moment involving $\phi$, $\Delta(\phi
p_{\phi})$, is fixed by the constraints, as we will see shortly.)  In the
terminology of \cite{EffTime}, these conditions implement the Zeitgeist during
which $\langle\phi\rangle$ as local internal time is current.  Imposing
(\ref{Zeitgeist}) initiates the transition to physical moments --- moments
computed for states in the physical Hilbert space on which $\phi$ does not act
as an operator.

Solving the effective constraints in the given Zeitgeist, we have
$\Delta(\phi p_{\phi})=-\frac{1}{2}i\hbar$ from $C_{\phi}=0$, which then
implies
\[
(\Delta p_{\phi})^2=
\frac{\langle\hat{p}\rangle^2}{\langle\hat{p}_{\phi}\rangle^2}
(\Delta p)^2+
\frac{1}{2}i\frac{W'(\langle\phi\rangle)\hbar}{\langle\hat{p}_{\phi}\rangle}
\]
from $C_{p_{\phi}}=0$.  Inserted in (\ref{Cnon}), this implies the reduced
constraint
\begin{equation} \label{Cimag}
C=\langle\hat{p}_{\phi}\rangle^2-\langle\hat{p}\rangle^2+
 \frac{\langle\hat{p}\rangle^2-
 \langle\hat{p}_{\phi}\rangle^2}{\langle\hat{p}_{\phi}\rangle^2}(\Delta p)^2+
 \frac{1}{2}i\frac{W'(\langle\phi\rangle)\hbar}{\langle\hat{p}_{\phi}\rangle} 
 + W(\langle\phi\rangle)
\end{equation}
amounting to the quantum constraint $C_Q=\langle\hat{C}\rangle$ on the space
on which $C_{\phi}$ and $C_{p_{\phi}}$ are solved in the given
Zeitgeist. Solving $C=0$ for $\langle\hat{p}_{\phi}\rangle$, we obtain the
time-dependent Hamiltonian for $\langle\phi\rangle$-evolution, including
quantum back-reaction. However, it still contains complex terms.

In (\ref{Cimag}), all terms except the last two are expected to be real-valued
because $\langle\hat{p}\rangle$ and $\Delta p$ are physical observables, and
$\langle\hat{p}_{\phi}\rangle$ can be interpreted physically as the local
energy value. The constraint can then be satisfied, only if we allow for an
imaginary part of $\langle\phi\rangle$, calculated from
\begin{equation} \label{Vtime}
\frac{1}{2}i\frac{W'(\langle\phi\rangle)\hbar}{\langle\hat{p}_{\phi}\rangle}+
W(\langle\phi\rangle)=0\,.
\end{equation}
For semiclassical states, to which this approximation of effective constraints
refers, we can Taylor expand the potential
\[
W(\langle\phi\rangle)=W({\rm Re}\langle\phi\rangle+i\,{\rm
Im}\langle\phi\rangle)= W({\rm
Re}\langle\phi\rangle)+i\,{\rm Im}\langle\phi\rangle\,
W'({\rm Re}\langle\phi\rangle)+ O(({\rm Im}\langle\phi\rangle)^2)
\]
by the imaginary term, expected to be at least of the order $\hbar$ because it
vanishes classically. To this order, the imaginary contribution ${\rm Im}C=0$
to $C$ in (\ref{Cimag}) implies that
\begin{equation}\label{imt}
 {\rm Im}\langle\phi\rangle=
 -\frac{\hbar}{2\langle\hat{p}_{\phi}\rangle}\,. 
\end{equation}
The remaining terms,
\begin{equation}
{\rm Re}C= \langle\hat{p}_{\phi}\rangle^2-\langle\hat{p}\rangle^2+
 \frac{\langle\hat{p}\rangle^2-
 \langle\hat{p}_{\phi}\rangle^2}{\langle\hat{p}_{\phi}\rangle^2}(\Delta p)^2
 + W({\rm Re}\langle\phi\rangle)=0
\end{equation}
provide the physical ${\rm Re}\langle\phi\rangle$-Hamiltonian upon solving the
constraint equation for $\langle\hat{p}_{\phi}\rangle$. At this stage, the
Hamiltonian and its solutions, corresponding to evolving observables with
respect to $\phi$, are all real: physical reality conditions are imposed and
we have solutions corresponding to states in the physical Hilbert space.

Although imaginary parts may be unexpected, a detailed analysis of this and
other models shows that they are fully consistent \cite{EffTimeLong}. In
models in which one can compute a physical Hilbert space, results equivalent
with those shown here are obtained.  Within the effective treatment of
constraints, if we transform to a different internal time such as $q$, which
is done by a gauge transformation in the effective constrained system so that
a new Zeitgeist --- the gauge-fixing (\ref{Zeitgeist}) --- is realized, the
imaginary parts are automatically transferred from $\langle\phi\rangle$ to
$\langle q\rangle$, in such a way that observables remain real. By successive
gauge transformations, one can evolve through turning points of local internal
times, without freezing the evolution of physical observables; see in
particular the cosmological example analyzed in \cite{EffTimeCosmo}. The
imaginary part of time can be seen as a remnant of non-unitarity problems of
evolution in local-time quantum systems, but unlike in Hilbert-space
treatments, it does not pose any problems at the effective level. 

Gauge transformations in effective constrained systems show that physical
results are independent of the choice of (local) internal time. One may
deparameterize the effective system in different ways to solve the resulting
equations, without affecting observables. This conclusion, one example for
effective solutions to the traditional problems of canonical quantum gravity,
indicates that deparameterization can be used consistently. However, in
complicated systems subject to ambiguities such as factor-ordering choices,
each deparameterization must be formulated in a specific way so that they all
can result from one non-deparameterized system, effective or not. At the
effective level, all quantum constraints and Zeitgeists must be computed and
implemented with the same operator $\hat{C}$ for different local time choices
to produce mutually consistent results.  In many constructions of physical
Hilbert spaces, however, one quantizes a system with a specific
deparameterization in mind, choosing factor orderings and using possible
simplifications. In such a case, there is no guarantee that results can agree
with those obtained from other parameterizations, and the independence of
physical results of the choice of time is put at risk.

\subsection{Modified Friedmann equations, or: the sins of sines}
\label{s:ModFried}

In quantum cosmology, a systematic derivation of quantum back-reaction is
required especially for reliable evaluations of holonomy corrections in the
Hamiltonian constraint, as they both are relevant in high-curvature regimes
and contribute to higher-curvature terms. Constraints appear in such systems,
but for simplicity we will refer to deparameterized toy models. Holonomy
corrections have provided a popular class of models within loop quantum
cosmology, in which the Friedmann equation is modified in a simple way
\cite{AmbigConstr}: The classical constraint equivalent to the Friedmann
equation is first modified to
\begin{equation} \label{Hmod}
  H_{\rm mod}=  -\frac{3}{8\pi G} \frac{\sin^2(\ell c)}{\gamma^2\ell^2}
  \sqrt{|p|}+ \rho 
  |p|^{3/2}=0
\end{equation}
with a holonomy parameter $\ell$ that could possibly depend on $p$.  According
to loop quantum cosmology \cite{LivRev,Springer}, the Hamiltonian is written
in canonical triad and connection variables, with $A_a^i=c\delta_a^i$ and
$E^b_j=p\delta^b_j$, $\{c,p\}=8\pi\gamma G/3$, under the assumption of
isotropy. The densitized-triad component $p$ can be positive and negative,
according to the orientation of the triad, and is related to the scale factor
by $|p|=a^2$. (Without loss of generality regarding effective equations, we
take $p$ to be positive in what follows.)  For spatially flat models, as
assumed here, $c=\gamma\dot{a}$ is proportional to the proper-time derivative
of the scale factor.  By the modification in (\ref{Hmod}), the periodic form
of holonomies is implemented, replacing the quadratic connection dependence of
the classical expression.

Computing Hamiltonian equations of motion for $p$ allows us to eliminate $c$
in  favor of $\dot{p}$, upon which the constraint equation takes the form of
some kind of Friedmann equation. We have $\dot{p}=\{p,H_{\rm mod}\}=
(\gamma\ell)^{-1} \sin(2\ell c)\sqrt{p}$. Using trigonometric identities, we
find $\sin^2(\ell c)= \frac{1}{2}(1-\sqrt{1-4\gamma^2\ell^2\dot{a}^2})$ with
$\dot{a}=\dot{p}/(2\sqrt{p})$. Inserting this in the modified constraint and
solving for $\dot{a}^2$, the modified Friedmann equation becomes
\begin{equation} \label{ModFried}
 \left(\frac{\dot{a}}{a}\right)^2= \frac{8\pi G}{3} \rho\left(1-\frac{8\pi
     G}{3} \gamma^2\ell^2 a^2\rho\right)\,.
\end{equation}
This simple and interesting equation, with just a quadratic correction to the
energy density, has served as the basis of many ad-hoc investigations of
potential effects of loop quantum cosmology.

Equation~(\ref{ModFried}) is clearly not an effective equation in the
generality written here. Quantum back-reaction is ignored, while all terms in
the complete series expansion of holonomy corrections
\begin{equation}\label{sinexp}
\frac{\sin^2(\ell c)}{(\ell
c)^2}-1=-\frac{1}{3}\ell^2c^2+ \frac{4}{45} \ell^4c^4+\cdots
\end{equation}
are taken into account for the calculation. A consistent treatment would
include only those higher-order terms in an expansion of holonomy
modifications that are larger than any quantum back-reaction or other term
that has been ignored. The relation of quantum back-reaction to
higher-curvature corrections indicates that $c^2$-corrections in
(\ref{sinexp}) should be of comparable size to $\dot{c}$-corrections from
quantum back-reaction, the latter of which are not included in
(\ref{ModFried}). Including holonomy corrections but ignoring quantum
back-reaction is therefore inconsistent, even at leading order in the
$c$-expansion, unless one considers only models in which quantum back-reaction
is weak. (Such models do indeed exist, as we will show later, but they are
very special.) Keeping all terms in the $c$-expansion to arbitrary orders then
leads to a questionable equation.

One could think that keeping small higher-order terms is harmless, but it
turns out that our cautionary considerations do matter for the form of
modified Friedmann equations. To see this concretely, let us look at a few
examples in which the holonomy modification in the Hamiltonian constraint is
expanded first, followed by a calculation of Hamiltonian equations of motion
and a modified Friedmann equation. The first order of $c$-corrections provides
a constraint
\begin{equation}
 H_1=-\frac{3}{8\pi G}\frac{c^2}{\gamma^2} \left(1-\frac{1}{3}
   \ell^2c^2\right) \sqrt{p}+\rho p^{3/2}=0\,.
\end{equation}
Proceeding as before, we compute $\dot{p}=\{p,H_1\}=
2\gamma^{-1}c(1-\frac{2}{3}\ell^2c^2)\sqrt{p}$, solve for $c$ in terms of
$\dot{p}$, insert the result in $H_1$, and rewrite as
\begin{equation}
 \left(\frac{\dot{a}}{a}\right)^2= \frac{8\pi G}{3} \rho\left(1-\frac{8\pi
     G}{3} \gamma^2\ell^2 a^2\rho\right)\,.
\end{equation}

Rather surprisingly, the result agrees exactly with the one obtained with the
full holonomy modification, (\ref{ModFried}). However, this outcome does not
mean that higher orders in the $c$-expansion do not matter. It rather shows
that the specific form of the full modification by $\sin^2(\ell c)$ is
arranged so delicately that all higher-order contributions beyond the
$c^2$-correction precisely cancel one another.  To confirm this, we go one
order beyond the quadratic correction, modifiying the Hamiltonian constraint
by
\begin{equation}
 H_2= -\frac{3}{8\pi G}\frac{c^2}{\gamma^2} \left(1-\frac{1}{3}
   \ell^2c^2 +\frac{4}{45}\ell^4c^4\right) \sqrt{p}+\rho p^{3/2}=0\,.
\end{equation}
Again we proceed as before. (The higher higher-order polynomial equations to
be solved for a relation of $c$ to $\dot{p}$ can be handled easily within the
perturbative  scheme of the $c$-expansion.) The result,
\begin{equation}
 \left(\frac{\dot{a}}{a}\right)^2= \frac{8\pi G}{3} \rho\left(1-\frac{8\pi
     G}{3} \gamma^2\ell^2 a^2\rho+ \frac{157}{45} \left(\frac{8\pi
       G}{3}\right)^2 \gamma^4\ell^4 a^4\rho^2\right)\,,
\end{equation}
now has a higher-than-quadratic correction in the energy density.  Going to
higher orders in $c$ and following this scheme shows that also the
energy-order increases to include all possible powers. The leading corrections
in $\rho$, such as $(8\pi G/3) \gamma^2\ell^2 a^2\rho$ with the same
coefficient in all modified Friedmann equations, do not change if one goes to
higher orders in the $c$-expansion and can therefore be used consistently ---
provided one stays in energy ranges in which it is the dominant term. When the
energy density approaches Planckian levels and holonomy corrections are
strong, however, $8\pi G\gamma^2\ell^2a^2 \rho$ is close to one and all terms
in the energy expansion are relevant. Bounce scenarios, for instance, cannot
be formulated with a consistent version of the equation. (In this context,
notice that the next term beyond $\rho^2$ enters with a positive sign. At high
density, it may well be larger than the correction in (\ref{ModFried}), in
which case no zero of $\dot{a}$ and no bounce would be reached.)

As anticipated, the sine-modification has its infinitely many higher-order
terms arranged such that all but the quadratic energy correction disappear. A
consistent perturbative treatment keeping only the relevant orders instead
produces a whole series expansion by the energy density. If one has reasons to
trust the whole sine function and to exclude all other corrections,
(\ref{ModFried}) is correct. But if there are additional corrections, however
weak, it is not consistent to keep all terms in a $c$-expansion of the sine
function; instead, one has a modified Friedmann equation with a perturbative
expansion in $\rho$. Those additional corrections then unhinge the fine
balance in the sine terms that eliminated all $\rho$-corrections beyond second
order, and their form must be known for a reliable derivation of correct
effective Friedmann equations. The main source of such extra terms is, of
course, quantum back-reaction, producing higher time derivatives that compete
with higher powers of $c$. (Higher orders are also sensitive to quantization
ambiguities, as analyzed for instance in \cite{AmbigBounce}.)

\subsection{Harmonic cosmology}

To understand the interrelation between different corrections, we should have
a more detailed look at quantum back-reaction in quantum cosmology.  In an
effective description of Wheeler--DeWitt minisuperspace models, one considers
the dynamics of expectation values $\langle\hat{a}\rangle$ and
$\langle\hat{p}_a\rangle$ coupled to fluctuations and higher moments
$\Delta(a^bp_a^c)$. The coupled dynamics, including quantum back-reaction, is
usually complicated and unruly, but it simplifies considerably if
perturbations around a simple model such as the harmonic oscillator in quantum
mechanics can be used.  As an analog of the harmonic oscillator in quantum
mechanics with simple effective equations, quantum cosmology has a harmonic
model given by a free, massless scalar in a spatially flat isotropic geometry
\cite{BouncePert}.

To realize the model as one with a Hamiltonian generating evolution, we must
pick a time variable which we do by parameterizing, using the scalar $\phi$ as
time. Since it is free and massless, the Hamiltonian constraint
\begin{equation}
  H(a,p_a,p_{\phi})= -\frac{2\pi G}{3} \frac{p_a^2}{a}+
  \frac{1}{2}\frac{p_{\phi}^2}{a^3} 
\end{equation}
implies that $p_{\phi}$ is a constant of motion and $\phi$ has no turning
points where $p_{\phi}$ would move through zero. The scalar therefore provides
a global internal time. Deparameterized models, as discussed before, cannot
produce reliable physical predictions unless one can show that results do not
depend on the choice of time. In the present context, we use the model merely
to illustrate properties of quantum cosmological dynamics. For realistic
effects, one can avoid deparameterization before quantization and the
dependence on time choices by using effective constraints instead of effective
deparameterized Hamiltonians \cite{EffCons,EffConsRel,EffTime,EffTimeLong}.

It is an interesting coincidence that the same model is easily
deparameterizable and at the same time, as we will see, harmonic, without
quantum back-reaction. Both features imply that the model is extremely special
even among symmetry-reduced isotropic systems; its implications must therefore
be interpreted with a great amount of care.

We perform deparameterization by solving the Hamiltonian constraint
$H(a,p_a,p_{\phi})=0$ for 
\begin{equation}
 p_{\phi}(a,p_a)= \pm \sqrt{\frac{4\pi G}{3}} |ap_a|\,,
\end{equation}
the Hamiltonian generating evolution with respect to $\phi$. Equations of
motion for $a(\phi)$ and $p_a(\phi)$ are then obtained via Poisson brackets
with $p_{\phi}(a,p_a)$. If solutions are to be transferred back to coordinate
time, such as proper time, we solve ${\rm d}\phi/{\rm
  d}\tau=\{\phi,H(a,p_a,p_{\phi})\}= p_{\phi}/a(\phi)^3$ for $\phi(\tau)$ (with
a constant $p_{\phi}$) and insert this function in our solutions for $a(\phi)$
and $p_a(\phi)$.

\subsubsection{Effective Wheeler--DeWitt equations}

Effective deparameterized equations are generated by the quantum Hamiltonian
$\langle p_{\phi}(\hat{a},\hat{p}_a)\rangle$,
\begin{equation}\label{dOdphi}
 \frac{{\rm d}\langle\hat{O}\rangle}{{\rm d}\phi}=
 \frac{\langle[\hat{O},p_{\phi}(\hat{a},\hat{p}_a)]\rangle}{i\hbar}=
 \{\langle\hat{O}\rangle,\langle p_{\phi}(\hat{a},\hat{p}_a)\rangle\}
\end{equation}
using the Poisson brackets (\ref{Poisson}).  The absolute value in
$p_{\phi}(a,p_a)$ makes a completely general expansion in moments complicated,
but for $|p_{\phi}|$ not close to zero, there is a simple effective
Hamiltonian. If we can ensure positivity of $\widehat{ap_a}$ in evolved
states, the absolute value can be dropped. This is possible in particular for
an initial state supported solely on the positive part of the spectrum of
$\widehat{ap_a}$ (an operator for which we will assume Weyl ordering). Since
$ap_a$ is preserved by the motion it generates, also after quantization, the
evolved state will remain supported on the positive part of the spectrum of
$\widehat{ap_a}$. Unless $|p_{\phi}|$ is close to zero, it is easy to find
initial states supported only on the positive part of the spectrum of
$\widehat{ap_a}$ and with specified initial expectation values for $\hat{a}$
and $\hat{p}_a$: Projecting out negative contributions will not change the
basic expectation values much. We are then allowed to write (\ref{dOdphi}) as
\begin{eqnarray}
 \frac{{\rm d}_+\langle\hat{O}\rangle_+}{{\rm d}\phi} = \pm \sqrt{\frac{4\pi
     G}{3}} \frac{_+\langle [\hat{O},\widehat{|ap_a|}]\rangle_+}{i\hbar}=  \pm
 \sqrt{\frac{4\pi 
     G}{3}} \frac{_+\langle
   [\hat{O},\widehat{ap_a}]\rangle_+}{i\hbar}\nonumber\\
  = \pm
 \sqrt{\frac{4\pi G}{3}} \{_+\langle\hat{O}\rangle_+, _+\langle
 \widehat{ap_a}\rangle_+\} 
\end{eqnarray}
using $\widehat{|ap_a|}|\psi\rangle_+= \widehat{ap_a}|\psi\rangle_+$ (and
$\widehat{|ap_a|}^{\dagger}|\psi\rangle_+=
\widehat{ap_a}^{\dagger}|\psi\rangle_+$) on states $|\psi\rangle_+$ with
support only on the positive part of the spectrum of $\widehat{ap_a}$.

On such positively supported states, the $\phi$-Hamiltonian is quadratic and
can easily be expanded in moments. We have the quantum Hamiltonian
\begin{equation}
 H_Q= \pm \sqrt{\frac{4\pi G}{3}} (\langle\hat{a}\rangle \langle\hat{p}_a\rangle
+\Delta(ap_a))\,,
\end{equation}
free of coupling terms of expectation values and moments: there is no quantum
back-reaction. We compute and solve equations of motion for expectation
values, resulting in
\begin{equation}
\langle\hat{a}\rangle(\phi)= \exp(\pm \sqrt{4\pi G/3}\: \phi) \quad
 \mbox{and} \quad \langle\hat{p}_a\rangle(\phi)= \exp(\mp
\sqrt{4\pi G/3}\: \phi)\,.
\end{equation}
To transform to proper time, we solve ${\rm d}\phi/{\rm d}\tau=
p_{\phi}\exp(\mp \sqrt{12\pi G}\phi)$ for 
\begin{equation}
 \phi(\tau)= \pm\frac{\log(\pm \sqrt{12\pi  G}p_{\phi}\tau)}{\sqrt{12\pi G}}
\end{equation}
and obtain $\langle\hat{a}\rangle(\tau)=
(\pm \sqrt{12\pi G} p_{\phi}\tau)^{1/3}$, the classical dependence on proper
time with a stiff matter source. (In particular, even after Wheeler--DeWitt
quantization the system remains singular: infinite density
$p_{\phi}^2/(2\langle\hat{a}\rangle^3)$ is reached at finite proper time.) 

In addition to these expectation-value solutions, the state evolves such that
its second-order moments change by
\begin{equation}
\frac{{\rm d}(\Delta a)^2}{{\rm d}\phi}= \pm
 2\sqrt{\frac{4\pi G}{3}} (\Delta a)^2 \quad,\quad \frac{{\rm
  d}(\Delta p_a)^2}{{\rm d}\phi}= \mp 2\sqrt{\frac{4\pi G}{3}} (\Delta p_a)^2
\end{equation}
and ${\rm d}\Delta(ap_a)/{\rm d}\phi=0$, with solutions such that $(\Delta
a)/\langle\hat{a}\rangle$ and $(\Delta p_a)/\langle\hat{p}_a\rangle$ are
constant. Semiclassicality is preserved exactly throughout evolution in this
harmonic model, even at high density. Note that $\Delta a$ and $\Delta p_a$
change nonetheless, but have often been assumed constant when state evolution
was modeled by Gaussians. Wrong quantum corrections then result, which is
especially significant in the presence of quantum back-reaction when the
harmonic model is generalized. Especially curvature fluctuations are important
because they grow when one evolves to high density, and they do show up in
effective constraints such as (\ref{Qconstraint}) as a simple example. This
issue is another illustration of the importance of complete effective
equations including the moment dynamics.

With additional ingredients such as spatial curvature, a cosmological
constant, a scalar mass or self-interaction, anisotropy or inhomogeneity, the
system is no longer harmonic and becomes subject to quantum
back-reaction. Deviations from the classical trajectory will then occur, to be
captured by effective equations. (Some of these ingredients also remove
deparameterizability, but at the effective level we can still use local
internal times.) With a cosmological constant, for instance, the
$\phi$-Hamiltonian becomes $p_{\phi}(a,p_a)= \pm\sqrt{4\pi G/3}\: a
\sqrt{p_a^2- 4\Lambda a^4}$, a non-quadratic expression that entails coupling
terms between expectation values and moments in a quantum Hamiltonian.

\subsubsection{Harmonic loop quantum cosmology}
\label{s:HarmLQC}

At first sight, it seems that quantum-geometry corrections in loop quantum
cosmology imply quantum back-reaction by deviations from the quadratic nature
if (\ref{Hmod}) is used. This expectation is correct for inverse-triad
corrections, but holonomy corrections, although they change the quadratic
nature by higher-order terms, still lead to a harmonic model free of quantum
back-reaction \cite{BouncePert}.

To see this, we first change to connection variables $c=\gamma \dot{a}=
-(4\pi\gamma G/3) p_a/a$ and $p=a^2$, with $\{c,p\}=8\pi\gamma G/3$. The
$\phi$-Hamiltonian is still quadratic in these variables, proportional to
$|cp|$, but the holonomy modification leads us to replace $c$ by $\sin(\ell
c)/\ell$ with some $\ell$ that may depend on $p$. After this, the Hamiltonian
$p_{\phi}(c,p)$ is no longer quadratic in $c$ and $p$. However, if we
introduce a new variable $J:= 3p\exp(i\ell c)/8\pi\gamma G$, we have a linear
$\phi$-Hamiltonian
\begin{equation} \label{pphi}
 p_{\phi}= \pm 2\sqrt{\frac{4\pi G}{3}} \frac{|{\rm Im}J|}{\ell}
\end{equation}
if $\ell$ is constant.  More generally, we can assume a power law for the
$p$-dependence of $\ell(p)=\ell_0 p^x$, and define new basic variables 
\begin{equation} \label{VJ}
 V:= \frac{3p^{1-x}}{8\pi \gamma G(1-x)} \quad,\quad J:=V\exp(i\ell_0 p^xc)
\end{equation}
so that the $\phi$-Hamiltonian remains linear (and is just multiplied with
$1-x$ compared to (\ref{pphi})).  Moreover, and importantly, we have a
(non-canonical) closed algebra of basic variables,
\begin{equation}
 \{V,J\}= -i\ell_0J\quad,\quad \{V,\bar{J}\}=i\ell_0\bar{J}\quad,\quad
 \{J,\bar{J}\}= 2i\ell_0V\,,
\end{equation}
with the Hamiltonian a linear combination of the generators. 

Given these properties, upon quantization the Ehrenfest equations still
provide closed equations for expectation values, without coupling to moments
and quantum back-reaction. The Hamiltonian operator 
\begin{equation} \label{pphiJ}
 \hat{p}_{\phi}= \pm \sqrt{\frac{4\pi G}{3}} (1-x)
 \left|\frac{\hat{J}-\hat{J}^{\dagger}}{i\ell_0}\right| 
\end{equation}
is linear in $\hat{J}$ and its adjoint, and the quantum Hamiltonian is linear
in $\langle\hat{J}\rangle$ and its complex conjugate. The only additional
condition to impose is a reality condition because we have used partially
complex variables. If we initially keep $J$ and $\bar{J}$ as independent
variables, valid solutions must satisfy $J\bar{J}=V^2$.

We quantize by turning $V$ and $J$ into operators, choosing an ordering of $J$
with the exponential to the right. The classical Poisson algebra is then
replaced by the closed commutator algebra
\begin{equation} \label{VJAlg}
[\hat{V},\hat{J}]= \ell_0\hbar\hat{J}\quad,\quad
 [\hat{V},\hat{J}^{\dagger}]= -\ell_0\hbar\hat{J}^{\dagger}\quad,\quad
 [\hat{J},\hat{J}^{\dagger}]= -\ell_0\hbar(2\hat{V}+\ell_0\hbar)\,,
\end{equation}
where the last $\hbar$ comes from reordering exponentials. (Note that this
non-canonical algebra implies $(V,J)$-moments not commuting with expectation
values on the quantum phase space.) The reality condition, with the same
ordering, reads $\hat{J}\hat{J}^{\dagger}-\hat{V}^2=0$, which implies
conditions on moments upon taking an expectation value, possibly preceded by
multiplication with basic operators. For second-order moments, we have
\begin{equation}\label{Reality}
 |\langle\hat{J}\rangle|^2-(\langle\hat{V}\rangle+\ell_0\hbar/2)^2= (\Delta
 V)^2- \Delta(J\bar{J})+\frac{1}{4}\ell_0^2\hbar^2\,.
\end{equation}
(For conditions on higher moments, see \cite{HighDens}.)  The reality
condition is a Casimir of the commutator algebra of type ${\rm sl}(2,{\mathbb
  R})$ and therefore commutes with the quantum Hamiltonian proportional to
$i(\hat{J}-\hat{J}^{\dagger})$. If reality holds for initial expectation
values and moments, it holds at all times. This statement is true not only for
the harmonic model but for any Hamiltonian because the Casimir commutes with
all $\hat{V}$, $\hat{J}$ and $\hat{J}^{\dagger}$ individually, and therefore
with any function of these variables.

Solutions for expectation values obtained from the linear quantum Hamiltonian
are 
\begin{eqnarray}
 \langle\hat{V}\rangle(\phi)&=& A\exp(C\phi)+B\exp(-C\phi)\quad\mbox{ and}\\
\langle\hat{J}\rangle(\phi)&=& A\exp(C\phi)-B\exp(-C\phi)+
\frac{i\ell_0}{C} p_{\phi}\,,
\end{eqnarray}
with two integration constants $A$ and $B$ as well as the constant
$C=\pm2\sqrt{4\pi G/3}(1-x)$. The reality condition then requires that
\begin{equation}
  |\langle\hat{J}\rangle|^2-\langle\hat{V}\rangle^2= -4AB+
  \frac{\ell_0^2p_{\phi}^2}{C^2} 
\end{equation}
is of the order $\langle\hat{V}\rangle\ell_0\hbar$ (the size of semiclassical
fluctuations $(\Delta V)^2$ and $\Delta(J\bar{J})$ in (\ref{Reality})), much
smaller than $p_{\phi}^2$ for a universe which has a large amount of matter
and is semiclassical at least once. (Recall that it is sufficient to impose
the reality condition at just one time, for instance when semiclassicality is
realized.) The product $AB$ must therefore be positive, close to
$\ell_0^2p_{\phi}^2/C^2$, and the function $\langle\hat{V}\rangle(\phi)\propto
\cosh(C\phi-C\phi_0)$ never becomes zero. Holonomy modifications in the
harmonic model replace the classical singularity by a bounce.

As already seen in the beginning of this section, the bounce property of
holonomy-modified equations is very sensitive to quantum back-reaction (or
other possible corrections). With quantum back-reaction, the equations become
more complicated but can still be analyzed numerically as long as moments do
not become large. The approach to high densities can therefore be studied, but
the Planck regime remains poorly controlled. In general, it is not
known whether loop models always exhibit a bounce.

\subsubsection{Quantum Friedmann equation}

As in the case of holonomy-modified constraints, one can put effective
equations into the form of quantum Friedmann equations. In the case of
harmonic loop quantum cosmology, we eliminate $\langle\hat{J}\rangle$ in favor
of $\dot{\langle\hat{V}\rangle}$ by using the effective equations and,
importantly, the reality condition. 

We have ${\rm d}\langle\hat{V}\rangle/{\rm d}\phi=\{\langle\hat{V}\rangle,
\langle\hat{p}_{\phi}\rangle\} \propto {\rm Re} \langle\hat{J}\rangle$ using
(\ref{pphiJ}) and (\ref{VJAlg}), which we turn into a proper-time derivative,
as needed for a Friedmann-type equation, using ${\rm d}\phi/{\rm d}\tau=
p_{\phi}/a(\phi)^3$ where $a(\phi)$ is taken as some power of
$\langle\hat{V}\rangle(\phi)$, depending on $x$ according to (\ref{VJ}). (More
precisely, we write the Friedmann equation in terms of $V$ by $\dot{a}/a=
\dot{V}/(2V(1-x))$ and then use $\langle\hat{V}\rangle(\phi)$. There is no
direct relation between $\langle\hat{a}\rangle$ and $\langle\hat{V}\rangle$
because these are expectation values of different powers of the basic
$\hat{V}$ unless $x=1/2$.) The real part of $\langle\hat{J}\rangle$, which is
proportional to ${\rm d}\langle\hat{V}\rangle/{\rm d}\phi$, is related to the
imaginary part by the reality condition (\ref{Reality}). The imaginary part,
finally, is proportional to the deparameterized quantum Hamiltonian
$p_{\phi}$. When all these relations are used and $p_{\phi}$ is expressed via
the energy density of the free massless scalar assumed here, we have
\begin{equation} \label{ReJ}
  {\rm Re}\langle\hat{J}\rangle=
  \pm\langle\hat{V}\rangle\sqrt{1-(8\pi G/3)\gamma^2(\ell a)^2\rho_Q}
\end{equation}
with
\begin{equation}
 \rho_Q=\rho_{\rm free}+ (3/8\pi G)(\gamma\ell a)^{-2}
 (\Delta(J\bar{J})-(\Delta V)^2)/\langle\hat{V}\rangle^2
\end{equation}
the free energy density corrected by moment terms. If (\ref{ReJ}) is squared
and suitable factors are inserted to express all terms by $a$, an equation for
$(\dot{a}/a)^2$ follows:
\begin{equation} \label{HarmFried}
 \left(\frac{\dot{a}}{a}\right)^2=
 \frac{8\pi G}{3}\rho_{\rm free} \left(1-\frac{8\pi G}{3}\gamma^2(\ell
   a)^2\rho_Q\right)\,. 
\end{equation}
Except for the fluctuation terms in $\rho_Q$, this is Eq.~(\ref{ModFried}).

There is another, more important difference: The derivation of
(\ref{HarmFried}) holds only in the harmonic model, in which moments enter
just by the reality condition, not by quantum back-reaction. The quantum
Friedmann equation is valid in this form only if the sole matter source is a
free, massless scalar, and there is no spatial curvature, a cosmological
constant, or deviations from isotropy. When any one of these conditions is
violated, quantum back-reaction results and there are additional corrections,
not just in $\rho_Q$ but also in the general form of (\ref{HarmFried}): As
already anticipated by considering holonomy expansions in
Section~\ref{s:ModFried}, a whole series of corrections in a density expansion
appears. The quantum Friedmann equation reads
\begin{equation} \label{EffFriedW}
\left(\frac{\dot{a}}{a}\right)^2 = \frac{8\pi G}{3}\left(\rho
 \left(1- \frac{8\pi G}{3}\gamma^2(\ell   a)^2 \rho_Q\right)
 \pm\frac{1}{2}\sqrt{1-\frac{8\pi G}{3}\gamma^2(\ell   a)^2\rho_Q} 
\:\eta W+ \frac{a^6W^2}{2p_{\phi}^2}\eta^2
\right)
\end{equation}
where $W(\phi)$ is a possible scalar potential, and we have a general quantum
parameter $\eta=\sum_k\eta_{k+1}(a^6W/p_{\phi}^2)^k$ with coefficients
$\eta_k$ that depend on the moments, especially correlation parameters
\cite{QuantumBounce,BounceSqueezed}. The expansion by $a^6W/p_{\phi}^2$ can be
interpreted as one by $(\rho-P)/(\rho+P)$ with pressure $P$. (For a free,
massless scalar, $\rho=P$.) A more-specific evaluation requires the detailed
computation of quantum back-reaction to analyze how the moments evolve and
what values they take especially at high density. Techniques for numerical
studies have been provided in \cite{HighDens}.

While it is difficult to find general information about the values of moments,
it is clear that they contribute, among other effects, the canonical analog of
higher-time derivatives as they appear in higher-curvature corrections. Moment
terms and quantum back-reaction should therefore be large in high-density
regimes, near the big bang. Reliable results in loop quantum cosmology can
only state that the singularity is avoided by a bounce when matter is kinetic
dominated, in which case $W\ll \rho_{\rm kin}$ and the $\eta$-dependent terms
in (\ref{EffFriedW}) can be ignored unless $\eta$ is extremely large. This
conclusion coincides with numerical investigations
\cite{APS,APSCurved,NegCosNum} of the underlying difference equation for wave
functions. However, such numerical studies suffer from the choices required
for wave functions, for instance by an initial state. With such methods, it is
difficult to capture general-enough effects, which in quantum cosmology with
its lack of distinguished states make robust conclusions difficult. As shown
by the generality of the quantum Friedmann equations displayed here, effective
techniques allow one to draw conclusions and confirm the regime-dependent
validity of some effects even when no specific states are chosen.

\subsubsection{Cosmic forgetfulness and signature change}

There are additional and more-surprising properties of the high-density regime
that invalidate a traditional bounce interpretation even in the harmonic
model. First, staying in the isotropic context, there is cosmic forgetfulness
\cite{BeforeBB,Harmonic}: When crossing the bounce in $\phi$-evolution, some
moments change in ways so sensitive to initial values that the pre-bounce
state cannot be recovered precisely from what may be known post bounce. Using
solutions of moment equations \cite{Harmonic} or considerations of
semiclassical wave functions \cite{LoopScattering}, one can derive an
inequality
\begin{equation} \label{Forget}
  \left|1-\frac{(\Delta V)_{\phi\to\infty}}{(\Delta V)_{\phi\to-\infty}}\right|\leq
  \frac{\Delta p_{\phi}/\langle\hat{p}_{\phi}\rangle}{(\Delta
    V/\langle\hat{V}\rangle)_{\phi\to\infty}}
\end{equation}
bounding the ratio of volume fluctuations at early and late times. For a state
with fluctuations symmetric around the high-density regime near the minimum of
$\langle\hat{V}\rangle(\phi)$, the left-hand side would be near zero. The
estimate can therefore be used to shed light on the question of how much a
quantum state may change while evolving through high density, and how much
possible changes can be controlled.

With matter fluctuations $\Delta p_{\phi}/\langle\hat{p}_{\phi}\rangle$
usually much larger than geometry fluctuations $(\Delta
V/\langle\hat{V}\rangle)_{\phi\to\infty}$ at large volume, where quantum field
theory on curved space-time should be a good approximation, the right-hand
side of the inequality is much larger than one, and $(\Delta
V)_{\phi\to-\infty}$ can differ significantly from $(\Delta
V)_{\phi\to\infty}$. The inequality can be saturated by highly squeezed
dynamical coherent states \cite{Harmonic}, showing that control on pre-bounce
fluctuations cannot be improved unless states are restricted further, more
strongly than by semiclassicality.

Several classes of specific states, especially ones with weak correlations of
the canonical variables --- the volume and the Hubble parameter --- show
more-symmetric behavior of volume fluctuations. Most wave functions that can
be constructed explicitly, for instance using ${\rm sl}(2,{\mathbb
  R})$-coherent states based on the algebra (\ref{VJAlg}) as used in
\cite{GroupLQC}, are only weakly correlated and do not show all possible
asymmetries. Again, the effective viewpoint using moments instead of wave
functions provides larger generality. And even though wave functions are not
provided in explicit terms, one can show that wave functions for the moment
solutions even at saturation of (\ref{Forget}) do indeed exist: examples for
such states are dynamical coherent states \cite{BounceCohStates} saturating
the uncertainty relation, whose existence can be shown by general methods
well-known from quantum mechanics.

The second feature preventing a bounce interpretation brings us back to
quantum space-time structure. For a reliable cosmological model, we must embed
a holonomy-modified isotropic version within a consistent deformation of the
constraint algebra. Only then can we be sure that the model describes
consistent evolution of quantum space-time. No complete extension of holonomy
corrections to inhomogeneity is known, but as we will see in the next section,
existing versions of holonomy modifications at high density imply drastic
modifications with signature change, turning space-time into a quantum version
of 4-dimensional Euclidean space. This happens right where the bounce would
be, but without time and evolution in Euclidean space, a bounce interpretation
is not valid even though the model remains non-singular.

\section{Quantum geometry and dynamics of space-time}
\label{s:QGST}

So far, in Section~\ref{QGS}, we have seen the quantum geometry of {\em
  space}, with its characteristic features of discrete structures in loop
quantum gravity. To fit these modifications into a covariant quantum {\em
  space-time} structure, we must find a consistent deformation of the
hypersurface-deformation algebra of which the modified Hamiltonian constraint
is a part. As has by now become clear from many examples, derived at different
levels of effective and operator calculations, the classical constraint
algebra is then indeed deformed: quantum-geometry corrections imply modified
quantum space-time structures \cite{ConstraintAlgebra}. The possibility of
such deformations has also been suggested based on Wheeler--DeWitt
quantization \cite{BohmEuclidean}. Instead of the classical commutator of two
time deformations, we have
\begin{equation}
 \{H[N_1],H[N_2]\}= D[\beta q^{ab}(N_1\nabla_bN_2-N_2\nabla_bN_1)]
\end{equation}
with a phase-space function $\beta\not=1$.

The typical and rather universal form of these deformations is as
follows. Holonomy corrections result in a curvature-dependent
$\beta(K)=\cos(2\ell K)$, where $\ell$, related to the quantum-gravity scale
$L$, is a holonomy parameter depending on the curves used to integrate the
connection, and $K$ is a curvature component such as the Hubble parameter for
perturbations around isotropic models \cite{ScalarHol} or the rate of change
of orbit areas in spherical symmetry \cite{JR,LTBII}. Such a deformation has
also been found by operator calculations in $2+1$-models
\cite{ThreeDeform}. For inverse-triad corrections, $\beta=\alpha^2$ depends on
the inverse-triad correction function $\alpha$ as in (\ref{alpha}), which in
turn depends on the quantum-gravity scale $L$
\cite{ConstraintAlgebra,JR,LTBII}. Also here, operator calculations have
provided supporting evidence \cite{TwoPlusOneDef}.

\subsection{Example: spherical symmetry}
\label{s:SphSymm}

Spherically symmetric models with their reduced number of free fields provide
an interesting testing ground for different quantum space-time structures, and
at the same time allow physical applications for instance to black-hole
physics. In Ashtekar--Barbero variables, used to compute deformations with
corrections from loop quantum gravity, we express the canonical structure by
four fields, two scalars $A_{\varphi}$ and $E^x$ and two densitized scalars
$A_x$ and $E^{\varphi}$. They appear as components of spherically symmetric
SU(2)-connections 
\begin{equation} \label{A}
A=A_x(x)\tau_3{\rm d} x+ A_{\varphi}(x)\bar{\Lambda}_A{\rm d} \vartheta+
A_{\varphi}(x)\Lambda_A\sin\vartheta{\rm d}\varphi+ 
\tau_3\cos\vartheta{\rm d}\varphi
\end{equation}
and densitized triads
\begin{equation} \label{E}
E=E^x(x)\tau_3\sin\vartheta\frac{\partial}{\partial x}+
E^{\varphi}(x)\bar{\Lambda}^E\sin\vartheta\frac{\partial}{\partial\vartheta}+
E^{\varphi}(x)\Lambda^E\frac{\partial}{\partial\varphi}\,.
\end{equation}
For the general derivation see \cite{SymmRed,SphSymm,Springer}.

The su(2)-valued fields $\Lambda^E=\tau_1\cos(\zeta(x))+\tau_2 \sin(\zeta(x))$
and $\bar{\Lambda}^E=\tau_3^{-1}\Lambda\tau_3$ describe a U(1)-gauge freedom
with gauge rotations by $\exp(\lambda(x)\tau_3)$, remnant from the initial
SU(2)-freedom. Similarly, the densitized triad has independent su(2)-matrices
$\Lambda_A$ and $\bar{\Lambda}_A$. Also $A_x$ is affected by these gauge
transformations, under which it changes to $A_x+{\rm d}\lambda/{\rm d}x$ like
a U(1)-connection, but the combination $A_x+{\rm d}\zeta/{\rm d}x$ is
invariant, and happens to agree with an extrinsic-curvature component $K_x$
(up to a factor of $\gamma$).

With different matrices $\Lambda_A$ and $\Lambda^E$, the components
$E^{\varphi}$ and $A_{\varphi}$, unlike $E^x$ and $A_x$ or $K_x$, are not
canonically conjugate. However, if we switch to extrinsic curvature also for
$\varphi$-components, we obtain canonical pairs $\{K_x(x_1),E^x(x_2)\}=
2\gamma G \delta(x_1,x_2)$ and $\{K_{\varphi}(x_1),E^{\varphi}(x_2)\}= \gamma
G \delta(x_1,x_2)$, as shown in \cite{SphSymmHam}. In these canonical
variables, we have the diffeomorphism constraint
\begin{equation} \label{DSph}
 D_{\rm grav}[N^x]=\int{\rm d} x N^x(2K_{\varphi}'E^{\varphi}-K_xE^x{}')
\end{equation}
and the Hamiltonian constraint 
\begin{equation} \label{HL}
H[N] = -\frac{1}{2G}\int {\rm d} x N |E^x|^{-1/2}\left(
 (1-\Gamma_{\varphi}^2+K_{\varphi}^2)E^{\varphi}+2
 |E^x|K_{\varphi}K_x+ 2|E^x|\Gamma_{\varphi}' \right)  \,.
\end{equation}
with $\Gamma_{\varphi}= -(E^x)'/2E^{\varphi}$ a spin-connection
component. (Primes denote derivatives by $x$.)

The appearance of inverses of $E^x$ and quadratic expressions in $K_x$ and
$K_{\varphi}$, the latter as canonical and gauge-invariant versions of the
connection, suggests inverse-triad and holonomy corrections from loop quantum
gravity. Hamiltonian constraint operators have been constructed in
\cite{SphSymmHam}, and more generally for Gowdy models in
\cite{EinsteinRosenAsh,EinsteinRosenQuant}, in which inverse-triad operators
and holonomy operators indeed appear. An effective Hamiltonian would be
obtained from expectation values of these operators, but the calculations are
complicated. Moreover, so far these Hamiltonians could not be ensured to be
anomaly-free at the operator level. Instead, we can parameterize effective
Hamiltonians by correction functions originating from inverse-triad and
holonomy operators, and compute Poisson brackets of the modified constraints
to see under which conditions they can be anomaly-free \cite{JR}.

We write a general modified Hamiltonian constraint as 
\begin{eqnarray} \label{HQSph}
H^Q_{\rm grav}[N]&=&-\frac{1}{2G}\int \md x\,
N\bigg(\alpha|E^x|^{-1/2} E^{\varphi}f_1(K_{\varphi},K_x)+
2\bar{\alpha} |E^x|^{1/2}f_2(K_{\varphi},K_x) \nonumber\\ 
&& \qquad\qquad+\alpha_{\Gamma}|E^x|^{-1/2}(1-\Gamma_{\varphi}^2)E^{\varphi}+
2\bar{\alpha}_{\Gamma}\Gamma_{\varphi}'|E^x|^{1/2} \bigg)\,,
\end{eqnarray}
with inverse-triad correction functions $\alpha$, $\bar{\alpha}$,
$\alpha_{\Gamma}$ and $\bar{\alpha}_{\Gamma}$ initially left independent of
one another, and holonomy correction functions
$f_1$ and $f_2$. All these functions may in principle depend on all canonical
variables, although the triad dependence of inverse-triad correction functions
and the curvature dependence of holonomy correction functions should be
primary. Moreover, an anomaly-free commutator with the diffeomorphism
constraint shows that inverse-triad correction functions can only depend on
$E^x$, not on the density-weighted $E^{\varphi}$.

Computing Poisson brackets of two modified Hamiltonian constraints with
different lapse functions, it turns out that anomaly freedom can be realized
if $f_1=F_1^2$ and $f_2=K_x F_2$ provided that $F_2= F_1(\partial F_1/\partial
K_{\varphi}) \alpha/\alpha_{\Gamma}$ \cite{JR}. Choosing a function $F_1$
periodic in $K_{\varphi}$, holonomy modifications for this component are
realized. The second correction function $F_2$ is then fixed, showing how
anomaly-freedom can put restrictions on possible modifications and quantum
corrections. In fact, the corrections seem even stronger for the
$K_x$-dependence, left unmodified in the function $f_2$ shown here. An
extension to a holonomy-corrected $K_x$-dependence appears more difficult than
one of the $K_{\varphi}$-dependence. Moreover, while $K_{\varphi}$ would give
rise only to pointwise exponentials $\exp(i\gamma\ell K_{\varphi})$ as
holonomies, the curve integration along angular directions, in which
$K_{\varphi}$ points, being trivial in spherically symmetric models, $K_x$
would be replaced by a holonomy $\exp(i\gamma\int_I{\rm d}x K_x)$ integrated
along some interval $I$. Derivative corrections should therefore result as
well, or the constraint would become non-local if integrations are left
unexpanded.  

If we take $F_1(K_{\varphi})=(\gamma\ell)^{-1} \sin(\gamma\ell K_{\varphi})$,
suitable for holonomy corrections as in the cosmological example (\ref{Hmod}),
we have $F_2(K_{\varphi},E^x)=(2\gamma\ell)^{-1}\sin(2\gamma\ell
K_{\varphi})\alpha/\alpha_{\Gamma}$. The algebraic deformation is then given
by $\beta(E^x,K_{\varphi})=\bar{\alpha}\bar{\alpha}_{\Gamma} \partial
F_2/\partial K_{\varphi}$ \cite{JR}. For the example provided, this means
$\beta(E^x,K_{\varphi})=
\alpha(E^x)\bar{\alpha}(E^x)(\bar{\alpha}_{\Gamma}(E^x)/\alpha_{\Gamma}(E^x))
\cos(2\gamma\ell K_{\varphi})$, a function that is negative for $\gamma\ell
K_{\varphi}\sim \pi/2$, at curvatures where the correction function $f_1$ is
near its maximum and a strong modification of the classical linear
function. This property is realized generically: Combining the previous
equations, we can write
\begin{equation}
 \beta(E^x,K_{\varphi})= \frac{1}{2}
 \alpha\bar{\alpha}\frac{\bar{\alpha}_{\Gamma}}{\alpha_{\Gamma}} \frac{\partial^2
f_1}{\partial K_{\varphi}^2}
\end{equation}
which is negative around maxima of $f_1$, irrespective of its
functional form. Consequences of negative $\beta$ will be discussed in more
detail soon. Note also that holonomy corrections and inverse-triad corrections
are rather independent of each other in their effect on the deformation,
affecting $\beta$ multiplicatively.

The inverse-triad correction functions must satisfy
\begin{equation} \label{alphaGamma}
 (\bar{\alpha}\alpha_{\Gamma}-\alpha\bar{\alpha}_{\Gamma}) (E^x)'
 + 2(\bar{\alpha}'\bar{\alpha}_{\Gamma}-
 \bar{\alpha}\bar{\alpha}_{\Gamma}') E^x=0
\end{equation}
for a closed constraint algebra \cite{JR}.  If $F_1$ is independent of $E^x$,
or at least depends on this triad variable in a way different from
inverse-triad corrections, one can show that both terms in (\ref{alphaGamma})
must vanish individually, and we have $\alpha_{\Gamma}=\alpha$ and
$\bar{\alpha}_{\Gamma}=\bar{\alpha}$. See also \cite{Action}.

For consistent deformations in the presence of cosmological perturbations,
anomaly-freedom is implemented in the same spirit, but with an extra
ingredient. Without any symmetry assumptions, requiring irregular lattices and
non-Abelian SU(2)-features, it is complicated to derive inverse-triad
operators or to parameterize holonomy corrections. One therefore starts using
all information about such correction functions that can be obtained in
tractable models, such as a homogeneous background, and inserts those
background functions just like $\alpha$, $\bar{\alpha}$, $\alpha_{\Gamma}$,
$\bar{\alpha}_{\Gamma}$, $f_1$ and $f_2$ in spherically symmetric
models. These functions refer only to the background variables, but depending
on the order of cosmological perturbations, also the dependence on
inhomogeneity is required. The corresponding terms, in many cases, cannot be
computed directly from operators; instead, one inserts ``counterterms'' in the
Hamiltonian constraint expanded by inhomogeneity, taking into account all
possible terms to the given order that could be generated by correction
functions depending on homogeneous fields \cite{ConstraintAlgebra}. Terms that
cannot be computed from operators are left unspecified as free functions. In
many cases, counterterms contribute derivative corrections, adding for
instance terms containing $\partial_aE^b_i$ for inverse-triad corrections. A
dependence of correction functions on integrated variables such as fluxes is
therefore realized even if $\alpha$ initially depends only on local triad
values. The condition of anomaly-freedom is often so restrictive that the
counterterms can be derived uniquely from known inverse-triad or holonomy
correction functions of the background. Consistent constraints can therefore
be computed even if not all quantum corrections are known in detail.

Another small difference between spherical symmetry and cosmological
perturbations is the treatment of the SU(2)-gauge. In spherical symmetry,
(\ref{DSph}) and (\ref{HQSph}) are manifestly invariant under these
transformations, even after deformation by quantum corrections. In
perturbative treatments, on the other hand, one usually fixes a background
triad, including its SU(2)-gauge. But then, one can see that physical results
and the deformation of the space-time algebra are independent of the specific
SU(2)-fixing chosen. The constructions are therefore consistent even if some
gauge has been fixed: All gauge fixings produce the same results. (The same
arguments can be applied to the time gauge used to descend from space-time
tetrads to spatial triads.) Such consistent derivations in the presence of
gauge fixing are possible in simple cases such as the Gauss constraint, which
moreover survives unmodified after quantization. Making gauge-fixings
consistent is much less trivial if attempted for complicated constraints such
as $H[N]$. First, it is difficult to find any good gauge fixing in general
terms; having to study even all possible gauge fixings and to make sure
that physical results do not depend on the choice is then nearly
impossible. In such cases, the only manageable approach is to forgo gauge
fixing before quantization or deformation, even if it makes derivations more
complicated than in one given gauge.

\subsection{The meaning of deformed hypersurface deformations}
\label{s:Deform}

\begin{figure}
\begin{center}
\includegraphics[height=5cm]{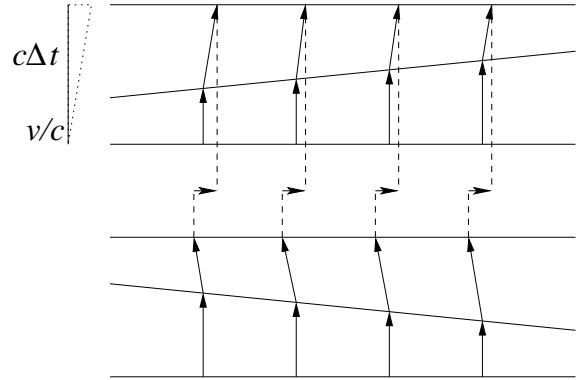}
\caption{Two linear deformations of spatial slices, first by
  $N_1=v\Delta x$ along the normals and then by  $N_2=\Delta
t-v\Delta x$ (top) and in the opposite ordering (bottom),
commute up to a spatial displacement by $\Delta x=v\Delta t$. Normals are
drawn according to Minkowski geometry, corresponding to Lorentzian signature.
\label{HypDefLinMink}}
\end{center}
\end{figure}

The hypersurface-deformation algebra encodes the space-time structure of
generally covariant theories just as the Poincar\'e algebra encodes special
relativity's structure. One can recover the Poincar\'e relations by using
functions $N$ and $N^a$ linear in some coordinates amounting to Minkowski
space-time or a local Minkowski patch. For instance, two linear lapse
functions of the form $N=\Delta t+ \vec{v}\cdot \Delta \vec{x}$, inserted in
(\ref{HH}), provide the commutator of Lorentz boosts by velocity $v$ and time
translations $\Delta t$. Inserting two such functions, $N_1=\vec{v}\cdot\Delta
\vec{x}$ and $N_2=\Delta t-\vec{v}\cdot\Delta\vec{x}$, on the right-hand side
of (\ref{HH}) shows a commutator that amounts to the displacement $\Delta
\vec{x}=\vec{v}\Delta t$. In terms of linear hypersurface deformations, the
relation follows from elementary geometry; see Fig.~\ref{HypDefLinMink}.

If the algebra is deformed, as in (\ref{HHbeta}), the same choice of linear
deformations along the normals gives rise to a rescaled relation $\Delta
x=\beta v\Delta t$. Quantum space-time, with its discrete structure that is
responsible for the algebraic deformation via holonomy and inverse-triad
corrections, changes the relation between velocity and displacement; discrete
space-time speeds up or slows down motion. Such a phenomenon is well-known
from condensed-matter physics and should not come as a surprise, although the
form in which it is realized here is rather different owing to the more-basic
notions of space and time involved.

\begin{figure}
\begin{center}
\includegraphics[height=5cm]{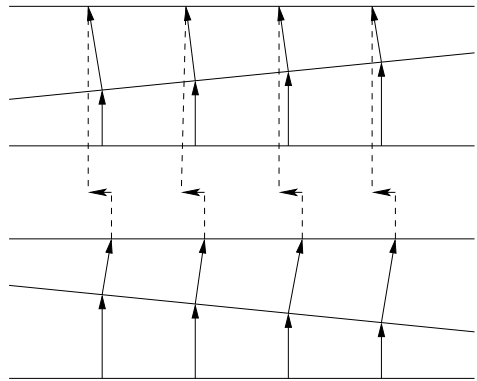}
\caption{With normals drawn according to Euclidean geometry, the displacement
  seen in Fig.~\ref{HypDefLinMink} points in the opposite direction: The sign
  in the commutator of two normal deformations is reversed.
\label{HypDefLin}}
\end{center}
\end{figure}

When $\beta$ becomes negative, as happens with holonomy corrections at high
density, the relation $\Delta x=\beta v\Delta t$ is rendered
counter-intuitive. However, the change of sign can be interpreted easily if
one redraws Fig.~\ref{HypDefLinMink} in Euclidean space, especially regarding
the directions of normals to spatial slices. As shown in Fig.~\ref{HypDefLin}
compared with Fig.~\ref{HypDefLinMink}, the displacement then indeed points in
the opposite direction (and does not change magnitude). The relation
$\Delta x=-v\Delta t$ does not describe motion but rather, despite the
notation of variables, a rotation. When $\beta$ turns negative, we have
Euclidean signature rather than Lorentzian. Even if $\beta$ is not exactly
$-1$ but negative, the structure is best described as Euclidean even though we
do not have classical Euclidean space (just as we do not have classical
Minkowski space if $\beta$ is positive but not exactly $+1$).

Signature change is not only a drastic reminder that we cannot take much of
our usual concepts for granted when space-time is quantized. It also shows the
limitations of approaches in which gauge-fixing or deparameterization is used
before quantization or modifications of the constraint. If one fixes the gauge
before quantization, one can only assume the classical space-time structure,
which does not allow signature change. After gauge fixing and quantization,
the full space-time structure can no longer be accessed, leaving one with the
conclusion that space-time is unmodified. However, since the constraints are
modified in quantization after gauge fixing and determine the gauge and
space-time structure, the procedure becomes intrinsically inconsistent.
Deparameterization cannot capture all quantum space-time effects either. When
one distinguishes a phase-space degree of freedom to measure change and
evolution, there is no guarantee that this degree of freedom actually behaves
like time in a space-time sense. Also Euclidean theories can formally be
deparameterized (internal time simply parameterizes gauge orbits of the
Hamiltonian constraint), showing that deparameterized ``evolution'' does not
necessarily imply evolution in a temporal sense. Again, only a complete
analysis of the off-shell constraint algebra, without eliminating some of its
more complicated ingredients by gauge-fixing or deparameterization, can show
what space-time structure is realized.

\begin{figure}
\begin{center}
\includegraphics[width=11cm]{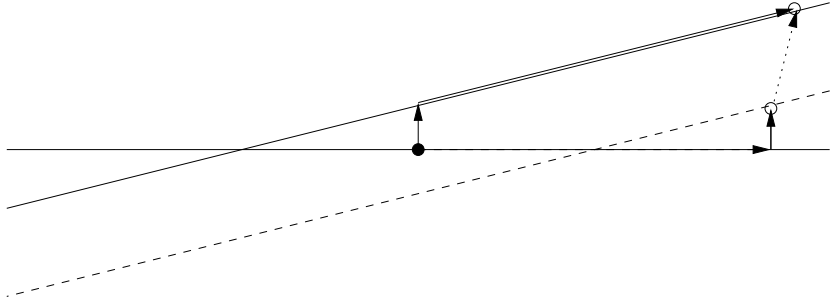}
\caption{The commutator of linear time and space deformations: A Lorentz boost
  produces the solid inclined line, the same boost performed after a spatial
  displacement the dashed line. A spatial point (full circle) is mapped to two
  different positions (open circles) depending on whether the spatial
  displacement is performed before or after the boost. The final positions
  differ by a time translation (dotted), according to $\{H[v x],D[\Delta x]\}=
  H[v\Delta x]$ using (\ref{HD}).
  \label{HDFig}}
\end{center}
\end{figure}

In our diagrams so far, we have assumed that normals are drawn using either
Minkowski or Euclidean geometry. We did not use quantum corrections of angles
even though distances and displacements did receive corrections as a
consequence of the algebra (\ref{HHbeta}). One may expect angles to change
too, in particular angles of normals to spatial slices drawn to visualize the
commutator of two time deformations. Such corrections could indeed happen, in
general deformations of space-time structures, but since the angle between
space and time directions would be involved, they would amount to deformations
of the commutator (\ref{HD}) of a time and a space deformation. A temporal and
a spatial deformation commute up to a temporal deformation, as illustrated in
Fig.~\ref{HDFig}. If there are quantum corrections to this relation, one might
interpret them as modified spatial displacements, rescaled compared to the
classical relations, or a modification in space-time angles used to define
temporal deformations along the normals. In the former case, also the purely
spatial commutator (\ref{DD}) should be modified, which is not the case. A
modified (\ref{HD}) therefore indicates quantum corrections to space-time
angles, not just to distances as indicated by a modified (\ref{HH}). For
instance, modifications which add a term of $D[\beta' NN^a]$ to the classical
result of $\{H[N],D[N^a]\}$, as possible for vector modes \cite{VectorHol} but
not with scalar modes \cite{ScalarHol}, would have an additional spatial shift
of the open circles in (\ref{HDFig}). (Note that Fig.~\ref{HDFig} also
illustrates the fact that (\ref{HD}) is not subject to a change in sign in
Euclidean signature.  If the normals are drawn according to Euclidean
geometry, just the rescaling of $\Delta x$ under boosts changes, but not the
direction of the temporal displacement.)

Based on algebraic calculations in effective loop quantum gravity,
modifications of (\ref{HD}) that seemed possible for vector modes subject to
holonomy corrections \cite{VectorHol} turned out not to be
consistent with scalar modes \cite{ScalarHol}. It therefore appears that
space-time angles are not affected by the corrections of loop quantum gravity,
but as discussed in Section~\ref{s:Diffeo}, stronger corrections to the
diffeomorphism constraint than used so far may be realized. For now, all
consistent deformations point to quantum corrections only in the commutator of
two time deformations.

\subsection{Consistency and space-time structures}
\label{s:SpaceTimeCons}

With a deformed hypersurface-deformation algebra the theory is fully
consistent. The full amount of gauge transformations exists to remove spurious
degrees of freedom, and constraints are guaranteed to be preserved by
evolution. Cosmological observables, for instance, can be computed by standard
Hamiltonian means, as developed for deformed algebras in
\cite{ConstraintAlgebra,ScalarGaugeInv}. (The classical Hamiltonian formalism
for cosmological perturbations, of \cite{HamGaugePert}, assumed
certain features of observables that are modified by deformations.)

However, other familiar notions used often in general relativity no longer
apply. Even the line element, one of the most basic mathematical objects of
differential geometry, must be treated with care --- or altogether
avoided. Constraints obeying a modified algebra generate gauge transformations
for metric components $q_{ab}$ that cannot agree with Lie derivatives or
space-time coordinate transformations --- otherwise, the form of gauge
transformations would imply the classical hypersurface-deformation
algebra. But if $q_{ab}$, or the space-time metric $g_{\mu\nu}$ completed in
the usual way using lapse and shift in
\begin{equation}
  {\rm d}s^2= g_{\mu\nu}{\rm d}x^{\mu}{\rm d}x^{\nu}= -N^2{\rm d}t^2+
  q_{ab}({\rm d}x^a+N^a{\rm d}t)({\rm d}x^b+N^b{\rm d}t)\,,
\end{equation} 
does not transform by classical
coordinate changes, the contraction $g_{\mu\nu}{\rm d}x^{\mu}{\rm d}x^{\nu}$
with standard coordinate differentials ${\rm d}x^{\mu}$ is not invariant and
cannot be used as a line element. One would have to modify transformations of
coordinate differentials as well to make the contraction invariant, in which
way one could possibly make contact with non-commutative \cite{Connes} or
fractional geometry \cite{Fractional}. (Modified hypersurface-deformation
algebras have also been found with higher-derivative dispersion relations for
matter \cite{TensorialDeform}, but the form does not appear related to what is
suggested by loop quantum gravity.)

Lacking a line element and related notions such as geodesics or trapped
surfaces, black-hole properties and even their definitions must be
rethought. One can only rely on canonical formulations of horizon conditions
that capture the well-known notions of standard space-time. Classically, there
are several different definitions of horizons which, at least in simple cases,
all provide the same results. However, when they are reformulated canonically
and adapted to quantum space-time, results may differ. As analyzed in
\cite{ModifiedHorizon}, the consistency of quantum space-time structures also
in this case helps to arrive at unambiguous answers.

For calculations in quantum space-time, one must rely on Hamiltonian methods,
using directly the modified constraints. First applied systematically in
\cite{ScalarGaugeInv}, one obtains unambiguous expressions for gauge-invariant
variables in cosmology as well as their evolution equations. Without using an
action, it may not always be obvious how to combine all equations obtained to
just one Mukhanov-type equation for the analog of the curvature perturbation,
but examples have been found for inverse-triad corrections \cite{LoopMuk} and
holonomy corrections \cite{ScalarHolMuk}. Quantum corrected wave equations are
then obtained, which directly show the modified speeds of modes as expected
from a deformed hypersurface-deformation algebra: electromagnetic and
gravitational waves obey the equation
\begin{equation} \label{Wave}
 -\frac{\partial^2w}{\partial t^2}+ \beta \Delta w+ f(a,\dot{a})w=0
\end{equation}
with the correction function $\beta$ from (\ref{HHbeta}) and a function $f$
that shows how the evolution of modes depends on an expanding
background. Density perturbations obey a similar equation, but with a
differently modified speed for inverse-triad corrections
\cite{LoopMuk}. Interesting phenomenological effects are therefore
suggested. (For holonomy corrections in currently existing versions, the
modified speeds of density perturbations and gravitational waves are identical
\cite{ScalarTensorHol}.)

Holonomy modifications at high density are especially drastic: they imply
signature change \cite{Action,SigChange}. We are no longer dealing with
quantum space-time but with a quantum version of 4-dimensional Euclidean
space, shown by the sign change of $\beta(K)$ for large $\ell K$. The
deformed hypersurface-deformation algebra (\ref{HHbeta}) then belongs to
Euclidean-type space, and (\ref{Wave}) shows the elliptic nature of linear
mode equations. For positive as well as negative $\beta$, we do not have
standard space or space-time unless the value is exactly $\pm1$; quantum
space-time effects always occur. But the distinction between positive and
negative values of $\beta$, as opposed to two different positive values, is
much more important because it changes the type of initial-value or boundary
problems in mode equations. Signature change defined by the sign of $\beta$ is
therefore physically relevant, even in the absence of a standard classical
space-time structure. In these high-density regimes, however, it is no longer
consistent to treat holonomy corrections in isolation because they mix with
quantum back-reaction, together forming higher-curvature corrections.

So far, no consistent deformation of the constraint algebra has been found for
quantum corrections caused by quantum back-reaction, adding moment-dependent
terms to the classical constraints. The problem is well-defined because we
know the Poisson algebra of moments, and even though canonical effective field
theory techniques remain incomplete, one could use those of quantum mechanical
models in loop quantum gravity restricted to fixed graphs. In any finite
region, there would then be a finite, though large, number of degrees of
freedom given by link holonomies and plaquette fluxes. (Most calculations in
the full theory are done with fixed graphs, anyway.) Since Poisson brackets of
moments always produce other moments, quantum corrections must be arranged so
that surplus terms delicately cancel in the constraint algebra. No such
version has been found yet, which is not altogether surprising because a
successful implementation would imply a consistent version of quantum
space-time, including moments of a state preserving covariance.

\subsection{Anomaly problem}

The problem of finding consistent deformations of the hypersurface-deformation
algebra is related to the long-standing anomaly issue of canonical quantum
gravity. If one can find an anomaly-free representation of quantum constraint
operators, turning the classical constraint algebra into some operator
version, effective constraints will automatically be first class and
consistent: If two constraint operators $\hat{C}_1$ and $\hat{C}_2$ commute up
to another constraint operator, the quantum constraints
$\langle\hat{C}_1\rangle$ and $\langle\hat{C}_2\rangle$ Poisson commute up to
another quantum constraint, thanks to
$\{\langle\hat{C}_1\rangle,\langle\hat{C}_2\rangle\}=
\langle[\hat{C}_1,\hat{C}_2]\rangle/i\hbar$. This statement also extends to
higher-order quantum constraints $\langle f(\hat{q},\hat{p})\hat{C}\rangle$
\cite{EffCons}. 

Structure functions in the constraint algebra cause several problems in trying
to find anomaly-free operator versions. But if such a version has been found,
there are no additional problems in the transition to effective
constraints. The product $[\hat{C}_1,\hat{C}_2]=\hat{f}\hat{C}$ with a
quantized structure function $\hat{f}$ will, after taking an expectation
value, be one of the higher-order quantum constraints.  However, there are
obstructions to simple attempts at solving the anomaly problem for constraint
operators, related for instance to the interrelation of ordering and
self-adjointness issues especially in the presence of structure functions
\cite{NonHerm}. If Hamiltonian constraint operators are self-adjoint, the
commutator $[\hat{H}[M],\hat{H}[N]]/i\hbar$ quantizing $\{H[M],H[N]\}$ in
(\ref{HH}) is self-adjoint too. But classically the bracket equals a product
of the local diffeomorphism constraint $D_a$ with the structure function
$q^{ab}$, two expressions that have a non-vanishing Poisson bracket because
$q^{ab}$ is not diffeomorphism invariant. Any quantization of the product
$q^{ab}D_a$ can then be self-adjoint and have a chance of agreeing with
$[\hat{H}[M],\hat{H}[N]]/i\hbar$ only if a symmetric ordering is used. But if
one simply reorders ``$\hat{q}^{ab}\hat{D}_a$'' symmetrically, a metric factor
would come to lie to the right of the diffeomorphism constraint, and the
product would no longer annihilate physical states. An anomalous version of
the constraints would be obtained.

In an effective description, the situation is more manageable. First, one can
use calculations of Poisson brackets instead of commutators of operators, even
in the presence of quantum corrections and ordering choices. Moreover, quantum
corrections may be suitably parameterized, to take into account different
ordering choices, quantization ambiguities in the representation of
inverse-triad and holonomy operators, and general classes of states in terms
of their moments. One can compute Poisson brackets with ambiguity functions,
such as (\ref{alpha}), or moment terms unspecified, and see what conditions
anomaly-freedom imposes on them. An operator calculation with free functions
or states, by comparison, would be much more involved. The self-adjointness
question can be left open at first by allowing for complex-valued constraints
--- kinematical moments appearing in effective constraints are complex-valued
anyway, even in the absence of structure-function issues, as seen in
Section~\ref{s:EffCons}. With this strategy the first consistent deformations
of the form (\ref{HHbeta}) have been found \cite{ConstraintAlgebra}, by now
confirmed also by operator calculations \cite{TwoPlusOneDef}. The effective
view is therefore reliable and powerful also in the context of the anomaly
problem.

\subsection{Quantum back-reaction and higher time derivatives}

Moments and their quantum back-reaction, if they appear in a consistent
deformation of the constraint algebra, are the canonical analog of
higher-curvature terms with their higher time derivatives. However, their form
is not directly one of higher time derivatives, and additional steps and
expansions, primarily an adiabatic one, are required to put canonical
effective equations in the form of higher-derivative ones; see
Section~\ref{s:Adia}. The adiabatic approximation is not always applicable. It
serves well for anharmonic oscillators expanded around the harmonic vacuum
with its constant moments. Moments of the anharmonic vacuum change, but
only slowly, and can be treated adiabatically. Solving for moments order by
order in the adiabatic expansion then shows how they are related to higher
time derivatives of expectation values. The same statements hold true in
quantum field theory, where one expands around the free vacuum in order to
describe excitations around the interacting vacuum using the low-energy
effective action.

These features have given rise to the expectation that quantum gravity should
be subject only to higher-curvature corrections. However, two hidden
assumptions are required for this conclusion. First, one assumes that quantum
gravity implies corrections only in the dynamics of gravitons, say, not in the
underlying space-time structure. Since gravity theories are fundamentally
about space-time structure, this assumption, valid in a perturbative context
of excitations on a fixed background, need not be true in general. 

The second assumption is that quantum gravity can be realized perturbatively
around some free theory of the usual form. Also this statement may be true for
perturbations around a fixed background and perhaps some other situations, but
it is not always valid. Quantum gravity is not expected to have a
non-perturbative ground state, and other distinguished states may be very
different from Gaussians as they appear in the harmonic-oscillator ground
state or free vacuum states. Such states may not allow adiabaticity of the
moments, and therefore cannot have a dynamics fully expressed by higher time
derivatives. Indeed, examples in quantum cosmology are known in which the
adiabaticity assumption is difficult to realize consistently
\cite{BouncePot}. Even for anharmonic oscillators, the adiabatic approximation
is not valid when one perturbs around correlated coherent states, such as
fully squeezed Gaussians, of the harmonic oscillator instead of the ground
state \cite{EffAc,HigherTime}.

Instead of a higher-curvature or other higher-derivative effective theory,
canonical quantum gravity has higher-dimensional effective systems in which
the moments play the role of independent degrees of freedom in addition to
expectation values. They are subject to their own equations of motion, and by
quantum back-reaction couple to them, as e.g.\ in
(\ref{dqdt})--(\ref{dppdt}). Their effect can be formulated by
higher-derivative terms only in certain regimes, but not in general.

\subsection{Effective actions}

So far, we have stayed in the canonical framework and dealt with effective
equations and constraints. Complementary information can be obtained if a
corresponding effective action is found. In principle, there is a one-to-one
correspondence between canonical and Lagrangian formulations, but in
perturbative settings, especially those that imply higher derivatives, the
transformation is far from obvious. A derivative expansion of a Lagrangian may
then correspond to a complicated resummation of a derivative-expanded
Hamiltonian. It is therefore of interest to look for and study effective
actions even if effective canonical equations are already known.

There are several examples in which effective actions have proven their
usefulness in quantum cosmology. Euclidean path-integral techniques have been
developed and applied to questions such as tunneling probabilities and some
semiclassical issues
\cite{EffAcOneLoop,EffAcTunneling,EffAcInflation,FermionDecoherence}. As
expected, in semiclassical regimes, corrections to Einstein's equation are of
higher-curvature form.  Also causal dynamical triangulations have led to
effective actions of cosmological systems by comparing detailed numerical
studies of volume fluctuations with the dynamics of minisuperspace models
\cite{CDTEffAc}. By matching volume correlation functions with results
expected from a higher-curvature effective action, quantum corrections can be
derived. A general comparison involving also possible modifications to
quantum space-time structure on top of higher-curvature corrections has not
yet been completed.  Effective actions of such forms often show more directly
than other types of effective equations when quantum effects become
significant.

Effective actions based on path integrals are subject to modified quantum
space-time structures just as canonical effective equations and
constraints. Path integration for gravitational theories requires an
integration over all metrics, with a suitable measure that preserves
covariance and does not introduce anomalies. Such a measure has not been found
in complete generality, indicating that one encounters in this approach the
same difficulties that appear when one tries to represent the canonical
constraint algebra consistently. The path-integral measure is indeed the place
to look at when one is interested in possible space-time modifications, but
its incomplete nature does not give many clues. Canonical theories, with
consistent deformations of the constraint algebra found in recent years, have
been able to make more progress in this direction. (The spin-foam approach
\cite{Rov:Loops,SpinFoamRev,PerezLivRev} attempts to take the results of loop
quantum gravity regarding background-independent representations to a level
comparable to path-integrals \cite{SurfaceSum}. However, while some aspects of
integrations over the space of metrics can be clarified, the issue of the
correct measure, or spin-foam face amplitudes, remains open also here
\cite{Anomaly}. Notwithstanding these problems, traditional effective-action
techniques have been applied to spin foams in \cite{EffAcSpinFoam}.)

Another question in path-integral based approaches is how the state dependence
enters effective actions. The low-energy effective action can be computed
quite conveniently with path integrals, but it hides the fact that one is
expanding around the ground state. If one tries to go beyond this limitation,
which in quantum cosmology with its lack of ground states is a severe one, the
required calculations become much more involved. Suitable states would have to
be implemented by additional wave-function factors in the path integral, and
integrations would no longer be Gaussian for general states. Moreover, since
path integrals in their most common form provide transition amplitudes between
two states separated by some finite time, one would have to put in the initial
and final state. Compared to effective canonical equations, which only require
initial moments to be specified, more a-priori information about the physical
system is therefore required. In quantum cosmology, this information is not
easily found by independent means, but is important for an analysis especially
of the Planckian regime.

\subsection{Regained dynamics: from canonical to Lagrangian}

In this situation, a combination of canonical and Lagrangian effective methods
is of interest. As already mentioned, it may be difficult to perform a
transformation for constrained systems, especially those with
higher-derivative terms. If the relation between the momenta and time
derivatives of configuration variables involves higher derivatives, as easily
happens with quantum-gravity corrections such as holonomy modifications, the
Lagrangian would appear as a complicated resummation of the higher-derivative
expansion of the Hamiltonian. Moreover, even if one can Legendre transform
the Hamiltonian constraint to arrive at a Lagrangian, one would, for a
comparison with path integrals, still have to look for a measure under which
the modified Lagrangian is covariant. Also integrating the Lagrangian to an
action requires new constructions even though just a classical measure on
space-time is needed. However, with space-time structures modified and
effective metrics non-existent --- see Section~\ref{s:SpaceTimeCons} ---
defining space-time integrations is non-trivial.  Measure issues, therefore,
cannot be resolved easily, but at least an effective Lagrangian for further
classical-type analysis may still be found.

Instead of attempting a Legendre transformation, the canonical methods of
\cite{Regained,LagrangianRegained} can be used to derive a Lagrangian (or the
constraints themselves) directly from a modified constraint algebra, to any
given order in derivatives. If only second-order derivatives are assumed and
the constraint algebra is classical, the Einstein--Hilbert action as a
two-parameter family with Newton's and the cosmological constant is obtained
as the unique solution. If higher derivative orders are allowed, one expects
higher-curvature corrections as well. And if even the constraint algebra is
quantum corrected, as suggested by loop quantum gravity, stronger corrections,
also to second derivative order, are obtained. With this procedure of
``regaining'' a Lagrangian from the constraint algebra one can sidestep the
complicated resummations that a Legendre transformation from modified
constraints to the higher-derivative Lagrangian would imply. Not all
coefficients in the Lagrangian may follow uniquely, especially if higher
derivatives are included leaving several options of higher-curvature
invariants of the same order. But the general form of modifications and
implications for quantum space-time structure can still be found.

\subsubsection{Functional equations}

To spell out the general procedure, let us assume that we have a Hamiltonian
or Hamiltonian constraint $H(q,p)$ depending on canonical fields $q(x)$ and
$p(x)$ in space. We introduce $\delta H/\delta p(x)=:v(x)$ as a new
independent variable in place of $p$, and then expand equations by this newly
defined $v$. If $H$ is a Hamiltonian generating evolution in some fixed time
parameter, we have $v(x)=\dot{q}(x)$ by Hamilton's equations.  If $H$ is a
Hamiltonian constraint in the absence of an absolute time, $v= (\delta
N)^{-1}\{q,H[\delta N]\}$ is the derivative of $q$ in a direction normal to
spatial slices. For gravitational variables with $q$ the metric or triad, $v$
would therefore be related to extrinsic curvature.  This change of variables
amounts to what is needed for a Legendre transformation from $(q,p)$ with
Hamiltonian $H$ to $(q,v)$ with Lagrangian $L=pv-H$, whose form will result as
a solution of the $v$-expanded equations.

Note that we cannot always assume the Hamiltonian to be local and free of
derivatives of $p$, which would imply that partial derivatives could be used
to compute $v$. Holonomy corrections in loop quantum gravity, for instance,
introduce higher spatial derivatives of the momentum conjugate to the
densitized triad. In such situations, there is no local relation between the
Hamiltonian and the Lagrangian, and explicitly performing a Legendre
transformation is complicated.

Instead of computing the transformation, we intend to calculate the Lagrangian
directly from the constraint algebra, assuming from now on the relations of
the (deformed) hypersurface-deformation algebra.  Using the definition of $v$
and
\begin{equation}
\left.\frac{\delta H}{\delta
    q(x')}\right|_{p(x)} = -\left.\frac{\delta L}{\delta q(x')}\right|_{v(x)} 
\end{equation}
for the unsmeared Hamiltonian constraint and the Lagrangian density, we write
the Poisson bracket (\ref{HHbeta}) of two Hamiltonian constraints as
\begin{eqnarray}
 \{H[N],H[M]\} &=& -\int{\rm d}^3x\int{\rm d}^3y \frac{\delta L(y)}{\delta
   q(x)}  v(x) N(y)M(x)- (N\leftrightarrow M)\\
&=& \int{\rm d}^3x \beta D^a(x) (N\nabla_aM-M\nabla_aN)
\end{eqnarray}
with the local diffeomorphism constraint $D^a$.  Taking functional derivatives
by $N$ and $M$, we arrive at the functional equation
\begin{equation} \label{FunctionalL}
 \frac{\delta L(x)}{\delta q(x')}v(x') +
 \beta(x)D^a(x)\nabla_a\delta(x,x')-(x\leftrightarrow  x')= 0
\end{equation}
for $L(x)$, which can be solved once an expression for the diffeomorphism
constraint $D^a$ is inserted, depending on whether $q$ refers to gravity or
some matter field. In all cases, $D^a$ is linear in the momenta. A linear
equation for $L$ is thus obtained \cite{LagrangianRegained}. If (\ref{DD}) and
(\ref{HD}) are unmodified, standard expressions for $D^a$ can be used. (See
Section~\ref{s:Diffeo} for a discussion of possible further modifications.)

\subsubsection{Matter}

To illustrate the regaining procedure, we look at a scalar matter field $\phi$
without derivative couplings, whose Hamiltonian obeys the (deformed)
hypersurface-deformation algebra on its own, without adding the gravitational
piece.  The Lagrangian density must be of the form ${\cal L}= \sqrt{\det g}
L(\phi,v,\psi)$ where $v=(\delta N)^{-1}\{\phi,H[\delta N]\}$, as before, is
the normal scalar velocity and $\psi= q^{ab} \nabla_a\phi\nabla_b\phi$ is the
only remaining scalar that can be formed from $\phi$ and its derivatives, to a
total derivative order of at most two. Higher derivatives may easily result in
interacting matter or quantum-gravity theories, with higher time derivatives
from quantum back-reaction and higher spatial ones, additionally, from
possible discretizations. (See e.g.\
\cite{ConsistDisc,UniformDisc,PerfectAction,CanSimp} for information about
discretized space-time theories.) For now, however, we look for modifications
implied by corrections that leave the classical derivative order unchanged.

With the canonical variables of a scalar field and its diffeomorphism
constraint $D^a=p_{\varphi}\nabla^a\phi$, Eq.~(\ref{FunctionalL}) assumes the
form
\begin{equation}
 \frac{\delta L(x)}{\delta\phi(x')} v(x')+ \beta \frac{\partial
   L(x)}{\partial v(x)} (\nabla^a\phi(x)) \nabla_a\delta(x,x') -
 (x\leftrightarrow x')=0\,.
\end{equation}
As in \cite{LagrangianRegained}, we write
\[
 \frac{\delta L(x)}{\delta \phi(x')}= \frac{\partial L(x)}{\partial \phi(x)}
 \frac{\delta\phi(x)}{\delta\phi(x')}+ 2\frac{\partial
   L(x)}{\partial\psi(x)} (\nabla^a\phi(x)) \nabla_a\delta(x,x')\,.
\]
It follows that
\[
 A^a:= (\nabla^a\phi) \left(\beta\frac{\partial L}{\partial v}+
   2v\frac{\partial L}{\partial\psi}\right)
\]
satisfies the equation $A^a(x)\nabla_a\delta(x,x')-(x\leftrightarrow x')=0$,
shown in \cite{LagrangianRegained} to imply $A^a=0$. Thus,
\[
 \beta\frac{\partial L}{\partial v}+
   2v\frac{\partial L}{\partial\psi}=0
\]
and $L$ must be of the form $L(\phi,\psi-v^2/\beta)$.  With non-trivial
deformation, $\beta\not=1$, the scalar field therefore obeys a modified
dispersion relation.  The kinetic term of the Lagrangian does not depend on
$\psi-v^2= g^{\mu\nu}(\nabla_{\mu}\phi)(\nabla_{\nu}\phi)$ in space-time
terms, but has its time derivatives in $\psi-v^2/\beta$ rescaled by the
correction function $\beta$. The resulting modified dispersion relation is in
agreement with the wave equation (\ref{Wave}).

At the canonical level, for comparison, we begin with a matter Hamiltonian
density of the form
\begin{equation}
 H= \nu \frac{p_{\phi}^2}{2\sqrt{\det q}}+ \frac{1}{2}\sigma \sqrt{\det
   q}\psi+ \sqrt{\det q} \: W(\phi)
\end{equation}
with general inverse-triad correction functions $\nu$ and $\sigma$,
and some potential $W(\phi)$. The corresponding Lagrangian density, with
$v=\nu p_{\phi}/\sqrt{\det q}$, takes the form
\begin{equation}
L= \sqrt{\det q}\left(\frac{v^2}{2\nu}-\frac{\sigma\psi}{2}-
   W(\phi)\right)= -\sqrt{\det q}\:\frac{\sigma}{2}\left(\psi-
\frac{v^2}{\beta}\right)- \sqrt{\det q} W(\phi)\,.
\end{equation}
This function has the same kinetic dependence as derived above, provided that
$\beta=\nu\sigma$, exactly the requirement for an anomaly-free constraint
algebra in the presence of inverse-triad corrections, where $\beta=\alpha^2$
\cite{ConstraintAlgebra}. Purely canonically, this consistency condition can
be seen to ensure causality in the sense that gravitational waves on quantum
space-time travel at the speed of light, modified by a factor of
$\sqrt{|\beta|}$ compared to the classical speed \cite{tensor}. No
super-luminal propagation happens when anomaly-freedom is taken into account,
even if propagation speeds may be larger than the classical speed of light for
$\alpha>1$.

\subsubsection{Effective action for inverse-triad corrections}
\label{s:EffInv}

For gravity, one example has been worked out in quite some detail with these
methods: inverse-triad corrections of loop quantum gravity. These corrections
have a characteristic component independent of higher derivatives, and
therefore can be analyzed already at the level of second-order equations. The
resulting second-order effective Lagrangian, regained from a modified
constraint algebra with a correction function $\beta$ independent of curvature
components, is
\begin{equation} \label{EffAc}
 L_{\beta}=\frac{1}{16\pi G} \sqrt{\det q}
\left(\frac{{\rm sgn} \beta}{\sqrt{|\beta|}} 
 \frac{v_{ab}v^{ab}- v^a_a v^b_b}{4}
 +  \sqrt{|\beta|} \,{}^{(3)}\!R-2\lambda\right) 
\end{equation}
with ``velocities'' $v_{ab}$, defined again as normal derivatives
\cite{Action}. For classical gravity, $v_{ab}=2K_{ab}$ would be proportional
to extrinsic curvature. Compared with the classical action obtained for
$\beta=1$, the notion of covariance has changed: Space and time derivatives
are corrected by different coefficients $\sqrt{|\beta|}$ of $^{(3)}R$ and
$|\beta|^{-1/2}{\rm sgn}(\beta)$ of $\frac{1}{4}(v_{ab}v^{ab}- v^a_av^b_b)$,
respectively. This result is consistent with the fact that the underlying
constraint algebra is modified, taking the form (\ref{HHbeta}). For this
reason, as already mentioned in the context of line elements, the manifold
structure required to integrate $L_{\beta}$ to an effective action is not
clear.

Nevertheless, the effective Lagrangian shows several characteristic
effects. First, inverse-triad corrections, having implications even without
higher-derivative or higher-order terms in $v_{ab}$, can easily be separated
from holonomy effects and higher-curvature corrections. They are especially
significant for small fluxes, where $\beta$ becomes small. With these
constructions, it is possible to use inverse-triad corrections even when
they imply strong modifications, regimes in which one would otherwise have to
include full holonomy and higher-curvature corrections as well. These latter
types of corrections are indeed present, but affect only higher orders in the
$v$-expansion. Inverse-triad effects up to quadratic order in $v$ are reliable
even if the other corrections are not known precisely.

Time derivatives are then dominant compared to the spatial Ricci scalar when
$\beta$ approaches zero at small fluxes, indicating that near-singular
geometries are controlled by homogeneous dynamics, strengthening the classical
BKL scenario \cite{BKL} by a no-singularity scenario in loop quantum gravity
\cite{NoSing}. While holonomy effects cannot easily be seen at this level
since they would manifest themselves only at higher orders in $v_{ab}$ and mix
with higher-derivative terms, they have one drastic implication at high
density if they are dominant. Then, the holonomy correction function $\beta$
becomes negative, a feature taken into account by the sign factors in
(\ref{EffAc}). When the sign changes, the signature does too: at the Lagrangian
level, one can interpret the effect by turning $t$ into $it$ in (\ref{EffAc})
with $v_{ab}$ interpreted as first-order time derivatives. (The constraint
algebra again provides a rigorous interpretation of this signature change; see
Section~\ref{s:Deform}.) At Planckian densities, holonomy effects are so
strong that they turn space-time into a quantum version of 4-dimensional
Euclidean space, lacking time and evolution. Temporal interpretations of
high-density holonomy implications, such as bounces, are incorrect. Only large
higher-derivative terms could prevent $\beta$ from turning negative, but then
other holonomy effects such as bounces would go away too.

\section{Implications for phenomenology and potential tests}

It is difficult to test quantum cosmology by any observational means, and
given the substantial lack of control over deep quantum regimes, devising
high-density scenarios of the universe remains a highly speculative
exercise. Higher-curvature corrections, as they always appear in quantum
gravity and cosmology except in the most simple harmonic models, are not
relevant at currently accessible scales. They are certainly important in the
Planckian regime, but then the present theories are so uncontrolled that it is
impossible to derive clear effects, and even if some could be suggested, they
would most likely be washed away by the immense amount of subsequent cosmic
expansion. 

If one goes beyond higher-curvature corrections as computed for
Wheeler--DeWitt quantum cosmology in \cite{WdWCMB}, using for instance
quantum-geometry effects from loop quantum gravity, the situation has a chance
of being more optimistic. Several investigations have been performed
\cite{SuperInflTensor,CCPowerLoop,TensorBackground,BounceTensor,SuperInflPowerSpec,TensorHalfII,TensorSlowRollInv,TensorSlowRoll,GravWaveLQC},
but not all are based on consistent implementations of inhomogeneity. The
general picture is therefore still incomplete. One type of corrections,
holonomy modifications, provides contributions of higher powers of the
connection or extrinsic curvature, very similar to some parts of
higher-curvature corrections. Holonomy corrections therefore cannot be
separated from higher-curvature effects, and cannot provide more-sizeable
consequences regarding observations.

Moreover, the quantum space-time structure corresponding to holonomy
corrections remains incompletely understood. Consistent deformations of the
classical constraint algebra with some holonomy-like effects are known in
spherically symmetric models \cite{JR,LTBII}, $2+1$-dimensional gravity
\cite{ThreeDeform} and for cosmological perturbations
\cite{ScalarHol}. However, in spherically symmetric and cosmological models,
only ``pointwise'' holonomy modifications have been implemented, replacing
connection components $c$ by $\exp(i\ell c)$ but not integrating over
curves. Curve integrations, on the other hand, provide additional terms which
are non-local or, when a derivative expansion (\ref{DerExp}) is used,
introduce higher spatial derivatives. Keeping only higher powers of $c$ as in
an expansion of the exponential, but ignoring spatial derivatives is not a
consistent approximation: In strong curvature regimes, where holonomy effects
should be significant, higher powers and higher derivatives of connection
components both contribute to the same order of curvature.

Inverse-triad corrections, fortunately, are much better-behaved. First, the
derivation of consistent deformations based on them is more complete, achieved
in the same kind of models --- cosmological perturbations
\cite{ConstraintAlgebra}, spherical symmetry \cite{JR,LTBII}, and
$2+1$-dimensional models \cite{TwoPlusOneDef} in which operator calculations
can be performed --- but with all crucial terms included. Also inverse-triad
corrections should come along with higher-derivative terms because they depend
on fluxes, or integrated densitized triads. However, the non-derivative
contribution is significant, as seen in Section~\ref{s:EffInv}, and, unlike
derivative terms, depends on parameters unrelated to curvature
components. Moreover, additional derivative terms are included by the
counterterms of \cite{ConstraintAlgebra}, while no connection-derivative
counterterms have been used for holonomy corrections in
\cite{ScalarHol}. Inverse-triad corrections are therefore more reliable than
holonomy corrections at the present stage of developments in loop quantum
cosmology.

Inverse-triad corrections are also more interesting from an observational
perspective. Because they do not directly refer to the curvature scale but
rather to the discrete quantum-gravity scale related to the Planck length,
there is no a-priori reason why they should be small at low curvature. They
can play a role in standard cosmological scenarios, for instance during
inflation. Indeed, in such a combined scenario, the window allowed for
inverse-triad effects is much smaller than the one for curvature or holonomy
modifications. For inverse-triad corrections, a parameter range of about four
orders of magnitude is consistent with observations
\cite{InflTest,InflConsist}, while curvature corrections have an allowed range
of about ten orders of magnitude, one compared to the ratio of densities in
observationally accessible regimes to the Planck density. There are also
indications of interesting and characteristic effects in non-Gaussianity
\cite{NonGaussInvVol}, although the required equations of motion second order
in inhomogeneity still have to be made consistent. By inverse-triad effects,
loop quantum gravity becomes falsifiable. 

For a more detailed review of phenomenological implications, see
\cite{GCPheno}. 

\section{Outlook}

Any quantum system can be evaluated consistently and reliably only when all
possible quantum effects and the relevant degrees of freedom are taken into
account.  For quantum cosmology, this means that one must go beyond the
traditional minisuperspace models and find consistent extensions to
inhomogeneity. Quantum-representation issues then become much more involved,
but can be handled for instance with methods of loop quantum gravity. In this
canonical setting, a large set of effective techniques, described in the main
part of this review, is now available. These methods allow one to forgo ad-hoc
assumptions, to implement full (but possibly deformed) space-time covariance,
and to derive a complete phenomenological setting in which all relevant
quantum effects are included. Unlike in traditional canonical quantizations and
derivations of wave functions, there do not appear to be major obstacles on
the way toward systematic comparisons with observations. Quantum cosmology is
therefore empirically testable.

Concretely working out all terms and studying the necessary parameterizations
of quantization and state ambiguities still remains to be completed. Even in
isotropic models beyond the harmonic one, not much is known about the
evolution of generic quantum states and the robustness of singularity
avoidance (see Sections~\ref{s:States} and \ref{s:QBR}). Control on
inhomogeneous modes, necessary for most physical questions in cosmology and an
understanding of quantum space-time, remains poor in strong quantum regimes,
suffering from quantization ambiguities and the difficult anomaly problem (see
Sections~\ref{s:Cova} and \ref{s:QGST}). Effective techniques, especially
effective constraints, have relieved some of the pressure caused by the
failure of traditional methods to address these problems, but they have not
been evaluated in sufficient detail to provide a reliable view on Planckian
stages in cosmology. Further in-depth investigations are required to change
this situation and to provide a complete and reliable phenomenology of quantum
cosmology.

\section*{Acknowledgements}

This work was supported in part by NSF grant PHY0748336.

\section*{References}


\begin{thebibliography}{100}

\bibitem{EffectiveGR}
J.~F.\ Donoghue,
\newblock General relativity as an effective field theory: The leading quantum
  corrections,
\newblock {\em Phys.\ Rev.\ D} 50 (1994) 3874--3888, [gr-qc/9405057]

\bibitem{BurgessLivRev}
C.~P.\ Burgess,
\newblock Quantum Gravity in Everyday Life: General Relativity as an Effective
  Field Theory,
\newblock {\em Living Rev.\ Relativity} 7 (2004), [gr-qc/0311082],
\newblock http://www.livingreviews.org/lrr-2004-5

\bibitem{QCFund}
M.\ Bojowald and C.\ Kiefer,
\newblock Quantum Cosmology: Fundamentals, [to appear]

\bibitem{Springer}
M.\ Bojowald,
\newblock {\em Quantum Cosmology: A Fundamental Theory of the Universe},
\newblock Springer, New York, 2011

\bibitem{CUP}
M.\ Bojowald,
\newblock {\em Canonical Gravity and Applications: Cosmology, Black Holes, and
  Quantum Gravity},
\newblock Cambridge University Press, Cambridge, 2010

\bibitem{GaugeInvTransPlanck}
S.\ Shankaranarayanan and M.\ Lubo,
\newblock Gauge-invariant perturbation theory for trans-Planckian inflation,
\newblock {\em Phys.\ Rev.\ D} 72 (2005) 123513, [hep-th/0507086]

\bibitem{Bardeen}
J.~M.\ Bardeen,
\newblock Gauge-invariant cosmological perturbations,
\newblock {\em Phys.\ Rev.\ D} 22 (1980) 1882--1905

\bibitem{CosmoPert}
V.~F.\ Mukhanov, H.~A.\ Feldman, and R.~H.\ Brandenberger,
\newblock Theory of cosmological perturbations,
\newblock {\em Phys.\ Rept.} 215 (1992) 203--333

\bibitem{LapseGauge}
J.~M.\ Pons, D.~C.\ Salisbury, and L.~C.\ Shepley,
\newblock Gauge transformations in the Lagrangian and Hamiltonian formalisms of
  generally covariant theories,
\newblock {\em Phys.\ Rev.\ D} 55 (1997) 658--668, [gr-qc/9612037]

\bibitem{DiracHamGR}
P.~A.~M.\ Dirac,
\newblock The theory of gravitation in Hamiltonian form,
\newblock {\em Proc.\ Roy.\ Soc.\ A} 246 (1958) 333--343

\bibitem{ConstraintAlgebra}
M.\ Bojowald, G.\ Hossain, M.\ Kagan, and S.\ Shankaranarayanan,
\newblock Anomaly freedom in perturbative loop quantum gravity,
\newblock {\em Phys.\ Rev.\ D} 78 (2008) 063547, [arXiv:0806.3929]

\bibitem{DSR}
J.\ Kowalski-Glikman,
\newblock Introduction to Doubly Special Relativity,
\newblock {\em Lect.\ Notes Phys.} 669 (2005) 131--159, [hep-th/0405273]

\bibitem{DSR2}
J.\ Magueijo and L.\ Smolin,
\newblock Lorentz Invariance with an Invariant Energy Scale,
\newblock {\em Phys.\ Rev.\ Lett.} 88 (2002) 190403

\bibitem{Rainbow}
J.\ Magueijo and L.\ Smolin,
\newblock Gravity's Rainbow,
\newblock {\em Class.\ Quant.\ Grav.} 21 (2004) 1725--1736, [gr-qc/0305055]

\bibitem{ScalarHolEv}
E.\ Wilson-Ewing,
\newblock Holonomy Corrections in the Effective Equations for Scalar Mode
  Perturbations in Loop Quantum Cosmology,
\newblock {\em Class.\ Quant.\ Grav.} 29 (2012) 085005, [arXiv:1108.6265]

\bibitem{Hybrid}
M.\ Mart{\'\i}n-Benito, L.~J.\ Garay, and G.~A.\ Mena~Marug\'an,
\newblock Hybrid Quantum Gowdy Cosmology: Combining Loop and Fock
  Quantizations,
\newblock {\em Phys.\ Rev.\ D} 78 (2008) 083516, [arXiv:0804.1098]

\bibitem{Blyth}
W.~F.\ Blyth and C.~J.\ Isham,
\newblock Quantization of a Friedmann universe filled with a scalar field,
\newblock {\em Phys.\ Rev.\ D} 11 (1975) 768--778

\bibitem{SelfAdFlat}
W.\ Kaminski and J.\ Lewandowski,
\newblock The flat FRW model in LQC: the self-adjointness,
\newblock {\em Class.\ Quant.\ Grav.} 25 (2008) 035001, [arXiv:0709.3120]

\bibitem{NonSelfAd}
W.\ Kaminski and T.\ Pawlowski,
\newblock The LQC evolution operator of FRW universe with positive cosmological
  constant,
\newblock {\em Phys.\ Rev.\ D} 81 (2010) 024014, [arXiv:0912.0162]

\bibitem{DensityOp}
W.\ Kaminski, J.\ Lewandowski, and T.\ Pawlowski,
\newblock Physical time and other conceptual issues of QG on the example of
  LQC,
\newblock {\em Class.\ Quantum Grav.} 26 (2009) 035012

\bibitem{KucharTime}
K.~V.\ Kucha\v{r},
\newblock Time and interpretations of quantum gravity,
\newblock In G.\ Kunstatter, D.~E.\ Vincent, and J.~G.\ Williams, editors, {\em
  Proceedings of the 4th Canadian Conference on General Relativity and
  Relativistic Astrophysics}, Singapore, 1992. World Scientific

\bibitem{Isham:Time}
C.~J.\ Isham,
\newblock Canonical Quantum Gravity and the Question of Time,
\newblock In J.\ Ehlers and H.\ Friedrich, editors, {\em Canonical Gravity:
  From Classical to Quantum}, pages 150--169. Springer-Verlag, Berlin,
  Heidelberg, 1994

\bibitem{AndersonTime}
E.\ Anderson,
\newblock The Problem of Time in Quantum Gravity, [arXiv:1009.2157]

\bibitem{ReducedKasner}
P.\ Malkiewicz,
\newblock Reduced phase space approach to Kasner universe and the problem of
  time in quantum theory,
\newblock {\em Class.\ Quantum Grav.} 29 (2012) 075008, [arXiv:1105.6030]

\bibitem{HigherMoments}
M.\ Bojowald, D.\ Brizuela, H.~H.\ Hernandez, M.~J.\ Koop, and H.~A.\
  Morales-T\'ecotl,
\newblock High-order quantum back-reaction and quantum cosmology with a
  positive cosmological constant,
\newblock {\em Phys.\ Rev.\ D} 84 (2011) 043514, [arXiv:1011.3022]

\bibitem{EffAcQM}
F.\ Cametti, G.\ Jona-Lasinio, C.\ Presilla, and F.\ Toninelli,
\newblock Comparison between quantum and classical dynamics in the effective
  action formalism,
\newblock In {\em Proceedings of the International School of Physics ``Enrico
  Fermi'', Course CXLIII}, pages 431--448, Amsterdam, 2000. IOS Press,
  [quant-ph/9910065]

\bibitem{MeasureUniverses}
G.~W.\ Gibbons, S.~W.\ Hawking, and J.~M.\ Stewart,
\newblock A Natural Measure On The Set Of All Universes,
\newblock {\em Nucl.\ Phys.\ B} 281 (1987) 736

\bibitem{MeasureInflation}
S.~W.\ Hawking and D.~N.\ Page,
\newblock How probable is inflation?,
\newblock {\em Nucl.\ Phys.\ B} 298 (1988) 789--809

\bibitem{MeasureCosmo}
J.~S.\ Schiffrin and R.~M.\ Wald,
\newblock Measure and Probability in Cosmology, [arXiv:1202.1818]

\bibitem{EffAc}
M.\ Bojowald and A.\ Skirzewski,
\newblock Effective Equations of Motion for Quantum Systems,
\newblock {\em Rev.\ Math.\ Phys.} 18 (2006) 713--745, [math-ph/0511043]

\bibitem{brackets}
M.\ Bojowald and T.\ Strobl,
\newblock Poisson Geometry in Constrained Systems,
\newblock {\em Rev.\ Math.\ Phys.} 15 (2003) 663--703, [hep-th/0112074]

\bibitem{Rov}
C.\ Rovelli,
\newblock {\em Quantum Gravity},
\newblock Cambridge University Press, Cambridge, UK, 2004

\bibitem{ThomasRev}
T.\ Thiemann,
\newblock {\em Introduction to Modern Canonical Quantum General Relativity},
\newblock Cambridge University Press, Cambridge, UK, 2007, [gr-qc/0110034]

\bibitem{ALRev}
A.\ Ashtekar and J.\ Lewandowski,
\newblock Background independent quantum gravity: A status report,
\newblock {\em Class.\ Quantum Grav.} 21 (2004) R53--R152, [gr-qc/0404018]

\bibitem{RS:Ham}
C.\ Rovelli and L.\ Smolin,
\newblock The physical Hamiltonian in nonperturbative quantum gravity,
\newblock {\em Phys.\ Rev.\ Lett.} 72 (1994) 446--449, [gr-qc/9308002]

\bibitem{QSDI}
T.\ Thiemann,
\newblock Quantum Spin Dynamics {(QSD)},
\newblock {\em Class.\ Quantum Grav.} 15 (1998) 839--873, [gr-qc/9606089]

\bibitem{EffCons}
M.\ Bojowald, B.\ Sandh\"ofer, A.\ Skirzewski, and A.\ Tsobanjan,
\newblock Effective constraints for quantum systems,
\newblock {\em Rev.\ Math.\ Phys.} 21 (2009) 111--154, [arXiv:0804.3365]

\bibitem{EffConsRel}
M.\ Bojowald and A.\ Tsobanjan,
\newblock Effective constraints for relativistic quantum systems,
\newblock {\em Phys.\ Rev.\ D} 80 (2009) 125008, [arXiv:0906.1772]

\bibitem{EffConsComp}
M.\ Bojowald and A.\ Tsobanjan,
\newblock Effective constraints and physical coherent states in quantum
  cosmology: A numerical comparison,
\newblock {\em Class.\ Quantum Grav.} 27 (2010) 145004, [arXiv:0911.4950]

\bibitem{EffTime}
M.\ Bojowald, P.~A.\ H\"ohn, and A.\ Tsobanjan,
\newblock An effective approach to the problem of time,
\newblock {\em Class.\ Quantum Grav.} 28 (2011) 035006, [arXiv:1009.5953]

\bibitem{EffTimeLong}
M.\ Bojowald, P.~A.\ H\"ohn, and A.\ Tsobanjan,
\newblock An effective approach to the problem of time: general features and
  examples,
\newblock {\em Phys.\ Rev.\ D} 83 (2011) 125023, [arXiv:1011.3040]

\bibitem{EffTimeCosmo}
P.~A.\ H\"ohn, E.\ Kubalova, and A.\ Tsobanjan,
\newblock Effective relational dynamics of the closed FRW model universe
  minimally coupled to a massive scalar field, [arXiv:1111.5193]

\bibitem{AshVar}
A.\ Ashtekar,
\newblock New Hamiltonian Formulation of General Relativity,
\newblock {\em Phys.\ Rev.\ D} 36 (1987) 1587--1602

\bibitem{AshVarReell}
J.~F.\ Barbero~G.,
\newblock Real Ashtekar Variables for Lorentzian Signature Space-Times,
\newblock {\em Phys.\ Rev.\ D} 51 (1995) 5507--5510, [gr-qc/9410014]

\bibitem{Immirzi}
G.\ Immirzi,
\newblock Real and Complex Connections for Canonical Gravity,
\newblock {\em Class.\ Quantum Grav.} 14 (1997) L177--L181

\bibitem{FluxAlg}
H.\ Sahlmann,
\newblock Some Comments on the Representation Theory of the Algebra Underlying
  Loop Quantum Gravity, [gr-qc/0207111]

\bibitem{Meas}
H.\ Sahlmann,
\newblock When Do Measures on the Space of Connections Support the Triad
  Operators of Loop Quantum Gravity?, [gr-qc/0207112]

\bibitem{LOST}
J.\ Lewandowski, A.\ Oko\l\'ow, H.\ Sahlmann, and T.\ Thiemann,
\newblock Uniqueness of diffeomorphism invariant states on holonomy-flux
  algebras,
\newblock {\em Commun.\ Math.\ Phys.} 267 (2006) 703--733, [gr-qc/0504147]

\bibitem{WeylRep}
C.\ Fleischhack,
\newblock Representations of the Weyl Algebra in Quantum Geometry,
\newblock {\em Commun.\ Math.\ Phys.} 285 (2009) 67--140, [math-ph/0407006]

\bibitem{ALMMT}
A.\ Ashtekar, J.\ Lewandowski, D.\ Marolf, J.\ Mour\~ao, and T.\ Thiemann,
\newblock Quantization of Diffeomorphism Invariant Theories of Connections with
  Local Degrees of Freedom,
\newblock {\em J.\ Math.\ Phys.} 36 (1995) 6456--6493, [gr-qc/9504018]

\bibitem{AreaVol}
C.\ Rovelli and L.\ Smolin,
\newblock Discreteness of Area and Volume in Quantum Gravity,
\newblock {\em Nucl.\ Phys.\ B} 442 (1995) 593--619, [gr-qc/9411005],
\newblock Erratum: {\em Nucl.\ Phys.\ B} 456 (1995) 753

\bibitem{Area}
A.\ Ashtekar and J.\ Lewandowski,
\newblock Quantum Theory of Geometry I: Area Operators,
\newblock {\em Class.\ Quantum Grav.} 14 (1997) A55--A82, [gr-qc/9602046]

\bibitem{Vol2}
A.\ Ashtekar and J.\ Lewandowski,
\newblock Quantum Theory of Geometry II: Volume Operators,
\newblock {\em Adv.\ Theor.\ Math.\ Phys.} 1 (1998) 388--429, [gr-qc/9711031]

\bibitem{NonExpLQC}
M.\ Varadarajan,
\newblock On the resolution of the big bang singularity in isotropic Loop
  Quantum Cosmology,
\newblock {\em Class. Quantum Grav.} 26 (2009) 085006, [arXiv:0812.0272]

\bibitem{TwoPlusOneDef}
A.\ Henderson, A.\ Laddha, and C.\ Tomlin,
\newblock Constraint algebra in LQG reloaded : Toy model of a ${\rm U}(1)^{3}$
  Gauge Theory I, [arXiv:1204.0211]

\bibitem{BouncePert}
M.\ Bojowald,
\newblock Large scale effective theory for cosmological bounces,
\newblock {\em Phys.\ Rev.\ D} 75 (2007) 081301(R), [gr-qc/0608100]

\bibitem{BounceCohStates}
M.\ Bojowald,
\newblock Dynamical coherent states and physical solutions of quantum
  cosmological bounces,
\newblock {\em Phys.\ Rev.\ D} 75 (2007) 123512, [gr-qc/0703144]

\bibitem{QSDV}
T.\ Thiemann,
\newblock {QSD V}: Quantum Gravity as the Natural Regulator of Matter Quantum
  Field Theories,
\newblock {\em Class.\ Quantum Grav.} 15 (1998) 1281--1314, [gr-qc/9705019]

\bibitem{InflTest}
M.\ Bojowald, G.\ Calcagni, and S.\ Tsujikawa,
\newblock Observational test of inflation in loop quantum cosmology,
\newblock {\em JCAP} 11 (2001) 046, [arXiv:1107.1540]

\bibitem{InvScale}
M.\ Bojowald,
\newblock Inverse Scale Factor in Isotropic Quantum Geometry,
\newblock {\em Phys.\ Rev.\ D} 64 (2001) 084018, [gr-qc/0105067]

\bibitem{Ambig}
M.\ Bojowald,
\newblock Quantization ambiguities in isotropic quantum geometry,
\newblock {\em Class.\ Quantum Grav.} 19 (2002) 5113--5130, [gr-qc/0206053]

\bibitem{DegFull}
M.\ Bojowald,
\newblock Degenerate Configurations, Singularities and the Non-Abelian Nature
  of Loop Quantum Gravity,
\newblock {\em Class.\ Quantum Grav.} 23 (2006) 987--1008, [gr-qc/0508118]

\bibitem{LM:Vertsm}
J.\ Lewandowski and D.\ Marolf,
\newblock Loop Constraints: A Habitat and their Algebra,
\newblock {\em Int.\ J.\ Mod.\ Phys.\ D} 7 (1998) 299--330, [gr-qc/9710016]

\bibitem{Consist}
R.\ Gambini, J.\ Lewandowski, D.\ Marolf, and J.\ Pullin,
\newblock On the Consistency of the Constraint Algebra in Spin Network Quantum
  Gravity,
\newblock {\em Int.\ J.\ Mod.\ Phys.\ D} 7 (1998) 97--109, [gr-qc/9710018]

\bibitem{DiffeoOp}
A.\ Laddha and M.\ Varadarajan,
\newblock The Diffeomorphism Constraint Operator in Loop Quantum Gravity,
  [arXiv:1105.0636]

\bibitem{KucharHypI}
K.~V.\ Kucha\v{r},
\newblock Geometry of hypersurfaces. I,
\newblock {\em J.\ Math.\ Phys.} 17 (1976) 777--791

\bibitem{InhomLattice}
M.\ Bojowald,
\newblock Loop quantum cosmology and inhomogeneities,
\newblock {\em Gen.\ Rel.\ Grav.} 38 (2006) 1771--1795, [gr-qc/0609034]

\bibitem{CosConst}
M.\ Bojowald,
\newblock The dark side of a patchwork universe,
\newblock {\em Gen.\ Rel.\ Grav.} 40 (2008) 639--660, [arXiv:0705.4398]

\bibitem{Hepp}
K.\ Hepp,
\newblock The Classical Limit for Quantum Mechanical Correlation Functions,
\newblock {\em Commun.\ Math.\ Phys.} 35 (1974) 265--277

\bibitem{HigherTime}
M.\ Bojowald, S.\ Brahma, and E.\ Nelson,
\newblock Higher time derivatives in effective equations of canonical quantum
  systems, [arXiv:1208.1242]

\bibitem{EffAcWKB}
N.~C.\ Dias, A.\ Mikovic, and J.~N.\ Prata,
\newblock Coherent States Expectation Values as Semiclassical Trajectories,
\newblock {\em J.\ Math.\ Phys.} 47 (2006) 082101, [hep-th/0507255]

\bibitem{RovelliTimeModel}
C.\ Rovelli,
\newblock Quantum Mechanics Without Time: A Model,
\newblock {\em Phys.\ Rev.\ D} 42 (1990) 2638--2646

\bibitem{RovelliTime}
C.\ Rovelli,
\newblock Time in Quantum Gravity: An Hypothesis,
\newblock {\em Phys.\ Rev.\ D} 43 (1991) 442--456

\bibitem{RovelliTimeReply}
C.\ Rovelli,
\newblock Quantum evolving constants: Reply to comment on `Time in quantum
  gravity: An Hypothesis',
\newblock {\em Phys.\ Rev.\ D} 44 (1991) 1339--1341

\bibitem{WaldTime}
R.~M.\ Wald,
\newblock A Proposal for solving the `problem of time' in canonical quantum
  gravity,
\newblock {\em Phys.\ Rev.\ D} 48 (1993) 2377--2381, [gr-qc/9305024]

\bibitem{WaldTimeModels}
A.\ Higuchi and R.~M.\ Wald,
\newblock Applications of a new proposal for solving the `problem of time' to
  some simple quantum cosmological models,
\newblock {\em Phys.\ Rev.\ D} 51 (1995) 544--561, [gr-qc/9407038]

\bibitem{Master}
T.\ Thiemann,
\newblock The Phoenix Project: Master Constraint Programme for Loop Quantum
  Gravity,
\newblock {\em Class.\ Quant.\ Grav.} 23 (2006) 2211--2248, [gr-qc/0305080]

\bibitem{MasterTesting}
B.\ Dittrich and T.\ Thiemann,
\newblock Testing the Master Constraint Programme for Loop Quantum Gravity I.
  General Framework,
\newblock {\em Class.\ Quant.\ Grav.} 23 (2006) 1025--1066, [gr-qc/0411138]

\bibitem{AmbigConstr}
K.\ Vandersloot,
\newblock On the Hamiltonian Constraint of Loop Quantum Cosmology,
\newblock {\em Phys.\ Rev.\ D} 71 (2005) 103506, [gr-qc/0502082]

\bibitem{LivRev}
M.\ Bojowald,
\newblock Loop Quantum Cosmology,
\newblock {\em Living Rev.\ Relativity} 11 (2008) 4, [gr-qc/0601085],
\newblock {\tt http://www.livingreviews.org/lrr-2008-4}

\bibitem{AmbigBounce}
O.\ Hrycyna, J.\ Mielczarek, and M.\ Szyd{\l}owski,
\newblock Effects of the quantisation ambiguities on the Big Bounce dynamics,
\newblock {\em Gen.\ Rel. Grav.} 41 (2009) 1025--1049, [arXiv:0804.2778]

\bibitem{HighDens}
M.\ Bojowald, D.\ Mulryne, W.\ Nelson, and R.\ Tavakol,
\newblock The high-density regime of kinetic-dominated loop quantum cosmology,
\newblock {\em Phys.\ Rev.\ D} 82 (2010) 124055, [arXiv:1004.3979]

\bibitem{QuantumBounce}
M.\ Bojowald,
\newblock Quantum nature of cosmological bounces,
\newblock {\em Gen.\ Rel.\ Grav.} 40 (2008) 2659--2683, [arXiv:0801.4001]

\bibitem{BounceSqueezed}
M.\ Bojowald,
\newblock How quantum is the big bang?,
\newblock {\em Phys.\ Rev.\ Lett.} 100 (2008) 221301, [arXiv:0805.1192]

\bibitem{APS}
A.\ Ashtekar, T.\ Pawlowski, and P.\ Singh,
\newblock Quantum Nature of the Big Bang: An Analytical and Numerical
  Investigation,
\newblock {\em Phys.\ Rev.\ D} 73 (2006) 124038, [gr-qc/0604013]

\bibitem{APSCurved}
A.\ Ashtekar, T.\ Pawlowski, P.\ Singh, and K.\ Vandersloot,
\newblock Loop quantum cosmology of k=1 FRW models,
\newblock {\em Phys.\ Rev.\ D} 75 (2007) 024035, [gr-qc/0612104]

\bibitem{NegCosNum}
E.\ Bentivegna and T.\ Pawlowski,
\newblock Anti-deSitter universe dynamics in LQC,
\newblock {\em Phys.\ Rev.\ D} 77 (2008) 124025, [arXiv:0803.4446]

\bibitem{BeforeBB}
M.\ Bojowald,
\newblock What happened before the big bang?,
\newblock {\em Nature Physics} 3 (2007) 523--525

\bibitem{Harmonic}
M.\ Bojowald,
\newblock Harmonic cosmology: How much can we know about a universe before the
  big bang?,
\newblock {\em Proc.\ Roy.\ Soc.\ A} 464 (2008) 2135--2150, [arXiv:0710.4919]

\bibitem{LoopScattering}
W.\ Kaminski and T.\ Pawlowski,
\newblock Cosmic recall and the scattering picture of Loop Quantum Cosmology,
\newblock {\em Phys.\ Rev.\ D} 81 (2010) 084027, [arXiv:1001.2663]

\bibitem{GroupLQC}
E.~R.\ Livine and M.\ Mart\'{\i}n-Benito,
\newblock Group theoretical Quantization of Isotropic Loop Cosmology,
  [arXiv:1204.0539]

\bibitem{BohmEuclidean}
N.\ Pinto-Neto and E.~S.\ Santini,
\newblock Must Quantum Spacetimes Be Euclidean?,
\newblock {\em Phys.\ Rev.\ D} 59 (1999) 123517, [arXiv:gr-qc/9811067]

\bibitem{ScalarHol}
T.\ Cailleteau, J.\ Mielczarek, A.\ Barrau, and J.\ Grain,
\newblock Anomaly-free scalar perturbations with holonomy corrections in loop
  quantum cosmology,
\newblock {\em Class.\ Quant.\ Grav.} 29 (2012) 095010, [arXiv:1111.3535]

\bibitem{JR}
J.~D.\ Reyes,
\newblock {\em Spherically Symmetric Loop Quantum Gravity: Connections to
  2-Dimensional Models and Applications to Gravitational Collapse},
\newblock PhD thesis, The Pennsylvania State University, 2009

\bibitem{LTBII}
M.\ Bojowald, J.~D.\ Reyes, and R.\ Tibrewala,
\newblock Non-marginal LTB-like models with inverse triad corrections from loop
  quantum gravity,
\newblock {\em Phys.\ Rev.\ D} 80 (2009) 084002, [arXiv:0906.4767]

\bibitem{ThreeDeform}
A.\ Perez and D.\ Pranzetti,
\newblock On the regularization of the constraints algebra of Quantum Gravity
  in $2+1$ dimensions with non-vanishing cosmological constant,
\newblock {\em Class.\ Quantum Grav.} 27 (2010) 145009, [arXiv:1001.3292]

\bibitem{SymmRed}
M.\ Bojowald and H.~A.\ Kastrup,
\newblock Symmetry Reduction for Quantized Diffeomorphism Invariant Theories of
  Connections,
\newblock {\em Class.\ Quantum Grav.} 17 (2000) 3009--3043, [hep-th/9907042]

\bibitem{SphSymm}
M.\ Bojowald,
\newblock Spherically Symmetric Quantum Geometry: States and Basic Operators,
\newblock {\em Class.\ Quantum Grav.} 21 (2004) 3733--3753, [gr-qc/0407017]

\bibitem{SphSymmHam}
M.\ Bojowald and R.\ Swiderski,
\newblock Spherically Symmetric Quantum Geometry: Hamiltonian Constraint,
\newblock {\em Class.\ Quantum Grav.} 23 (2006) 2129--2154, [gr-qc/0511108]

\bibitem{EinsteinRosenAsh}
K.\ Banerjee and G.\ Date,
\newblock Loop quantization of polarized Gowdy model on $T^3$: Classical
  theory,
\newblock {\em Class.\ Quantum Grav.} 25 (2008) 105014, [arXiv:0712.0683]

\bibitem{EinsteinRosenQuant}
K.\ Banerjee and G.\ Date,
\newblock Loop Quantization of Polarized Gowdy Model on $T^3$: Quantum Theory,
\newblock {\em Class.\ Quantum Grav.} 25 (2008) 145004, [arXiv:0712.0687]

\bibitem{Action}
M.\ Bojowald and G.~M.\ Paily,
\newblock Deformed General Relativity and Effective Actions from Loop Quantum
  Gravity, [arXiv:1112.1899]

\bibitem{VectorHol}
J.\ Mielczarek, A.\ Cailleteau, Barrau, T.\, and J.\ Grain,
\newblock Anomaly-free vector perturbations with holonomy corrections in loop
  quantum cosmology,
\newblock {\em Class.\ Quant.\ Grav.} 29 (2012) 085009, [arXiv:1106.3744]

\bibitem{ScalarGaugeInv}
M.\ Bojowald, G.\ Hossain, M.\ Kagan, and S.\ Shankaranarayanan,
\newblock Gauge invariant cosmological perturbation equations with corrections
  from loop quantum gravity,
\newblock {\em Phys.\ Rev.\ D} 79 (2009) 043505, [arXiv:0811.1572]

\bibitem{HamGaugePert}
D.\ Langlois,
\newblock Hamiltonian formalism and gauge invariance for linear perturbations
  in inflation,
\newblock {\em Class.\ Quant.\ Grav.} 11 (1994) 389--407

\bibitem{Connes}
A.\ Connes,
\newblock Formule de trace en geometrie non commutative et hypothese de
  Riemann,
\newblock {\em C.R.\ Acad.\ Sci.\ Paris} 323 (1996) 1231--1235

\bibitem{Fractional}
G.\ Calcagni,
\newblock Fractal universe and quantum gravity,
\newblock {\em Phys.\ Rev.\ Lett.} 104 (2010) 251301, [arXiv:0912.3142]

\bibitem{TensorialDeform}
K.\ Giesel, F.~P.\ Schuller, C.\ Witte, and M.~N.~R.\ Wohlfahrt,
\newblock Gravitational dynamics for all tensorial spacetimes carrying
  predictive, interpretable and quantizable matter,
\newblock {\em Phys.\ Rev.\ D} 85 (2012) 104042, [arXiv:1202.2991]

\bibitem{ModifiedHorizon}
M.\ Bojowald, G.~M.\ Paily, J.~D.\ Reyes, and R.\ Tibrewala,
\newblock Black-hole horizons in modified space-time structures arising from
  canonical quantum gravity,
\newblock {\em Class.\ Quantum Grav.} 28 (2011) 185006, [arXiv:1105.1340]

\bibitem{LoopMuk}
M.\ Bojowald and G.\ Calcagni,
\newblock Inflationary observables in loop quantum cosmology,
\newblock {\em JCAP} 1103 (2011) 032, [arXiv:1011.2779]

\bibitem{ScalarHolMuk}
T.\ Cailleteau and A.\ Barrau,
\newblock Gauge invariance in Loop Quantum Cosmology : Hamilton-Jacobi and
  Mukhanov-Sasaki equations for scalar perturbations, [arXiv:1111.7192]

\bibitem{ScalarTensorHol}
T.\ Cailleteau, A.\ Barrau, J.\ Grain, and F.\ Vidotto,
\newblock Consistency of holonomy-corrected scalar, vector and tensor
  perturbations in Loop Quantum Cosmology, [arXiv:1206.6736]

\bibitem{SigChange}
J.\ Mielczarek,
\newblock Signature change in loop quantum cosmology, [arXiv:1207.4657]

\bibitem{NonHerm}
A.\ Komar,
\newblock Consistent Factor Ordering Of General Relativistic Constraints,
\newblock {\em Phys.\ Rev.\ D} 20 (1979) 830--833

\bibitem{BouncePot}
M.\ Bojowald, H.\ Hern\'andez, and A.\ Skirzewski,
\newblock Effective equations for isotropic quantum cosmology including matter,
\newblock {\em Phys.\ Rev.\ D} 76 (2007) 063511, [arXiv:0706.1057]

\bibitem{EffAcOneLoop}
A.~O.\ Barvinsky and A.~Yu.\ Kamenshchik,
\newblock One loop quantum cosmology: The Normalizability of the Hartle-Hawking
  wave function and the probability of inflation,
\newblock {\em Class.\ Quantum Grav.} 7 (1990) L181--L186

\bibitem{EffAcTunneling}
A.~O.\ Barvinsky and A.~Yu.\ Kamenshchik,
\newblock Tunneling geometries. 1. Analyticity, unitarity and instantons in
  quantum cosmology,
\newblock {\em Phys.\ Rev.\ D} 50 (1994) 5093--5114, [gr-qc/9311022]

\bibitem{EffAcInflation}
A.~O.\ Barvinsky and A.~Yu.\ Kamenshchik,
\newblock Quantum origin of the early universe and the energy scale of
  inflation,
\newblock {\em Int.\ J.\ Mod.\ Phys.\ D} 5 (1996) 825--844, [gr-qc/9510032]

\bibitem{FermionDecoherence}
A.~O.\ Barvinsky, A.~Yu.\ Kamenshchik, and C.\ Kiefer,
\newblock Effective action and decoherence by fermions in quantum cosmology,
\newblock {\em Nucl.\ Phys.\ B} 552 (1999) 420--444, [gr-qc/9901055]

\bibitem{CDTEffAc}
J.\ Ambj{\o}rn, A.\ Gorlich, J.\ Jurkiewicz, R.\ Loll, J.\ Gizbert-Studnicki,
  and T.\ Trzesniewski,
\newblock The Semiclassical Limit of Causal Dynamical Triangulations,
\newblock {\em Nucl.\ Phys.\ B} 849 (2011) 144--165, [arXiv:1102.3929]

\bibitem{Rov:Loops}
C.\ Rovelli,
\newblock Loop Quantum Gravity,
\newblock {\em Living Rev.\ Rel.} 1 (1998) 1, [gr-qc/9710008],
\newblock {\tt http://www.livingreviews.org/Articles/Volume1/1998-1rovelli}

\bibitem{SpinFoamRev}
A.\ Perez,
\newblock Spin Foam Models for Quantum Gravity,
\newblock {\em Class.\ Quantum Grav.} 20 (2003) R43, [gr-qc/0301113]

\bibitem{PerezLivRev}
A.\ Perez,
\newblock The Spin Foam Approach to Quantum Gravity, [arXiv:1205.2019]

\bibitem{SurfaceSum}
M.\ Reisenberger and C.\ Rovelli,
\newblock ``Sum over Surfaces'' form of Loop Quantum Gravity,
\newblock {\em Phys.\ Rev.\ D} 56 (1997) 3490--3508, [gr-qc/9612035]

\bibitem{Anomaly}
M.\ Bojowald and A.\ Perez,
\newblock Spin Foam Quantization and Anomalies,
\newblock {\em Gen.\ Rel.\ Grav.} 42 (2010) 877, [gr-qc/0303026]

\bibitem{EffAcSpinFoam}
A.\ Mikovic and M.\ Vojinovic,
\newblock Effective action and semiclassical limit of spin foam models,
\newblock {\em Class.\ Quantum Grav.} 28 (2011) 225004, [arXiv:1104.1384]

\bibitem{Regained}
S.~A.\ Hojman, K.\ Kucha\v{r}, and C.\ Teitelboim,
\newblock Geometrodynamics Regained,
\newblock {\em Ann.\ Phys.\ (New York)} 96 (1976) 88--135

\bibitem{LagrangianRegained}
K.~V.\ Kucha\v{r},
\newblock Geometrodynamics regained: A Lagrangian approach,
\newblock {\em J.\ Math.\ Phys.} 15 (1974) 708--715

\bibitem{ConsistDisc}
R.\ Gambini and J.\ Pullin,
\newblock Canonical quantization of general relativity in discrete space-times,
\newblock {\em Phys.\ Rev.\ Lett.} 90 (2003) 021301, [gr-qc/0206055]

\bibitem{UniformDisc}
M.\ Campiglia, C.\ Di~Bartolo, R.\ Gambini, and J.\ Pullin,
\newblock Uniform discretizations: a new approach for the quantization of
  totally constrained systems,
\newblock {\em Phys.\ Rev.\ D} 74 (2006) 124012, [gr-qc/0610023]

\bibitem{PerfectAction}
B.\ Bahr and B.\ Dittrich,
\newblock Improved and Perfect Actions in Discrete Gravity,
\newblock {\em Phys.\ Rev.\ D} 80 (2009) 124030, [arXiv:0907.4323]

\bibitem{CanSimp}
B.\ Dittrich and P.~A.\ H\"ohn,
\newblock Canonical simplicial gravity, [arXiv:1108.1974]

\bibitem{tensor}
M.\ Bojowald and G.\ Hossain,
\newblock Quantum gravity corrections to gravitational wave dispersion,
\newblock {\em Phys.\ Rev.\ D} 77 (2008) 023508, [arXiv:0709.2365]

\bibitem{BKL}
V.~A.\ Belinskii, I.~M.\ Khalatnikov, and E.~M.\ Lifschitz,
\newblock A general solution of the Einstein equations with a time singularity,
\newblock {\em Adv.\ Phys.} 13 (1982) 639--667

\bibitem{NoSing}
M.\ Bojowald and G.~M.\ Paily,
\newblock A no-singularity scenario in loop quantum gravity, [arXiv:1206.5765]

\bibitem{WdWCMB}
C.\ Kiefer and M.\ Kraemer,
\newblock Quantum Gravitational Contributions to the CMB Anisotropy Spectrum,
\newblock {\em Phys.\ Rev.\ Lett.} 108 (2012) 021301, [arXiv:1103.4967]

\bibitem{SuperInflTensor}
E.~J.\ Copeland, D.~J.\ Mulryne, N.~J.\ Nunes, and M.\ Shaeri,
\newblock The gravitational wave background from super-inflation in Loop
  Quantum Cosmology,
\newblock {\em Phys.\ Rev.\ D} 79 (2009) 023508, [arXiv:0810.0104]

\bibitem{CCPowerLoop}
G.\ Calcagni and M.~V.\ Cort\^es,
\newblock Inflationary scalar spectrum in loop quantum cosmology,
\newblock {\em Class.\ Quantum Grav.} 24 (2007) 829--853, [gr-qc/0607059]

\bibitem{TensorBackground}
J.\ Mielczarek and M.\ Szyd\l{}owski,
\newblock Relic gravitons as the observable for Loop Quantum Cosmology,
\newblock {\em Phys.\ Lett.\ B} 657 (2007) 20--26, [arXiv:0705.4449]

\bibitem{BounceTensor}
J.\ Mielczarek,
\newblock Gravitational waves from the Big Bounce,
\newblock {\em JCAP} 0811 (2008) 011, [arXiv:0807.0712]

\bibitem{SuperInflPowerSpec}
M.\ Shimano and T.\ Harada,
\newblock Observational constraints of a power spectrum from super-inflation in
  Loop Quantum Cosmology,
\newblock {\em Phys.\ Rev.\ D} 80 (2009) 063538, [arXiv:0909.0334]

\bibitem{TensorHalfII}
A.\ Barrau and J.\ Grain,
\newblock Cosmological footprint of loop quantum gravity,
\newblock {\em Phys.\ Rev.\ Lett.} 102 (2009) 081301, [arXiv:0902.0145]

\bibitem{TensorSlowRollInv}
J.\ Grain, A.\ Barrau, and A.\ Gorecki,
\newblock Inverse volume corrections from loop quantum gravity and the
  primordial tensor power spectrum in slow-roll inflation,
\newblock {\em Phys.\ Rev.\ D} 79 (2009) 084015, [arXiv:0902.3605]

\bibitem{TensorSlowRoll}
J.\ Grain, T.\ Cailleteau, A.\ Barrau, and A.\ Gorecki,
\newblock Fully LQC-corrected propagation of gravitational waves during
  slow-roll inflation,
\newblock {\em Phys.\ Rev.\ D} 81 (2010) 024040, [arXiv:0910.2892]

\bibitem{GravWaveLQC}
J.\ Mielczarek, T.\ Cailleteau, J.\ Grain, and A.\ Barrau,
\newblock Inflation in loop quantum cosmology: Dynamics and spectrum of
  gravitational waves,
\newblock {\em Phys.\ Rev.\ D} 81 (2010) 104049, [arXiv:1003.4660]

\bibitem{InflConsist}
M.\ Bojowald, G.\ Calcagni, and S.\ Tsujikawa,
\newblock Observational constraints on loop quantum cosmology,
\newblock {\em Phys.\ Rev.\ Lett.} 107 (2011) 211302, [arXiv:1101.5391]

\bibitem{NonGaussInvVol}
L.-F.\ Li, R.-G.\ Cai, Z.-K.\ Guo, and B.\ Hu,
\newblock Non-Gaussian features from the inverse volume corrections in loop
  quantum cosmology, [arXiv:1112.2785]

\bibitem{GCPheno}
G.\ Calcagni,
\newblock Observational Effects from Quantum Cosmology,
\newblock {\em Annalen Phys.} (2012) to appear, [arXiv:1209.0473]

\end{thebibliography}

\end{document}